%% file: main.tex
\documentclass[10pt,letterpaper]{article}
\usepackage{specguard-report}
\input{authors}
\hypersetup{
  pdftitle={The Revision Is the Evidence: Evolution-Aware Retrieval for Auditing 5G Security Standards},
  pdfauthor={Guangyuan Dong; Zidong Yu; Guohao Qi; Zhengqi Feng; Han Yang Alexander Lim; Diqing Xu; Bingchen Liu; Jinxu Zhang; Zhen Tian; Bohan Li},
  pdfsubject={SpecGuard technical report},
  pdfkeywords={SpecGuard, 3GPP, evolution-aware retrieval, temporal knowledge graphs, security auditing}
}

\begin{document}
\twocolumn[{\begin{@twocolumnfalse}
\begin{minipage}{\textwidth}
  \vspace*{-25.819pt}
  \noindent\input{logo-row}\par
  \vspace{6pt}{\color{ReportBlue}\hrule height 0.5pt}
  \vspace{6mm}
  {\centering\sffamily\bfseries\fontsize{17}{20}\selectfont
    The Revision Is the Evidence: Evolution-Aware Retrieval\\
    for Auditing 5G Security Standards\par}
  \vspace{6mm}{\color{ReportBlue}\hrule height 0.5pt}
  \vspace{6mm}\vspace{0.35pt}
  {\centering\sffamily\bfseries\fontsize{10}{12}\selectfont\reportauthors\par}
  \vspace{3mm}
  {\centering\sffamily\mdseries\fontsize{10}{12}\selectfont\reportaffiliations\par}
  \vspace{9mm}
  \begin{tcolorbox}[colback=AbstractGray,colframe=AbstractGray,
    boxrule=0pt,arc=8pt,left=17pt,right=17pt,top=11pt,bottom=10pt]
    {\centering\sffamily\bfseries\fontsize{12}{12}\selectfont\color{ReportBlue} Abstract\par}
    \vspace{7pt}
    {\fontsize{9.5}{11.5}\selectfont\input{sections/abstract}\par}
    \vspace{8pt}
    {\sffamily\fontsize{8}{10}\selectfont
      \textbf{Keywords:} Evolution-aware retrieval; temporal knowledge graphs; retrieval-augmented generation; benchmark construction; security protocol analysis.\par
      \vspace{5pt}
      \textbf{Contact:} \href{mailto:guangyuand@acm.org}{\texttt{guangyuand@acm.org}}\quad
      \href{mailto:bhli@nuaa.edu.cn}{\texttt{bhli@nuaa.edu.cn}}\par
      \vspace{4pt}\textbf{Date:} November, 2026\par}
  \end{tcolorbox}
  \vspace{8pt}
\end{minipage}
\end{@twocolumnfalse}}]
\begin{NoHyper}
\begingroup
\renewcommand{\thefootnote}{*}\footnotetext{Equal contribution.}
\renewcommand{\thefootnote}{\textdagger}\footnotetext{Corresponding authors.}
\endgroup
\end{NoHyper}
\setcounter{footnote}{0}

\input{sections/body}

\bibliographystyle{unsrtnat}
\bibliography{references}
\balance

\onecolumn
\appendix
\setcounter{table}{0}
\setcounter{figure}{0}
\setcounter{equation}{0}
\setcounter{algorithm}{0}
\renewcommand{\thetable}{A\arabic{table}}
\renewcommand{\thefigure}{A\arabic{figure}}
\renewcommand{\theequation}{A\arabic{equation}}
\renewcommand{\thealgorithm}{A\arabic{algorithm}}
\renewcommand{\theHtable}{appendix.\arabic{table}}
\renewcommand{\theHfigure}{appendix.\arabic{figure}}
\renewcommand{\theHequation}{appendix.\arabic{equation}}
\section*{Appendices}
\addcontentsline{toc}{section}{Appendices}
\input{sections/appendix_intro}
\input{sections/appendices}

\nocite{hou2026riskediting,hou2026aiotrecognition,11301014}
\nocite{11130528,zhao2026seal}
\nocite{chen2026port}
\nocite{zhao2025poisoning}
\nocite{liu2026should,liu2025toward,liu2024tsla}
\nocite{liu2025rora,liu2026structured,liu2023scalable}
\nocite{liu2023interpretable,liu2022efficient,liu2022explainable}
\end{document}

%% file: authors.tex
\newcommand{\reportauthors}{Guangyuan Dong$^{1,*}$, Zidong Yu$^{2,*}$, Guohao Qi$^{1}$, Zhengqi Feng$^{1}$, Han Yang Alexander Lim$^{1}$,\\
Diqing Xu$^{1}$, Bingchen Liu$^{3,\dagger}$, Jinxu Zhang$^{4,\dagger}$, Zhen Tian$^{5,\dagger}$, Bohan Li$^{6,\dagger}$}
\newcommand{\reportaffiliations}{$^{1}$National University of Singapore, $^{2}$Syracuse University,\\
$^{3}$Shandong University, $^{4}$Nanyang Technological University,\\
$^{5}$University of Glasgow, $^{6}$Nanjing University of Aeronautics and Astronautics}

%% file: logo-row.tex
\newcommand{\reportlogo}[2][]{\raisebox{-0.5\height}{\includegraphics[#1]{#2}}}%
\noindent\mbox{%
\reportlogo[width=62pt,trim=42.75bp 81bp 47.25bp 111bp,clip]{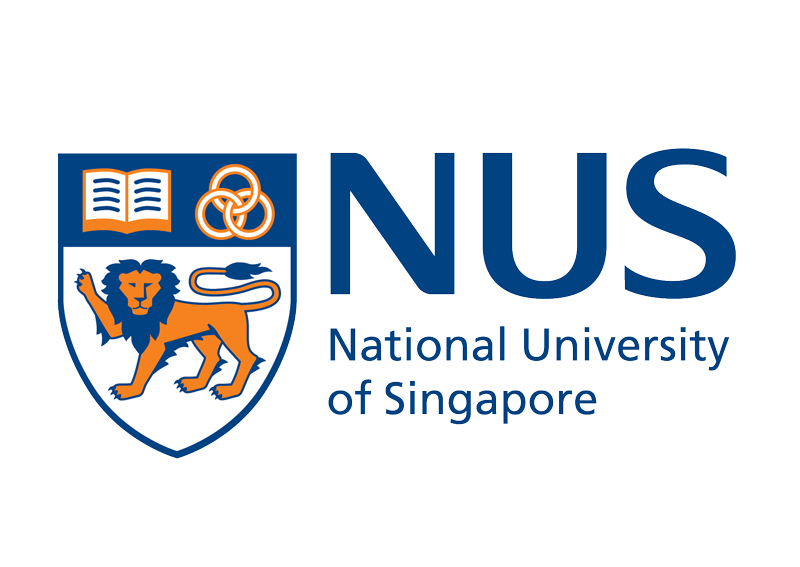}\hspace{4pt}%
\reportlogo[width=60pt]{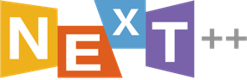}\hspace{4pt}%
\reportlogo[width=78pt]{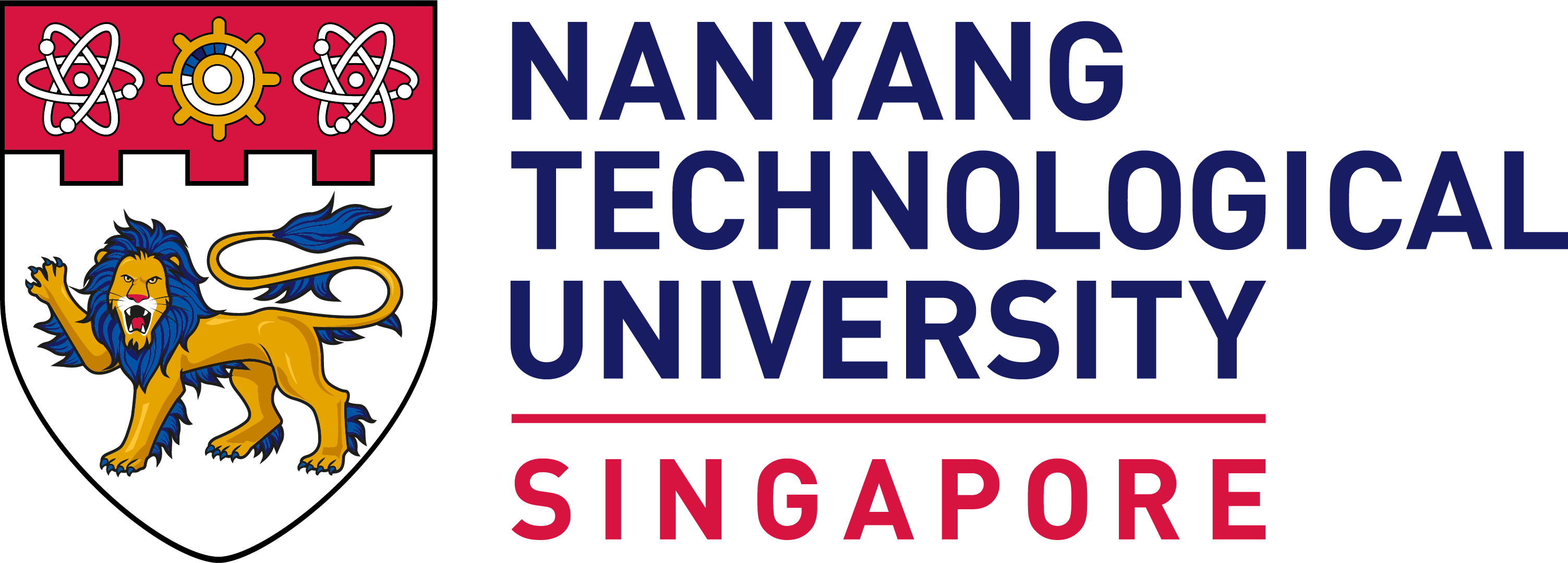}\hspace{4pt}%
\reportlogo[height=28pt]{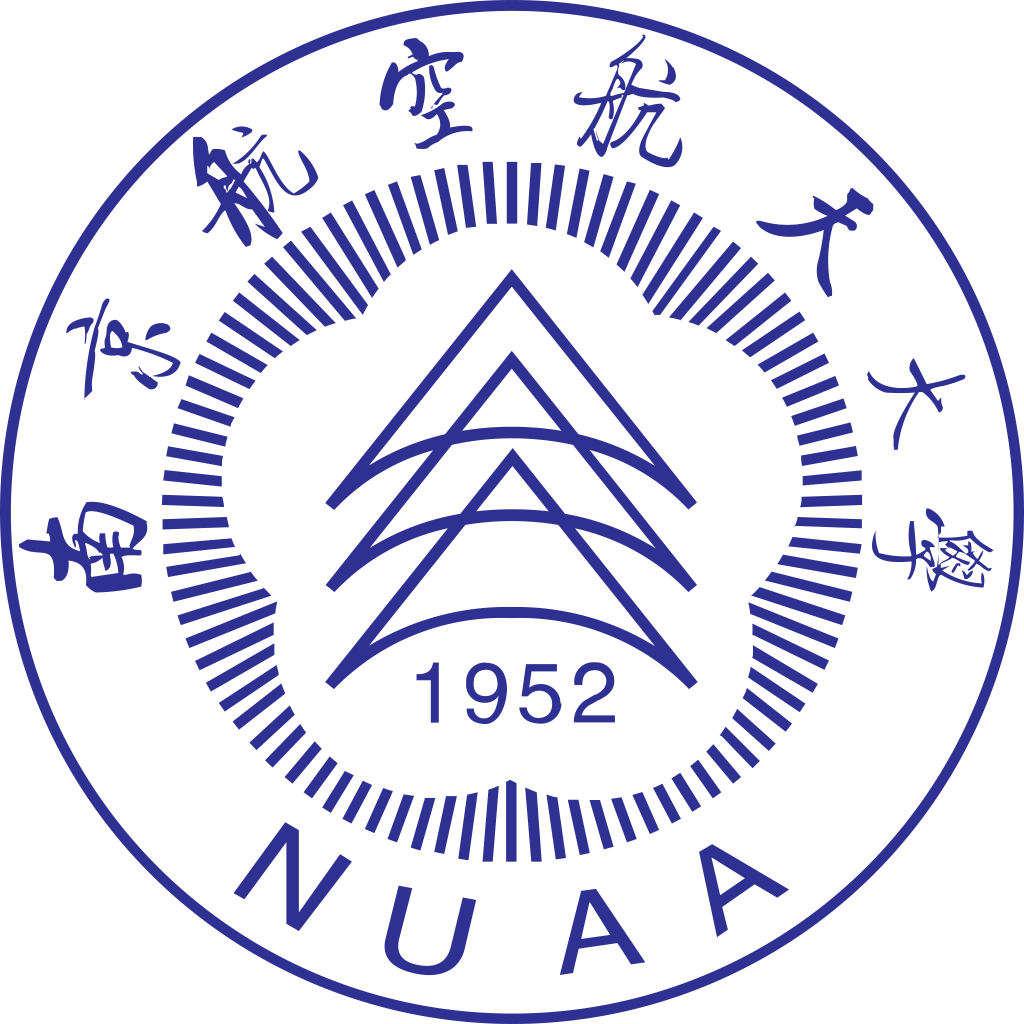}\hspace{4pt}%
\reportlogo[height=28pt]{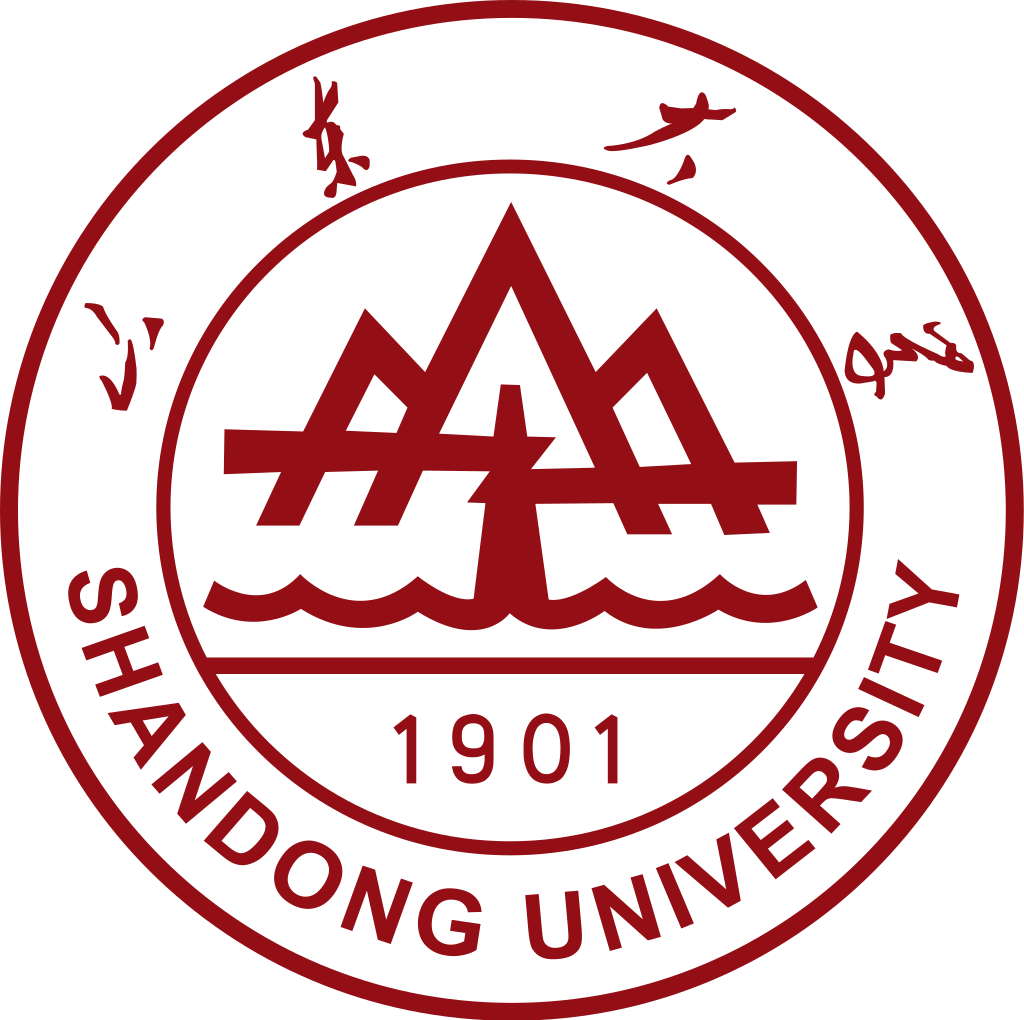}\hspace{4pt}%
\reportlogo[width=100pt]{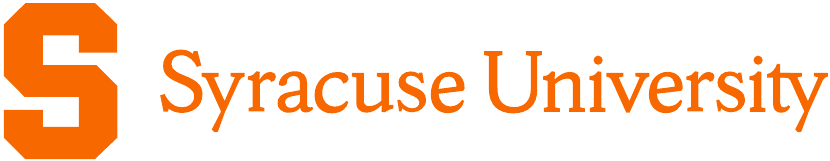}\hspace{4pt}%
\reportlogo[width=90pt,trim=39.84bp 39.84bp 39.84bp 39.36bp,clip]{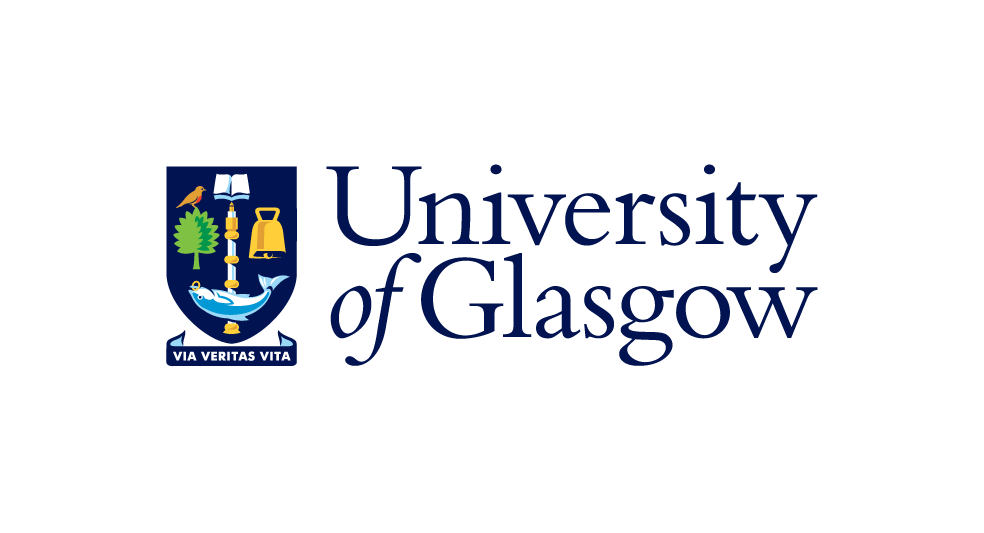}%
}

%% file: sections/abstract.tex
Standards are versioned corpora: every change to a 3GPP specification is a public Change Request stating what changed, why, and which clauses it touches. An audit asks which clause version holds at a date, what a revision touched, and which text supports each claim; document retrieval cannot tell superseded from current versions. We treat the revision history as the evidence. \textbf{3GPP-EvoCorp} organizes 198,553 parsed Change Requests, 26,932 of them security-relevant, into an audited bitemporal Protocol Evolutionary Graph with 351K nodes and 1.04M typed edges, with three expert-checked benchmarks. \textbf{SpecGuard} retrieves, classifies and traverses the graph under a snapshot frozen at the query date and writes audits in which every claim cites a Change Request or clause version and rules stated as in force are checked against the version current then. Across 39 configurations and four backbones under chronological splits, it lifts structural exact match from 0.672 to 0.752 and cuts unsupported claims from 9.4 to 6.2 percent against the strongest baseline, and most of this gain transfers to other retrievers through the revision-chain organisation. In a retrospective, anchor-consistent check on CREEK's public sets it recovers 12 and explains 11 of 14 weaknesses confirmed in commercial networks.

%% file: sections/body.tex
\section{Introduction}
\label{sec:intro}
Many corpora are \emph{versioned}: they change through explicit revision events whose history is itself evidence. This paper studies the clearest large-scale case, the 3GPP mobile-network standards. Specifications are maintained by working groups (SA3 security, CT1 signalling, RAN2 radio), versioned per \emph{Release}, and every modification is a \emph{Change Request} (CR): a temporary document (tdoc) agreed in a working group and approved at a TSG plenary that records the reason for the change, the consequences if it is not approved, the affected clauses and the version step. Security CRs often close a weakness long before the field catches up. For the 14 weaknesses that CREEK~\cite{chen2022seeing} confirmed in commercial systems, CREEK's minimum estimates put the weakness in the field for 0.7--8.7 years after the fix (median 5.6), and six were still present in CREEK's 2021 tests, up to 11.7 years after the update (Table~\ref{tab:cases}).

\begin{figure*}[!tb]
\centering
\includegraphics[width=\textwidth]{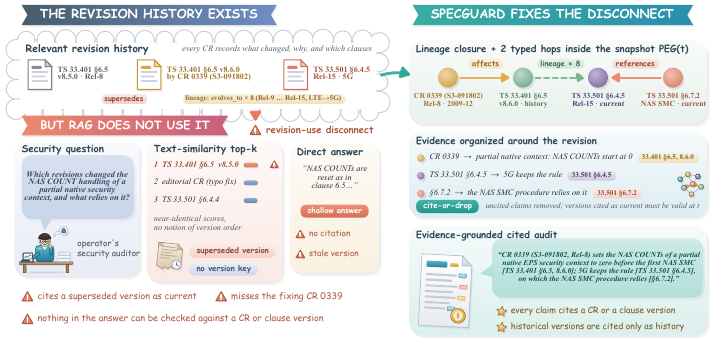}
\Description{Left: the revision history exists but text-similarity RAG retrieves a superseded clause version and answers without a checkable citation. Right: SpecGuard reaches the 5G clause through the clause lineage and two typed hops on the Protocol Evolutionary Graph and emits a cited audit in which historical versions are cited only as history.}
\caption{Motivating example, CREEK case 2: tdoc S3-091802 = TS~33.401 CR~0339 (Rel-8, Dec.\ 2009; Appendix~\ref{s:case}) fixes the NAS COUNT initialisation of a \emph{partial native} EPS security context (\S6.5); 5G inherits the rule (TS~33.501 \S6.4.5), on which the NAS security mode command (\S6.7.2) relies. \textbf{Left:} text-similarity RAG retrieves the superseded version and cites nothing checkable. \textbf{Right:} \sys reaches the 5G clause through the clause's \emph{lineage} (Def.~\ref{def:task}) and two typed hops and cites each version with its role.}
\label{fig:motivation}
\end{figure*}

\begin{figure*}[!tb]
\centering
\includegraphics[width=\textwidth]{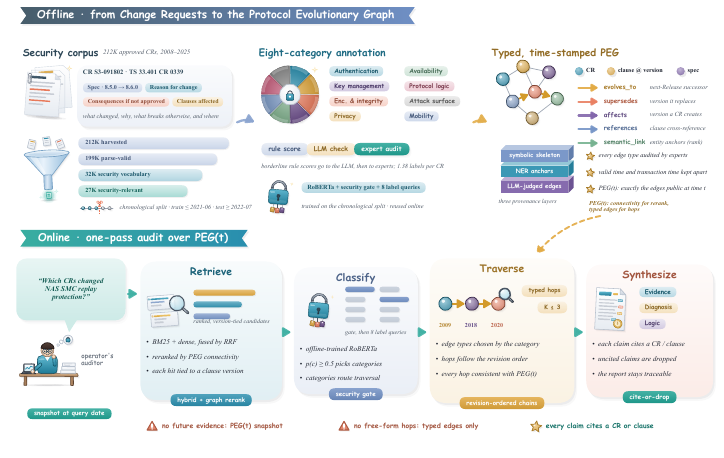}
\Description{Design view: offline construction of the Protocol Evolutionary Graph from 3GPP CRs; online one-pass retrieve, classify, traverse, synthesize.}
\caption{Design view of \sys. \textbf{Offline:} CRs are parsed, filtered, labelled and organised into the audited \peg. \textbf{Online:} one pass retrieves, classifies, traverses typed edges in revision order under the snapshot and synthesises a cited report under cite-or-drop (\S\ref{sec:method}).}
\label{fig:framework}
\end{figure*}

A security question about a clause (Fig.~\ref{fig:motivation}) asks \textbf{\emph{which version holds at the query date}}, \textbf{\emph{what else the change touched}} and \textbf{\emph{whether each claim can be checked}}. Retrieval over independent documents cannot tell a superseded version from its replacement, time-aware retrievers rank by date without modelling supersession, and graph-RAG compresses cross-clause links into summaries that drop clause numbers and version boundaries~\cite{edge2024local,guo2024lightrag}. \thesis{The evidence for a security question is distributed over a revision graph, not a document set.} We therefore make the revision history a traversable bitemporal object. \corpus builds an audited \emph{Protocol Evolutionary Graph} (\peg) over 198,553 parsed CRs, 26,932 of them security-relevant (\S\ref{sec:peg}), and \sys (Fig.~\ref{fig:framework}) composes, from standard components, hybrid retrieval reranked by graph connectivity, multi-label classification, typed traversal in revision order and synthesis under a \emph{cite-or-drop} rule. Retrieval here includes the traversal that assembles the evidence; most of the end-to-end gain comes from how that evidence is organised and version-checked, and the organisation transfers to other retrievers. Every system sees a snapshot frozen at the query date, and writer, answerer and judge are separate models.

\noindent\textbf{Contributions.} \contrib\textbf{Resource:} \corpus, an evolution-aware, audited corpus and bitemporal graph with three benchmarks and expert-written held-out slices. \contrib\textbf{Problem and method:} \emph{snapshot-consistent, version-aware auditing}, with current/historical roles, lineages, chains and run-time-enforced properties measured at revision boundaries, and \sys, standard components plus a version-checked citation rule. \contrib\textbf{Evidence:} across 39 configurations and four backbones, revision-graph grounding improves structural QA, faithfulness and blinded expert judgment over graph-RAG, temporal graph-RAG, telecom RAG and long-context readers, improves retrieval over learned rerankers on PEG-derived, judgment-independent and expert-written queries, and leaves token overlap nearly unchanged; organising any retriever's evidence as revision chains captures most of the gain, and the full system keeps a smaller, significant margin beyond it. A retrospective, anchor-consistent check on CREEK's public sets recovers 22/25 expert-labelled CRs and 12/14 confirmed weaknesses and explains 11.

\section{Related Work}
\label{sec:related}
Time-travel search indexes document versions~\cite{berberich2007time}; bitemporal databases~\cite{jensen1999temporal} and temporal graph systems~\cite{debrouvier2021temporal} supply the valid-time/transaction-time and branching semantics we adopt. Time-sensitive QA~\cite{chen2021timeqa} and temporal KGQA~\cite{saxena2021cronkgqa,mavromatis2022tempoqr} track time over curated graphs; TempRetriever~\cite{abdallah2026tempretriever} fuses time into dense retrieval. Graph-RAG builds entity or community graphs~\cite{edge2024local,gutierrez2025hipporag2,guo2024lightrag}; T-GRAG~\cite{li2025tgrag} time-stamps extracted facts; Graphiti~\cite{rasmussen2025zep} keeps a bitemporal graph of extracted facts with graph-distance reranking. RACER uses reinforcement learning to select reasoning paths and combines evidence from multiple paths for knowledge-graph question answering~\cite{lou2026racer}. Our departure is representational: \peg's nodes are clause versions and its edges normative relations fixed by the standard, so a hop is a citable revision, not an extracted fact. Telecom benchmarks~\cite{maatouk2023teleqna,karim2023spec5g} and RAG systems~\cite{nikbakht2024tspec,yilma2024telecomrag,bornea2024telcorag} (Telco-RAG is compared here) target one specification state. In cellular security, formal and testing approaches~\cite{hussain2018lteinspector,al2024hermes} and specification mining~\cite{rahman2024cellularlint} need hand-built models; CREEK~\cite{chen2022seeing} mines security CRs and supplies our external check; CR-level refinement~\cite{llmcr2025} treats over 200K approved CRs as standalone documents; CellularSpecSec-Bench~\cite{cellularspecsec2026} pairs a benchmark with a taxonomy graph carrying clause provenance; none models version lineage, Release semantics or revision edges.

In sequential state tracking, PAR supervises each intermediate prediction against the state reached after a known sequence of operations~\cite{wang2026par}. Agent memory raises a related question about the history of a record. MemGuard retains verifier judgments to resolve stale or conflicting records~\cite{wang2026memguard}, while MEMMARK encodes ownership in memory writes so that an agent's snapshot can later be attributed~\cite{zhang2026memmark}. In 3GPP the public change request identifies the update that introduced a clause version.

Retrieval systems differ in what a representation must preserve. DeepInterestGR uses interests inferred from reviews to retrieve items~\cite{zeng2026deepinterestgr}. Xu et al. address the imbalance between an image and its caption in cross-modal retrieval~\cite{xu2026lever}, while PriMD separates shared cues from modality-specific details when an input is missing in emotion recognition~\cite{zhang2026primd}. Two image-fusion studies check what survives generation. CERES re-indexes the output to test which source concepts remain retrievable~\cite{dong2026ceres}, and LANCE compares the output with a structured description of the inputs~\cite{dong2026lance}. For a standards audit, preserving the wording is insufficient if the generated claim cites a superseded clause.

RCR-Router selects shared-memory items for each agent according to its role and token budget~\cite{liu-etal-2026-rcr}. In \sys, the budgeted unit is a change record or clause version, together with the links that establish which rule was in force.

\section{3GPP-EvoCorp: Corpus, Graph, and Benchmarks}
\label{sec:dataset}

\subsection{From Change Requests to a Security Corpus}
\label{sec:corpus}
We harvested \textbf{212,406} approved CRs from the official 3GPP CR database (2008 to 2025-06-30; Appendix~\ref{s:corpus}); removing 3,120 superseded revisions and 10,733 unparsable covers leaves 198,553 parse-valid CRs, whose cover fields $T_{\text{struct}}$ are read from rendered cover pages with Qwen3-VL-30B (98.3\% field accuracy), followed by the redlined text. A CR is \emph{security-relevant} if its \emph{Reason}/\emph{Consequences} fields invoke the TS~33-series security vocabulary (412 terms) and the adjudication of \S\ref{sec:annotation} confirms it: 31,864 candidates, \textbf{26,932} confirmed (1.5K per year on average). Filter recall is 0.956 (95\% CI 0.91--0.98; 6 of 800 expert-labelled discards are security-relevant). As a coverage check, 1,181 (93.0\%) of the 1,270 CRs that CREEK's classifier labelled security-relevant at confidence above 0.99 are among ours, a share that the set's own reported precision of 91.6\% bounds. Mirror CRs (27.1\% of security CRs) are separate nodes grouped into families. With 7,445 hard negatives (4,932 rejected candidates, 2,513 near-misses) the 34,377-CR classification corpus is split \emph{chronologically} by approval date: 26,431 CRs up to 2021-06-30 (train), 2,118 up to 2022-06-30 (validation), 5,828 after (test); the 412 families straddling a boundary follow their latest member (626 earlier members withheld from training and query anchors), so no training CR postdates the cut-off or has a mirror in validation or test.

\subsection{Eight-Category Multi-Label Annotation}
\label{sec:annotation}
Eight categories grounded in TS~33 (Protocol Logic \& State Machine, Authentication, Key Management, Encryption \& Integrity, Privacy, Service Availability (DoS), Attack Surface \& Threats, Mobility Security; Appendix~\ref{s:algo}) cover all but 0.6\% of security CRs (resolved by a second expert pass), with 1.38 labels per CR. A weighted-lexicon rule score over the \emph{Reason} and \emph{Consequences} fields (term weights $\omega_w\in\{1,2,3\}$ set by three experts, ordinal $\alpha=0.79$) accepts at $\ge 0.5$ and rejects at $\le 0.3$; the ambiguous band (21.4\% of decisions) is adjudicated by Qwen3-8B~\cite{yang2025qwen3} with a category-aware template (Appendix~\ref{s:corpus}). On 1,500 test-period CRs re-annotated by three experts (Fleiss' $\kappa=0.83$), automated labels reach micro-F1 0.89.

\subsection{Protocol Evolutionary Graph (PEG)}
\label{sec:peg}
\peg's 351K nodes are 198,553 CR nodes (keyed by tdoc and CR number), 121K clause-version nodes, 31K security-entity, 112 specification and 17 Release nodes. Clause-version nodes cover the 112 specifications that carry at least one security CR (all 26,932 lie in them): one node per (specification, clause, Release, version) that some CR touches, plus the copy created when a Release is opened; the 101,204 CRs of these specifications carry the \texttt{affects} edges (2.0 per CR, 1.6 of them deterministic from the listed clauses), while the 97,349 CRs of the other 1,138 specifications are retrievable nodes without edges, never traversed. Edges are directed and typed: \texttt{affects} (CR$\rightarrow$the clause version it \emph{creates}; deterministic for the listed clauses, LLM-inferred when the body names another clause), \texttt{supersedes} (version$\rightarrow$the version of the same clause and Release it replaces: new to old), \texttt{evolves\_to} (version$\rightarrow$the earliest version of the same clause in the \emph{next} Release approved on or after it, i.e., the Release-opening copy or a mirror CR: old to new; plus 9K LTE$\rightarrow$5G edges such as TS~33.401 \S6.5$\rightarrow$TS~33.501 \S6.4.5, proposed by the 5G clause's explicit cross-reference to its LTE counterpart or by text similarity and confirmed by the succession prompt; LLM-inferred, 60 audited at precision 0.950, provenance in Appendix~\ref{s:peg}), \texttt{references} (explicit cross-reference) and \texttt{semantic\_link} (text$\rightarrow$security entity, entity co-reference). A version has at most one \texttt{supersedes} edge (none for the 24.3K segment-opening versions) and one same-specification \texttt{evolves\_to} edge (none for 21.5K versions: the lineage ends in that Release, the next Release is not yet open, or a late correction postdates every version of the next Release without a mirror CR), which with the 9K cross-specification edges reconciles 121K nodes with 97K \texttt{supersedes} and 109K \texttt{evolves\_to} edges; $\tau$ (Def.~\ref{def:peg}) is non-decreasing along \texttt{evolves\_to} and non-increasing along \texttt{supersedes}. The 7,650 lineages span 3.2 Release segments and 15.9 versions on average (p90 36, maximum 118). CRs are public from their working-group meeting, before approval, so $\mathcal{G}_{\le t}$ never contains information unpublished at $t$.

Construction is coarse to fine: (1)~a \emph{symbolic skeleton} (Releases, version keys, publication dates, listed \texttt{affects}, successions, explicit \texttt{references}) extracted deterministically; (2)~a RoBERTa-base NER model (six entity types, F1 0.912); (3)~Qwen3-32B judging candidate \emph{implicit} edges proposed by text, identifier and embedding similarity (Appendix~\ref{s:peg}). The 1.04M edges are \texttt{references} 549K, \texttt{affects} 206K (44K inferred), \texttt{evolves\_to} 109K (31K, including the 9K cross-specification edges), \texttt{supersedes} 97K (9K) and \texttt{semantic\_link} 79K (inferred). \textbf{Audit.} Three experts labelled 1,200 edges stratified by type and provenance (Fleiss' $\kappa=0.81$): precision 0.873--0.985 by type, micro \textbf{0.937} (symbolic edges 0.969, LLM-inferred 0.882; Appendix~\ref{s:peg}). Against the lineage and \texttt{affects} subgraph that experts reconstructed among the security-relevant clauses of the five most CR-dense specifications (2,145 gold edges), \peg's recall is \textbf{0.893}, and 0.871 on five random specifications; the ten hold 48\% of the clause-version nodes.

\subsection{Benchmarks with Expert-Written Held-Out Slices}
\label{sec:benchmarks}
Each benchmark pairs LLM-generated queries with an \emph{expert-written held-out slice} by three experts independent of the authors who saw CRs and specifications but \emph{not} \peg or any prompt (Appendix~\ref{s:bench}); queries are split by the approval date of their latest gold CR, and all numbers use the test splits (719 / 1,240 / 1,129 queries). Every query carries its audit date $t_q$ and target Release $R_q$, given to every system in a fixed header line (generated queries: 30--180 days after the latest gold CR's approval, and its Release). \textbf{5G-Retrieval} (300 comparative, 175 analytical, 150 temporal, 94 factual; +300 expert-written): GPT-5 writes each query from a \emph{set} of 1--5 gold CRs linked by shared clauses or evolution edges, with graded relevance (Appendix~\ref{s:bench}). Because generated queries and gold sets inherit \peg structure, we also collect \emph{judgment-independent pooled relevance} from the top-20 of every other retriever (two engineers without \peg access, $\kappa=0.81$), and treat the 300 expert-written queries, graded by their authors over shuffled pools from every retriever (31 CRs per query; 100 queries double-graded, $\kappa=0.79$), as the evidence that inherits nothing from \peg (Appendix~\ref{s:bench}). \textbf{5G-KGQA} (362 temporal, 314 relation, 258 multi-hop, 306 entity-linking; +150 expert-written): subgraphs with 2--4 cross-lineage edges and lineage segments of 1--11 Releases are sampled from \peg, compiled to SPARQL, executed and paraphrased by GPT-5; EM is set equality after alias normalisation. Because this gold inherits \peg's errors, \emph{every} test question was re-verified by two annotators against the specifications without \peg (1,194 of 1,240 agree, 46 corrected); all EM values use the verified gold, which lowers \sys by 1.9 points and the strongest baseline by 0.4, the measured size of the circularity. \textbf{5G-SecQA} (243 factual, 567 causal, 319 counterfactual; +100 expert-written) evaluates end-to-end audit synthesis: GPT-5 writes questions and 80-word references from the CR text and the cited clauses only (never from \peg chains); a counterfactual reference restates the one-line \emph{Consequences} field in terms of the affected procedure, so it cannot be copied from the cover; two senior engineers review all items double-blind ($\kappa=0.78$); writer, answerer (Qwen3-8B), judge (DeepSeek-V3.1) and embedder (BGE-M3) are distinct.

\section{Problem Definition and the SpecGuard Framework}
\label{sec:method}
\begin{defn}[Bitemporal Protocol Evolutionary Graph]\label{def:peg}
$\mathcal{G}=(V,E,\tau,\iota)$ is the typed graph of \S\ref{sec:peg}. The \emph{transaction time} $\tau(x)$ is the plenary approval date of the CR that created $x$ (other node types in Appendix~\ref{s:algo}). The \emph{valid time} $\iota(v)=[p(v),p(v'))$ of a clause version $v$ opens when the specification version carrying $v$ is published and closes when the next version $v'$ of the same clause and Release is published; $p(v)\ge\tau(v)$ (median lag 24 days), 11.8\% of versions are retroactive corrections to a frozen Release, and a newer Release never closes an older one's interval. The \emph{snapshot} $\mathcal{G}_{\le t}$ is induced by $\{x:\tau(x)\le t\}$ with valid-time endpoints later than $t$ masked to open, so it reveals nothing approved or published after $t$.
\end{defn}
\begin{defn}[Lineage, chain, auditing task]\label{def:task}
The \emph{lineage} $L(c)$ of a clause $c$ is the set of its versions across Releases and in its successor specification with the \texttt{supersedes} edges inside each Release and the \texttt{evolves\_to} edges between Releases; because Releases are maintained in parallel, a lineage is a branching graph whose branches are Release segments, not a path. A \emph{revision chain} in $\mathcal{G}_{\le t}$ is a connected subgraph formed by the versions of one lineage inside the snapshot with the \texttt{supersedes} edges of each segment and one \texttt{evolves\_to} edge into the opening version of each later segment, the \texttt{affects} edges of the CRs that created those versions and at most one \texttt{references} edge at either end; it is emitted segment by segment, each segment's versions in $\tau$ order, every version annotated with its creating CRs (Appendix~\ref{s:algo}). For a query $q=(\text{text},t_q,R_q)$ a version $v$ is \emph{current} iff $\mathrm{rel}(v)=R_q$ and $t_q\in\iota(v)$; every other version in $\mathcal{G}_{\le t_q}$, including the still-open last version of an older Release, is \emph{historical}, and each clause has at most one current version. \textbf{Snapshot-consistent version-aware auditing} returns (i) the current versions relevant to $q$, (ii) chains in $\mathcal{G}_{\le t_q}$ connecting them to the CRs that created them or their predecessors, and (iii) a report whose every claim cites an element of (i) or (ii) with its version key, historical versions only in their revision role, under \textbf{P1} (snapshot consistency: nothing cited or shown to the generator has $\tau>t_q$ or an endpoint after $t_q$), \textbf{P2} (order preservation: every edge of an emitted chain is an edge of $\mathcal{G}_{\le t_q}$ of the stated type and every segment is listed in non-decreasing $\tau$) and \textbf{P3} (version validity: a claim tagged \cur cites, for every clause it names, a version of that clause current for $q$; a \cur claim naming no clause is retained only if every version its citations resolve to is current).
\end{defn}
\noindent $(t_q,R_q)$ is how a deployed specification state enters the task (the Rel-8 state of October 2009: $R_q=$ Rel-8, $t_q=2009$-10); vendor behaviour outside the standard is out of scope. P1 and P2 hold by construction and are verified mechanically on every emitted chain; P3 is the substantive guarantee, measured at revision boundaries (\S\ref{sec:kgqa_exp}). \peg lives in a property-graph store (Neo4j~5) with $\tau$ and $\iota$ as properties, so a snapshot is a predicate evaluated during traversal; the text indices apply $\tau\le t_q$ \emph{before} scoring, so the candidate set equals that of an index frozen at $t_q$ (Appendix~\ref{s:algo}).

\para{Retrieve and classify.}
The online index covers all 198,553 CRs; Reciprocal Rank Fusion of BM25 and dense (bge-large-en-v1.5) ranks ($k=60$) gives $\mathrm{score}_{\text{hybrid}}$, and non-security candidates are rejected by the gate below. A RoBERTa-base encoder reads cover and body in 512-token chunks (stride 128) and max-pools token states; a security gate $p_{\text{sec}}=\sigma(\mathrm{MLP}(\mathbf{h}_{[\mathrm{CLS}]}))$ filters editorial changes, and eight \emph{learnable label queries} attend over the token states to give one vector per category, trained with the gate loss, per-label BCE and a Jaccard-weighted multi-label supervised contrastive loss (Appendix~\ref{s:bench}); every encoder baseline receives the identical chunked input. ARBOR uses specialty and question-type labels to select low-rank adapter components for medical QA~\cite{dong2026arbor}. In \sys, the predicted labels determine which relations the traversal can follow.

\para{Traverse and rerank.}
Traversal (pseudo-code in Appendix~\ref{s:algo}) starts from the lineages of the gated hits (at most $B$ seeds; a CR seed contributes the lineages it \texttt{affects}) and runs $K{=}3$ hops. Predicted categories with $p(c)\ge0.5$ (top-1 if none) share the per-hop budget $B{=}50$ by largest-remainder apportionment, each selecting an edge-type allowlist and entity filter (Appendix~\ref{s:algo}); a hop follows the cross-lineage \texttt{affects} and \texttt{references} edges of the current lineages in both directions, keeps targets with $\tau\le t_q$ (P1), scores each candidate by $p(c)\cdot\mathrm{sim}(w,q)$ and retains the top $b_c$ per category (at most $B$ lineages per hop), each emitted as a chain (Def.~\ref{def:task}). Lineages are expanded at no hop cost, so $K$ counts cross-lineage hops (Fig.~\ref{fig:motivation}: two hops across eight \texttt{evolves\_to} edges). The traversal scans at most $B\,K\,d_{\max}$ cross-lineage edges (1,840 per query on average; Appendix~\ref{s:algo}). Connectivity then reranks the evidence:
\begin{equation}
\begin{aligned}
S_{\text{final}}(d)={}&\alpha\cdot\mathrm{score}_{\text{hybrid}}(d)\\
&+(1-\alpha)\cdot\mathrm{Conn}_{\mathrm{PEG}}(d,q),
\end{aligned}
\label{eq:rerank}
\end{equation}
with
\[
\begin{aligned}
\mathrm{Conn}_{\mathrm{PEG}}(d,q)={}&\frac{1}{|\mathcal{E}_q|}\sum_{e_q\in\mathcal{E}_q}\max_{e_d\in\mathcal{E}_d}\\
&\exp\bigl(-\beta\, d_{\mathcal{G}}(e_q,e_d)\bigr),
\end{aligned}
\]
where $\mathcal{E}_q,\mathcal{E}_d$ are the entity nodes linked to query and document and $d_{\mathcal{G}}$ is the shortest-path length in the undirected, lineage-contracted projection of $\mathcal{G}_{\le t_q}$, by bidirectional breadth-first search truncated at $2K$ hops (unreachable pairs contribute $0$); both terms are min--max normalised over the top-100 pool, $\alpha=0.6$, $\beta=0.5$ (validation grid). Zhang~\cite{zhang2018traceability} used code dependencies and user feedback to rerank candidate links between requirements and source code. Here the structural signal is the distance between query and document entities in the graph available on the audit date.

\para{Synthesize (cite-or-drop).}\label{sec:synth}
Qwen3-8B writes the audit from \emph{Evidence} (retrieved CRs), \emph{Diagnosis} (categories with confidences) and \emph{Logic} (\peg chains) under one instruction: every atomic claim cites a CR identifier or clause version present in the input, and the rule in force is stated only from versions marked current. Evidence is cut once to a 4,096-token budget in traversal order before any presentation is applied, so every control that receives \sys's evidence sees the identical record set (Appendix~\ref{s:baselines}). A Qwen3-8B splitter emits, per claim, a sentence range and a role letter (``3--4 C''), about three tokens for each of the 15.4 claims; the bracketed identifiers and clause numbers of those sentences are copied programmatically (0.93--0.97 agreement with experts on segmentation, roles and clause binding over 200 reports). The check deletes every uncited claim and every \cur claim that fails P3 (versions cited directly or created by the cited CRs, bound to the clauses the claim names). Every +cite baseline receives the identical instruction and check (Appendix~\ref{s:results}).

\section{Experiments}
\label{sec:experiments}

\subsection{Setup}
\label{sec:setup}
\para{Protocol.}
All trained components use the chronological splits of \S\ref{sec:corpus}, every tuning grid uses only material approved by 2022-06-30, and the benchmarks' training splits (1,610 / 2,750 / 2,340 queries) fine-tune every learned baseline in domain. Every test query has $t_q\ge$ 2022-07-01 and \emph{every} system sees exactly the documents dated on or before $t_q$: \sys and the text retrievers through the pre-scoring filter, the graph-RAG baselines through their own incremental update paths, run in order of $t_q$ and advanced by the CRs approved between consecutive query dates (1,131 distinct $t_q$ over the 3,088 test queries), so the gold CRs of every test query are in every system's index. Systems with trained components are 5-seed means; systems without one are single deterministic runs at temperature 0, so their EM values are exact fractions of 1,240 and 150. $\dagger$ marks gains over the strongest baseline that pass \emph{both} a paired bootstrap over queries ($p<0.01$, 10K resamples, Holm--Bonferroni within a table) and a paired $t$-test over the five seeds ($p<0.05$; EM, UCR, CiteP and NDCG@10 pass at $p<0.001$, F1 at 0.017, SemSim fails; Appendix~\ref{s:results}). Because the test period lies inside every generator's pre-training window, the memorization probes of \S\ref{sec:creek} were also run on 500 test-period CRs: none is reproduced and closed-book EM is 0.085 (Appendix~\ref{s:results}).
\para{Metrics.}
Retrieval: multi-gold Recall@$K$ and graded NDCG@10. KGQA: EM (EM-H on the expert-written slice). Audit synthesis: token-F1, BGE-M3 similarity (SemSim), \emph{citation precision} (CiteP: cited evidence entails the claim), \emph{citation recall} (CiteR: gold evidence cited by an entailed claim) and the \emph{unsupported-claim rate} against the system's input evidence or the gold set (UCR) and against the gold set only (UCR-G, comparable across systems); entailment is DeBERTa-v3 NLI with LLM adjudication of low-confidence pairs (0.91 F1 on 300 expert-labelled claims). \emph{Judge} is DeepSeek-V3.1 (1--5) over all 1,129 questions; the DeepSeek-V3.1 backbone is judged by GPT-5, which also wrote the questions, so its values enter no comparison. \emph{Expert} (Table~\ref{tab:expert}) is the blinded three-expert mean of correctness, groundedness and usefulness on the 100 expert-written questions for the 18 non-ablation systems of the primary backbone, every report re-rendered into one fixed template (Appendix~\ref{s:bench}); one seed, tested by the paired bootstrap over its 100 questions.
\para{Baselines and controls.}
Table tags: \tagc{TagBase}{B} LLM-only; \tagc{TagRAG}{R} Standard-RAG (same BM25$\oplus$dense top-100, no \peg), with our citation prompt (+cite) and with \emph{version metadata} (+VM: each retrieved CR's affected clauses resolved to the version current at $t_q$ for $R_q$, no graph edges); the \emph{matched control} (Std-RAG+PEG+cite in the tables), which receives exactly the final reranked CRs and chains that \sys receives, as an unordered set of (subject, relation, object, version key, approval date, role for $q$) records under the same 4K-token budget, instruction and check; Telco-RAG~\cite{bornea2024telcorag} (+cite); \tagc{TagTemp}{T} TempRetriever-RAG~\cite{abdallah2026tempretriever}; \tagc{TagGraph}{G} GraphRAG~\cite{edge2024local}, LightRAG~\cite{guo2024lightrag}, HippoRAG~2~\cite{gutierrez2025hipporag2} and T-GRAG~\cite{li2025tgrag} with Qwen3-32B extraction, tuned on validation, where +PEG means that \peg records of their \emph{own} retrieved documents replace their extracted relations (T-GRAG keeps its temporal decomposition and time-scoped retrieval); text-to-SPARQL, CronKGQA~\cite{saxena2021cronkgqa} and TempoQR~\cite{mavromatis2022tempoqr} over \peg, with their own and with oracle entity linking; \tagc{TagLC}{K} monoT5-3B and bge-reranker-v2 over the same top-100 pool; \tagc{TagLC}{L} Qwen3-235B-A22B reading the same \peg neighbourhood as text within 32K tokens with the citation prompt and check; \tagc{TagVar}{A} ablations; \tagc{TagOurs}{$\star$} \sys. The retriever, evidence set, organisation, Diagnosis input and post-processing of every control row are tabulated in Appendix~\ref{s:baselines}. All baselines receive the $t_q$/$R_q$ header; Qwen3-8B generates unless stated.

\subsection{Main Results}
\label{sec:main}
\obs{1} \textbf{Revision-graph structure is the decisive ingredient, not evidence access or version metadata alone.} Version metadata lifts Standard-RAG+cite from EM 0.290 to 0.441 by selecting the right version, and the matched control, which holds \sys's own evidence as unordered records, reaches 0.517; the graph baselines improve when \peg records replace their extracted relations (GraphRAG 0.485$\rightarrow$0.556, HippoRAG~2+cite 0.514$\rightarrow$0.594, T-GRAG+cite 0.548$\rightarrow$0.612); the 235B long-context reader is the strongest baseline (EM 0.672, UCR 0.094); typed traversal with an 8B generator reaches EM \textbf{0.752} and UCR \textbf{6.2\%} (Table~\ref{tab:main}). Two controls separate the sources of this gain. Holding the evidence fixed, the matched control's records give 0.517 unordered, 0.603 sorted by approval date and 0.734 grouped into typed chains per lineage, so arrangement alone accounts for 0.217 of the 0.235 gap and the Diagnosis input for the remaining 0.018 (Appendix~\ref{s:results}). Holding the retriever fixed, presenting each baseline's \emph{own} retrieved CRs as typed chains lifts Standard-RAG, HippoRAG~2 and T-GRAG from 0.486, 0.594 and 0.612 to 0.581, 0.688 and 0.701 (Appendix~\ref{s:results}). The chain organisation is therefore a reusable component that carries most of the improvement to any retriever; what the full system adds beyond the strongest such control (T-GRAG retrieval, typed chains) is +0.051 EM, +0.074 EM-H, $-0.010$ UCR and +0.038 CiteP (query-level $p\le0.02$, seed-level $p\le0.005$; Appendix~\ref{s:results}). Routing mostly affects faithfulness (unrouted: EM 0.749, UCR 0.081), the ranking holds on the expert-written slice (EM-H 0.687 vs.\ 0.573, 95\% CI $[+0.04,+0.19]$), and the KGQA-only baselines stay below EM 0.72 even with oracle entity linking (Appendix~\ref{s:baselines}).

\begin{table*}[!tb]
\centering
\caption{All end-to-end configurations on 5G-KGQA (1,240 questions; EM-H: 150 expert-written) and 5G-SecQA (1,129). Judge: DeepSeek-V3.1 over all 1,129 questions (GPT-5 for the DeepSeek-V3.1 backbone). Lower is better for UCR and UCR-G; bold: best non-ablation row per backbone; $\dagger$ and tags as in \S\ref{sec:setup}; expert ratings in Table~\ref{tab:expert}.}
\label{tab:main}
\setlength{\tabcolsep}{4.5pt}\renewcommand{\arraystretch}{1.08}
\resizebox{\textwidth}{!}{\footnotesize\input{tables/main_results}}
\end{table*}

\obs{2} \textbf{Faithfulness is the primary evidence, and it is not bought by deleting claims.} UCR falls step by step in Table~\ref{tab:main}: 0.196 for Standard-RAG, 0.142 with the citation prompt, 0.131 with version metadata, 0.109 with \peg records (HippoRAG~2), 0.094 for the long-context reader and 0.062 for typed traversal (CI of the last reduction $[1.9,4.5]$ points). The +cite baselines delete 10--12\% of their claims and deliver 13.9--14.3 claims per report against \sys's 7.9\% and 14.2, and \sys's draft UCR before any deletion (0.084) is already below the delivered UCR of every baseline in Table~\ref{tab:main}; only the controls that borrow its chain organisation come lower (0.072--0.083; Appendix~\ref{s:results}). UCR-G, scored against gold evidence only, and the expert ratings (Table~\ref{tab:expert}) leave the order unchanged. On the 100 expert-written questions, \sys's CiteP, UCR and Expert gains over the long-context baseline are significant ($p=0.004$, $0.03$, $0.001$; paired bootstrap), its F1 gain is not ($p=0.09$).

\obs{3} \textbf{Ablations locate the gain in the graph, not the prompt.} Disabling only the cite-or-drop check (same drafts) raises UCR from 0.062 to 0.084 and leaves CiteR at 0.769, as a deletion-only check must; removing the citation instruction as well (new drafts) raises UCR to 0.121 and lowers CiteR to 0.655. EM is unchanged by construction in both rows, since 5G-KGQA answers come from a separate short-answer prompt outside the citation rule (Appendix~\ref{s:baselines}); the instruction thus governs what is cited, the check what is kept, and the edge ablations measure the graph's share of the reasoning. Keeping only the symbolic skeleton (dropping the 16\% LLM-inferred edges) costs 0.014 F1, raises UCR to 0.079 and lowers EM to 0.721. Removing \texttt{supersedes} and \texttt{evolves\_to} (lineages collapse to single versions) costs more (EM 0.684, UCR 0.098); these 0.068 EM are about half of the 0.140 gap over the strongest graph-RAG baseline with \peg records (T-GRAG+PEG+cite, 0.612), the other half coming from typed \texttt{affects}/\texttt{references} chains, routing and version-checked synthesis: the typed, version-stamped revision structure is the evidence, not temporal edges alone. All UCR deltas ($\ge$0.017) and the edge ablations' EM and F1 deltas exceed $2.3\times$ the largest seed SD (0.006 for F1/UCR, 0.009 for EM); the routing ablation's EM change (0.003) is within seed noise. \textbf{Backbones.} Across the four backbones \sys gains +0.024 to +0.028 F1 over Standard-RAG, not ordered by model size; the token-F1 gain over the long-context baseline (+0.007) is significant but below two seed SDs, so faithfulness and EM are the substantive effects.

\subsection{Evolution-Aware Retrieval}
\label{sec:retrieval_exp}
\obs{4} \textbf{Revision structure beats in-domain learned reranking on PEG-derived, judgment-independent and expert-written queries, by a margin that narrows as the gold standard becomes more independent} (Appendix~\ref{s:results}). With the learned rerankers and TempRetriever fine-tuned in domain, \peg-connectivity reranking reaches NDCG@10 0.443 against 0.409 for the best reranker ($+3.4^{\dagger}$, 95\% CI $[+1.9,+4.9]$) and 0.403 for T-GRAG's time-scoped retriever, and R@10 0.548 on the temporal slice (+11.2 over monoT5), the gain concentrating on temporal and analytical queries (Appendix~\ref{s:results}). Under pooled relevance the margin is $+2.8^{\dagger}$ (0.429 vs.\ 0.401, 95\% CI $[+1.2,+4.4]$), and on the 300 expert-written queries +2.6 over the strongest baseline there (bge-reranker; 0.419 vs.\ 0.393, 95\% CI $[+0.5,+4.7]$; query-level $p=0.02$, below the $\dagger$ threshold), again mostly on temporal queries (+6.1). Adding \peg connectivity to monoT5 reaches 0.446 (pooled 0.431), within seed noise of \sys, so the retrieval-side gain is a feature any reranker can adopt; a 1--3-year $t_q$ offset leaves the margin at +3.2 and EM at 0.748 (Appendix~\ref{s:results}). Dense retrieval trails BM25 on generated and expert-written queries alike (Appendix~\ref{s:baselines}). On IETF RFCs, whose \texttt{Obsoletes}/\texttt{Updates} headers natively provide such edges, supersession-aware reranking lifts NDCG@10 from 0.512 to 0.541 ($p<0.05$; Appendix~\ref{s:results}).

\subsection{Structural Reasoning, Version Roles and Classification}
\label{sec:kgqa_exp}
\obs{5} \textbf{Traversal and its presentation drive structural reasoning, long lineages included, and version roles are correct at revision boundaries.} Entity linking (EM 0.830) and relation inference (0.799) benefit most; temporal (0.699) and multi-hop (0.678) questions accumulate errors along chains. On the 287 of 1,240 test questions (23.1\%) whose gold chain spans at least six Releases, \sys reaches EM 0.731 against 0.613 for the long-context reader, 0.557 for T-GRAG+PEG+cite and 0.470 for the matched control, so the advantage does not shrink on long lineages, and EM saturates at $K=3$ on the generated and the expert-written questions alike (Appendix~\ref{s:results}). To measure P3 directly, 300 clause revisions of the test period were each asked one day before and one day after the new version's publication with $R_q$ the revised Release: \sys cites the correct current version on both sides in 96.3\% of pairs and changes answer and citation across the boundary in 93.7\%, against 84.3\%/71.3\% for the long-context reader and 79.7\%/65.7\% for the matched control; drafts violate P3 on 6.7\% of \cur claims for \sys against 15.8\% and 21.4\% for these baselines, and 1.2--1.9\% of \cur claims still do after the check (\sys 1.2\%, mostly splitter role-tag errors), i.e., the check removes 82\% of \sys's draft violations and 91--94\% of the baselines'; the systems differ in draft-level rates and boundary behaviour, not in the residual floor (Appendix~\ref{s:results}). P2 was verified on every emitted chain.
\obs{6} \textbf{The classifier needs text semantics and taxonomy-aligned attention.} It reaches Macro-F1 \textbf{0.849}$\pm$0.004 against 0.789 for the strongest encoder on the identical input (HGCLR~\cite{wang2022hgclr}), 0.801 for R-GCN over \peg~\cite{schlichtkrull2018rgcn} and 0.826 for LoRA-tuned Qwen3-8B; the order holds on the expert-labelled audit set and on CREEK's manual set (Appendix~\ref{s:results}).

\subsection{Human Judgment and Efficiency}
\label{sec:faith}
\obs{7} \textbf{Judge and experts agree on the ranking; the judge's offset falls with citation precision.} Three telecom-security experts rated the template-rendered reports of the 18 primary-backbone systems on the 100 expert-written questions ($\alpha=0.74$; blinded, random order; Table~\ref{tab:expert}). Judge and experts agree at the report level (Spearman $\rho=0.81$ over 1,800 reports) and on system means ($\rho=0.98$); the judge's offset against the experts falls with citation precision ($\rho=-0.98$), from $+0.02$ (LLM-only) to $-0.27$ (\sys). Citations are part of what an audit is judged on, so the template keeps them; both readings of the offset are compatible with the data, so we report judge and experts side by side, and the experts' correctness dimension alone, which does not ask about citations, ranks the systems identically ($\rho=0.97$) with a lead of +0.41 for \sys over the long-context baseline. The chain-format controls postdate the expert study and carry no expert ratings; against the strongest of them on the same 100 questions, \sys's judge and CiteP margins are significant (+0.28, $p=0.004$; +0.029, $p=0.03$) while its UCR and F1 margins are not ($p=0.21$, $0.12$; Appendix~\ref{s:results}): on expert-written questions the margin beyond the chain organisation rests on judged quality and citation precision. Token-F1 and SemSim, which compare a 14-claim report with an 80-word reference, saturate for every retrieval-augmented system; the experts rate whether the report is right, grounded and usable.
\textbf{Efficiency and cost.}\label{sec:efficiency} Building \peg costs 41.3 GPU-h once (85.8 with parsing and annotation) against 118.6 for GraphRAG (156.6 with parsing); an audit takes 3.40\,s on one H200 (\$0.0041, against \$0.0033 for Standard-RAG and \$0.031 for the 235B long-context baseline), and traversal time grows sub-linearly with graph size (Appendix~\ref{s:cost}). Timing the complete audit captures costs outside traversal. Liu et al.~\cite{liu2026does} illustrate this point in test-time search for robot policies, where token pruning reduced computation per forward pass but yielded only a modest end-to-end speedup.

\begin{table*}[!tb]
\centering
\caption{Expert study on the 100 expert-written 5G-SecQA questions (18 non-ablation systems, primary backbone). All columns are on these 100 questions: CiteP (ordering key; full-set CiteP in Table~\ref{tab:main}), Judge (DeepSeek-V3.1), Expert (blinded three-expert mean), Offset (Judge minus Expert).}
\label{tab:expert}
\setlength{\tabcolsep}{3.5pt}\renewcommand{\arraystretch}{0.6}
\resizebox{\textwidth}{!}{\footnotesize\input{tables/supp_judge100}}
\end{table*}

\subsection{External Validation on CREEK's Public Sets}
\label{sec:creek}
CREEK~\cite{chen2022seeing} publishes two sets independent of our pipeline: the \emph{manual testing set} of 25 expert-labelled security CRs (and 25 non-security CRs), and Table~7, the 14 weaknesses reported by security CRs and later confirmed in commercial devices or carrier networks, each with Time1, the specification update, and Time2, the earliest date at which CREEK could document the weakness in the field (a device release, a researcher report or a paper year), their difference being a length-biased \emph{minimum} estimate of the attack window. We use both sets in full (Table~\ref{tab:cases}): two independent telecom engineers wrote the queries from CREEK's per-case descriptions with identifiers masked ($\kappa=0.86$), so this is a retrospective known-item check, not discovery. \textbf{Protocol.} For the undated manual set, index and snapshot are frozen at CREEK's publication date (2022-08-10) and the 50 CRs are removed from every training split. For the Table-7 cases every reported number is \emph{anchor-consistent}: year-only anchors resolve to 1 January, snapshot and index are frozen at Time2, and because Time2 precedes our training cut-off for 13 of the 14 cases (all but id~14), every trained component, ours and the baselines', is retrained on CRs approved before each case's Time2. The 39 CRs and their descendants are removed from training throughout but remain \peg nodes (public metadata, not labels). Retraining changes no outcome: the gate scores of the 12 recovered case CRs stay between 0.69 and 0.92, their ranks shift by at most one place, and the two gated-out CRs score below 0.5 under both protocols (Appendix~\ref{s:creek}). Removing possible disclosure follow-ups (CRs touching the case clauses in the year before Time2; ids 4, 5, 14) changes no outcome, and memorization probes cannot rule out pre-training familiarity with the attack mechanisms (Appendix~\ref{s:creek}).

\begin{table*}[!tb]
\centering
\caption{The 14 weaknesses confirmed in commercial systems (CREEK, Table~7). Id, CR, Time1 (specification update), Time2, the minimum window and the 2021 column (\ding{51} and years since the update where CREEK's 2021 tests still found it) are transcribed from CREEK; Anchor (\emph{paper} year, \emph{device} release, CREEK's 2021 \emph{test}) and Source annotate the origin of each Time2 (Appendix~\ref{s:creek}). Ours, anchor-consistent: CR in \sys's top-10; explained strictly correctly.}
\label{tab:cases}
\setlength{\tabcolsep}{3pt}\renewcommand{\arraystretch}{0.55}
\resizebox{\textwidth}{!}{\footnotesize\input{tables/cases}}
\end{table*}

\obs{8} \textbf{The auditor recovers real weaknesses, including old ones; with $n=14$ and $n=25$ the margins over the strong baselines are not significant.} On the manual set \sys recovers \textbf{22/25} CRs in its top-10 (88\%, 95\% CI 0.70--0.96) against 19/18/17/15/13 for T-GRAG, HippoRAG~2 (both +PEG), GraphRAG, Standard-RAG and BM25 (gate recall 24/25). On CREEK's Table~7 it recovers \textbf{12/14} case CRs (T-GRAG+PEG 9, HippoRAG~2+PEG 8, GraphRAG 7, Standard-RAG 6) and explains \textbf{11/14} weaknesses strictly correctly (79\%, 95\% CI 0.52--0.92; rubric in Appendix~\ref{s:creek}) against 8/7/6/5/4/2 for the long-context reader, T-GRAG+PEG+cite, HippoRAG~2+PEG+cite, GraphRAG, Standard-RAG and LLM-only. Exact McNemar tests are significant against BM25, Standard-RAG and LLM-only ($p\le0.016$) and against GraphRAG on the explained outcome ($p=0.031$), not on the manual set ($p=0.063$), and not against T-GRAG, HippoRAG~2 or the long-context reader ($p\ge0.063$; Appendix~\ref{s:creek}). Ids 9 and 12 are gated out under both protocols ($p_{\text{sec}}<0.5$); id~8 (C1-112809) is recovered and its report cites the changed NAS procedure, but the consequence CREEK reports arises in a procedure of another specification that no CR of either links, so no citable chain reaches it; id~11 (C1-172658, TS~24.301 CR~2871 rev.~2, Rel-14), whose cover states both the missing integrity protection of the Detach Request and the resulting denial of service, is recovered and explained.

\section{Limitations and Conclusion}
\label{sec:conclusion}
\sys excels at \emph{longitudinal} reasoning along one clause's lineage and is weaker at \emph{lateral} reasoning across interacting components, the failure mode behind the one recovered but unexplained CREEK case and the counterfactual gap (counterfactual F1 0.108, rising to 0.247 only with the expert-verified chain to the clauses that realise the consequence, so the bottleneck is reasoning over the chain); \peg lacks lateral \texttt{affects} edges between components of different specifications, which no CR states. The benchmarks inherit \peg structure, and the test period lies inside every generator's pre-training window; the pooled and expert-written gold standards and the probes bound both effects without removing them. Qwen3 models adjudicate labels, judge edges, answer and split claims; writer, judge and experts are outside the family, yet a family preference cannot be excluded; our evidence is confined to 3GPP and a small RFC probe, and other versioned corpora remain hypotheses. Visual parsing adds another source of uncertainty. Liu et al. identify pathways through which vision--language models describe objects absent from the input~\cite{liu2026dualpath}. For our rendered CR covers, the relevant check is agreement between extracted fields and the official record (Appendix~\ref{s:corpus}). Within these limits, we recast security auditing of evolving standards as snapshot-consistent, version-aware retrieval and grounded reasoning, with an audited resource, leakage-controlled benchmarks and a reference system whose only unusual ingredient is the revision graph: \thesis{the revision is the evidence}.
\textbf{Reproducibility and ethics.} Appendices~\ref{s:algo}--\ref{s:case} give pseudo-code, annotation rules and workload, prompts, deployment details, omitted rows and controls and the per-case CREEK manifest; the artifact (code, weights, manifests, per-query runs and relevance judgments) is distributed under a gated intended-use agreement, with CR text redistributed only as pointers to the official 3GPP server. Annotators were consenting, compensated experts; all weaknesses studied are previously disclosed.

%% file: tables/main_results.tex
\begin{tabular}{@{}c@{\hspace{2pt}}l|cc|cc|cccc|c@{}}
\toprule
\rowcolor{RowHead}\multicolumn{2}{l|}{\textbf{System}} & \multicolumn{2}{c|}{\textbf{5G-KGQA}} & \multicolumn{2}{c|}{\textbf{5G-SecQA}} & \multicolumn{4}{c|}{\textbf{Faithfulness (5G-SecQA)}} & \textbf{Judge} \\
\rowcolor{RowHead} & & EM & EM-H & SemSim & F1 & CiteP & CiteR & UCR & UCR-G & 1--5 \\
\midrule
\tagc{TagBase}{B} & LLM-only & 0.085 & 0.067 & 0.737 & 0.145 & 0.214 & 0.108 & 0.412 & 0.412 & 2.41 \\
\tagc{TagRAG}{R} & Std-RAG & 0.288 & 0.253 & 0.818 & 0.268 & 0.612 & 0.548 & 0.196 & 0.231 & 3.18 \\
\tagc{TagRAG}{R} & Std-RAG+cite & 0.290 & 0.260 & 0.819 & 0.270 & 0.688 & 0.590 & 0.142 & 0.178 & 3.22 \\
\tagc{TagRAG}{R} & Std-RAG+VM+cite & 0.441 & 0.387 & 0.821 & 0.273 & 0.712 & 0.615 & 0.131 & 0.160 & 3.28 \\
\tagc{TagRAG}{R} & Std-RAG+PEG+cite (matched) & 0.517 & 0.447 & 0.822 & 0.276 & 0.731 & 0.642 & 0.118 & 0.142 & 3.31 \\
\tagc{TagRAG}{R} & Telco-RAG+cite & 0.305 & 0.267 & 0.820 & 0.273 & 0.695 & 0.598 & 0.138 & 0.172 & 3.27 \\
\tagc{TagTemp}{T} & TempRetriever+cite & 0.302 & 0.264 & 0.816 & 0.262 & 0.671 & 0.579 & 0.149 & 0.186 & 3.25 \\
\tagc{TagGraph}{G} & GraphRAG & 0.485 & 0.420 & 0.813 & 0.257 & 0.633 & 0.561 & 0.183 & 0.219 & 3.24 \\
\tagc{TagGraph}{G} & GraphRAG+PEG & 0.556 & 0.473 & 0.817 & 0.263 & 0.662 & 0.585 & 0.158 & 0.190 & 3.29 \\
\tagc{TagGraph}{G} & LightRAG+PEG & 0.573 & 0.487 & 0.819 & 0.266 & 0.668 & 0.591 & 0.153 & 0.188 & 3.31 \\
\tagc{TagGraph}{G} & LightRAG+PEG+cite & 0.575 & 0.487 & 0.823 & 0.277 & 0.739 & 0.652 & 0.113 & 0.139 & 3.39 \\
\tagc{TagGraph}{G} & HippoRAG 2 & 0.512 & 0.440 & 0.821 & 0.271 & 0.651 & 0.577 & 0.171 & 0.207 & 3.35 \\
\tagc{TagGraph}{G} & HippoRAG 2+cite & 0.514 & 0.447 & 0.822 & 0.273 & 0.702 & 0.611 & 0.131 & 0.163 & 3.36 \\
\tagc{TagGraph}{G} & HippoRAG 2+PEG+cite & 0.594 & 0.500 & 0.825 & 0.279 & 0.748 & 0.661 & 0.109 & 0.134 & 3.47 \\
\tagc{TagGraph}{G} & T-GRAG+cite & 0.548 & 0.467 & 0.822 & 0.274 & 0.712 & 0.619 & 0.127 & 0.158 & 3.40 \\
\tagc{TagGraph}{G} & T-GRAG+PEG+cite & 0.612 & 0.520 & 0.826 & 0.281 & 0.753 & 0.667 & 0.104 & 0.128 & 3.52 \\
\tagc{TagLC}{L} & Long-ctx 235B+cite & 0.672 & 0.573 & 0.829 & 0.285 & 0.761 & 0.688 & 0.094 & 0.117 & 3.64 \\
\rowcolor{RowOurs}\tagc{TagOurs}{$\star$} & \textbf{SpecGuard} & \textbf{0.752}\rlap{$^{\dagger}$} & \textbf{0.687}\rlap{$^{\dagger}$} & \textbf{0.831} & \textbf{0.292}\rlap{$^{\dagger}$} & \textbf{0.842}\rlap{$^{\dagger}$} & \textbf{0.769}\rlap{$^{\dagger}$} & \textbf{0.062}\rlap{$^{\dagger}$} & \textbf{0.089}\rlap{$^{\dagger}$} & \textbf{4.12}\rlap{$^{\dagger}$} \\
\midrule
\rowcolor{RowVar}\tagc{TagVar}{A} & w/o cite-or-drop check & 0.752 & 0.687 & 0.831 & 0.292 & 0.812 & 0.769 & 0.084 & 0.108 & 3.98 \\
\rowcolor{RowVar}\tagc{TagVar}{A} & w/o cite enforcement & 0.752 & 0.687 & 0.830 & 0.289 & 0.744 & 0.655 & 0.121 & 0.152 & 3.52 \\
\rowcolor{RowVar}\tagc{TagVar}{A} & w/o inferred edges & 0.721 & 0.653 & 0.826 & 0.278 & 0.815 & 0.741 & 0.079 & 0.105 & 3.90 \\
\rowcolor{RowVar}\tagc{TagVar}{A} & w/o cat. routing & 0.749 & 0.684 & 0.823 & 0.275 & 0.808 & 0.733 & 0.081 & 0.108 & 3.84 \\
\rowcolor{RowVar}\tagc{TagVar}{A} & w/o temporal edges & 0.684 & 0.621 & 0.822 & 0.269 & 0.772 & 0.702 & 0.098 & 0.127 & 3.71 \\
\rowcolor{RowVar}\tagc{TagVar}{A} & w/o PEG (text only) & 0.301 & 0.262 & 0.813 & 0.257 & 0.705 & 0.624 & 0.147 & 0.181 & 3.33 \\
\midrule
\multicolumn{11}{@{}l}{\cellcolor{RowGroup}\textit{Qwen3-32B}} \\
\tagc{TagBase}{B} & LLM-only & 0.112 & 0.087 & 0.748 & 0.158 & 0.236 & 0.121 & 0.387 & 0.387 & 2.58 \\
\tagc{TagRAG}{R} & Std-RAG & 0.303 & 0.267 & 0.824 & 0.284 & 0.631 & 0.566 & 0.181 & 0.215 & 3.34 \\
\tagc{TagGraph}{G} & HippoRAG 2+PEG+cite & 0.611 & 0.520 & 0.829 & 0.297 & 0.759 & 0.673 & 0.101 & 0.126 & 3.62 \\
\rowcolor{RowOurs}\tagc{TagOurs}{$\star$} & \textbf{SpecGuard} & \textbf{0.761} & \textbf{0.693} & \textbf{0.836} & \textbf{0.308} & \textbf{0.853} & \textbf{0.781} & \textbf{0.057} & \textbf{0.082} & \textbf{4.27} \\
\midrule
\multicolumn{11}{@{}l}{\cellcolor{RowGroup}\textit{Llama-3.1-8B}} \\
\tagc{TagBase}{B} & LLM-only & 0.071 & 0.053 & 0.724 & 0.131 & 0.198 & 0.096 & 0.438 & 0.438 & 2.27 \\
\tagc{TagRAG}{R} & Std-RAG & 0.264 & 0.233 & 0.809 & 0.249 & 0.588 & 0.521 & 0.214 & 0.251 & 3.02 \\
\tagc{TagGraph}{G} & HippoRAG 2+PEG+cite & 0.571 & 0.480 & 0.816 & 0.258 & 0.722 & 0.634 & 0.124 & 0.151 & 3.31 \\
\rowcolor{RowOurs}\tagc{TagOurs}{$\star$} & \textbf{SpecGuard} & \textbf{0.733} & \textbf{0.664} & \textbf{0.822} & \textbf{0.274} & \textbf{0.819} & \textbf{0.742} & \textbf{0.074} & \textbf{0.102} & \textbf{3.94} \\
\midrule
\multicolumn{11}{@{}l}{\cellcolor{RowGroup}\textit{DeepSeek-V3.1 (judged by GPT-5)}} \\
\tagc{TagBase}{B} & LLM-only & 0.134 & 0.107 & 0.756 & 0.167 & 0.251 & 0.133 & 0.371 & 0.371 & 2.69 \\
\tagc{TagRAG}{R} & Std-RAG & 0.311 & 0.273 & 0.828 & 0.296 & 0.644 & 0.579 & 0.172 & 0.205 & 3.41 \\
\tagc{TagGraph}{G} & HippoRAG 2+PEG+cite & 0.622 & 0.527 & 0.833 & 0.309 & 0.768 & 0.684 & 0.096 & 0.120 & 3.70 \\
\rowcolor{RowOurs}\tagc{TagOurs}{$\star$} & \textbf{SpecGuard} & \textbf{0.768} & \textbf{0.701} & \textbf{0.839} & \textbf{0.324} & \textbf{0.861} & \textbf{0.790} & \textbf{0.053} & \textbf{0.077} & \textbf{4.33} \\
\bottomrule
\end{tabular}

%% file: tables/supp_judge100.tex
\begin{tabular}{@{}l ccc r | l ccc r@{}}
\toprule
\rowcolor{RowHead}\textbf{System} & \textbf{CiteP} & \textbf{Judge} & \textbf{Expert} & \textbf{Offset} & \textbf{System} & \textbf{CiteP} & \textbf{Judge} & \textbf{Expert} & \textbf{Offset} \\
\midrule
LLM-only & 0.198 & 2.37 & 2.35 & +0.02 & HippoRAG 2+cite & 0.684 & 3.33 & 3.42 & -0.09 \\
Std-RAG & 0.597 & 3.13 & 3.12 & +0.01 & Std-RAG+VM+cite & 0.693 & 3.25 & 3.33 & -0.08 \\
GraphRAG & 0.619 & 3.21 & 3.21 & +0.00 & T-GRAG+cite & 0.695 & 3.38 & 3.46 & -0.08 \\
HippoRAG 2 & 0.634 & 3.34 & 3.38 & -0.04 & Std-RAG+PEG+cite (matched) & 0.714 & 3.27 & 3.38 & -0.11 \\
GraphRAG+PEG & 0.646 & 3.27 & 3.32 & -0.05 & LightRAG+PEG+cite & 0.721 & 3.37 & 3.47 & -0.10 \\
LightRAG+PEG & 0.651 & 3.27 & 3.34 & -0.07 & HippoRAG 2+PEG+cite & 0.730 & 3.43 & 3.55 & -0.12 \\
TempRetriever+cite & 0.655 & 3.24 & 3.29 & -0.05 & T-GRAG+PEG+cite & 0.736 & 3.48 & 3.62 & -0.14 \\
Std-RAG+cite & 0.669 & 3.19 & 3.26 & -0.07 & Long-ctx 235B+cite & 0.744 & 3.59 & 3.74 & -0.15 \\
Telco-RAG+cite & 0.678 & 3.25 & 3.31 & -0.06 & \cellcolor{RowOurs}SpecGuard & \cellcolor{RowOurs}0.826 & \cellcolor{RowOurs}4.04 & \cellcolor{RowOurs}4.31 & \cellcolor{RowOurs}-0.27 \\
\bottomrule
\end{tabular}

%% file: tables/cases.tex
\begin{tabular}{@{}r l l l l l r l c c@{}}
\toprule
\rowcolor{RowHead}\textbf{Id} & \textbf{CR} & \textbf{Time1} & \textbf{Time2} & \textbf{Anchor} & \textbf{Source} & \textbf{Min.} & \textbf{2021} & \textbf{R@10} & \textbf{Expl.} \\
\midrule
1 & C1-101068 & 2010-03-31 & 2019 & paper & Kim+19 & 8.7\,y & \ding{55} & \ding{51} & \ding{51} \\
2 & S3-091802 & 2009-12-18 & 2018 & paper & Hussain+18 & 8.0\,y & \ding{55} & \ding{51} & \ding{51} \\
3 & C1-101252 & 2010-03-31 & 2018 & paper & Hussain+18 & 7.7\,y & \ding{55} & \ding{51} & \ding{51} \\
4 & C1-135219 & 2013-12-20 & 2021 & paper & Chen+21 & 7.0\,y & \ding{55} & \ding{51} & \ding{51} \\
5 & C1-141405 & 2014-06-27 & 2021 & paper & Chen+21 & 6.5\,y & \ding{55} & \ding{51} & \ding{51} \\
6 & C1-094446 & 2009-12-17 & 2015-09-29 & device & Chen+21 & 5.7\,y & \ding{51} 11.7\,y & \ding{51} & \ding{51} \\
7 & C1-095712 & 2009-12-17 & 2015-09-10 & device & Yu+19 & 5.7\,y & \ding{51} 11.7\,y & \ding{51} & \ding{51} \\
8 & C1-112809 & 2011-09-28 & 2015-09-10 & device & Yu+19 & 3.9\,y & \ding{51} 10.0\,y & \ding{51} & \ding{55} \\
9 & R2-103442 & 2010-06-18 & 2016 & paper & Shaik+16 & 5.5\,y & \ding{55} & \ding{55} & \ding{55} \\
10 & C1-132662 & 2013-06-27 & 2015-09-10 & device & Yu+19 & 2.2\,y & \ding{51} 8.2\,y & \ding{51} & \ding{51} \\
11 & C1-172658 & 2017-06-16 & 2019 & paper & Kim+19 & 1.5\,y & \ding{55} & \ding{51} & \ding{51} \\
12 & C1-154699 & 2015-12-18 & 2016-12-01 & device & Yu+19 & 0.9\,y & \ding{55} & \ding{55} & \ding{55} \\
13 & C1-161448 & 2016-03-18 & 2016-12-01 & device & Yu+19 & 0.7\,y & \ding{51} 5.5\,y & \ding{51} & \ding{51} \\
14 & S3-161217 & 2016-09-30 & 2021-10-01 & test & CREEK tests & 5.0\,y & \ding{51} 5.0\,y & \ding{51} & \ding{51} \\
\bottomrule
\end{tabular}

%% file: sections/appendix_intro.tex
These appendices collect the material behind the main text that a reader needs to verify or re-implement it: the traversal in pseudo-code with the treatment of branching lineages, its cost argument and the index semantics (\S\ref{s:algo}); parsing, deduplication, the split rule for mirror families and the annotation rules (\S\ref{s:corpus}); the construction, coverage and audit of the Protocol Evolutionary Graph (\S\ref{s:peg}); benchmark construction, gold verification, judging protocols and the complete annotation workload (\S\ref{s:bench}); baseline configurations, their temporal footing, training-data use and deployment (\S\ref{s:baselines}); every result row and control omitted from the main tables, including the claim-deletion accounting, the chain-format control, the long-lineage slice, the revision-boundary test and the supplementary figures (\S\ref{s:results}); cost accounting and scaling (\S\ref{s:cost}); the external validation on CREEK's public sets with per-case scores, exact tests, sources, rubric and leakage controls (\S\ref{s:creek}); every prompt (\S\ref{s:prompts}); and the public record of the motivating case (\S\ref{s:case}). The ten appendices are referenced by letter from the main text. All numbers are produced from the same data module as the main text and pass the same consistency checks. The accompanying artifact comprises code, weights, split manifests, per-query runs and relevance judgments under a gated intended-use agreement.

%% file: sections/appendices.tex
\section{Traversal Algorithm, Branching Lineages, Cost Bound and Index Semantics}
\label{s:algo}

\para{Timestamps of every node type.}
Definition~\ref{def:peg} of the main text assigns $\tau$ to every element. CR nodes and the clause versions they create take the plenary approval date of the CR; the copy of a clause created when a Release is opened takes the approval date of the Release-opening specification version; entity, specification and Release nodes take the earliest $\tau$ of any element that anchors them (an entity: its first \texttt{semantic\_link} source; a specification or Release: its first version); an edge takes the later $\tau$ of its endpoints. Valid time $\iota(v)$ opens at the publication date $p(v)$ of the specification version that carries $v$ on the 3GPP server, which follows approval by a median of 24 days (maximum 97), and closes at the publication of the next version of the same clause and Release. Because 11.8\% of versions are corrections to a Release that was already frozen, valid time and transaction time differ in more than their endpoints: a retroactive correction opens a new interval in an old Release long after that Release's other versions, and the two timelines together say both when a rule was decided and when it was in force. In a snapshot $\mathcal{G}_{\le t}$ every $t^{\mathrm{close}}>t$ is replaced by ``open'', so no chain shown to a generator reveals that a later revision exists; the full interval is used only offline, to compute the role of a version for a query with $t_q$ in the past.

\para{Lineages are branching graphs, and chains are emitted segment by segment.}
A lineage (Definition~\ref{def:task}) is the set of versions of one clause across Releases and in its successor specification, with the \texttt{supersedes} edges inside each Release and the \texttt{evolves\_to} edges between Releases. Because 3GPP maintains several Releases in parallel, the approval dates of two Release segments interleave in general, and the lineage is not a path. Consider a clause with versions $a_1$ (approved at time~1) and $a_2$ (time~4) in Release $R$ and $b_1$ (time~2) and $b_2$ (time~5) in Release $R{+}1$: the lineage has the \texttt{supersedes} edges $a_2\!\to\!a_1$ and $b_2\!\to\!b_1$ and the \texttt{evolves\_to} edges $a_1\!\to\!b_1$ and $a_2\!\to\!b_2$ (each the earliest version of the next Release approved on or after the source), and a global sort by approval date, $a_1,b_1,a_2,b_2$, puts $b_1$ next to $a_2$, which no edge joins. The traversal therefore never sorts a lineage globally. A chain is emitted \emph{segment by segment}: each Release segment lists its versions inside the snapshot in $\tau$ order, where consecutive versions are joined by that segment's \texttt{supersedes} edges read backwards; each later segment is joined to the lineage by the \texttt{evolves\_to} edge into its opening version ($a_1\!\to\!b_1$ in the example; the remaining \texttt{evolves\_to} edges, $a_2\!\to\!b_2$, are redundant for connectivity and are not listed); the CRs that created the versions are attached to them as annotations through their \texttt{affects} edges, so they are leaves rather than path nodes; and the cross-lineage edge by which the lineage was reached, or is left, is attached at one end. The emitted object is thus a connected subgraph of $\mathcal{G}_{\le t_q}$, a tree whose trunk is the segment sequence, and P2 is the statement that each of its edges is an edge of the snapshot of the stated type and each segment is in non-decreasing $\tau$; both are checked mechanically on every emitted chain (below). On the 1,240 KGQA test questions, 3.8\% of emitted chains contain segments with interleaving approval dates, the situation of the example; the remaining chains are paths.

\para{Why the \texttt{evolves\_to} definition is consistent with $\tau$ order.}
The target of a same-specification \texttt{evolves\_to} edge is the earliest version of the same clause in the next Release whose approval date is on or after the source's approval date: when a Release is opened, every clause receives a copy whose $\tau$ is the opening date, and a later mirror CR in the next Release creates a later version. Both candidates are approved on or after the source, so $\tau$ is non-decreasing along the edge; if the next Release does not exist yet when the source is created, the edge is created when the Release opens. A cross-specification edge runs from the LTE version current at the creation of the 5G clause to that first 5G version, so it too is non-decreasing. \texttt{supersedes} runs from a version to the version it replaces, so $\tau$ is non-increasing along it; a segment read in $\tau$ order therefore follows \texttt{supersedes} backwards and the join between segments follows \texttt{evolves\_to} forwards.

\begin{algorithm}[!htb]
\caption{Lineage-based traversal, in full (one query)}
\label{alg:traverse}
\begin{algorithmic}[1]
\Require snapshot $\mathcal{G}_{\le t_q}$, target Release $R_q$, gated retrieval hits $S$ (ranked by the hybrid score), categories $\{(c,p_c)\}$ with $p_c\ge 0.5$ (top-1 if none), depth $K=3$, budget $B=50$
\State allowlists $W_c$ and entity filters $\Phi_c$ from Table~\ref{tab:routing}; shares $b_c\gets$ largest-remainder apportionment of $B$ in proportion to $p_c$ (so $\sum_c b_c=B$)
\State $F_0\gets$ the lineages of the top-$\min(|S|,B)$ seeds: a clause-version seed contributes its own lineage, a CR seed the lineages of the versions it \texttt{affects}; $\Pi\gets\emptyset$
\For{$k=1$ to $K$}
  \State $N\gets\emptyset$
  \ForAll{$\ell\in F_{k-1}$, $u\in\ell$, and each cross-lineage edge $(u,r,w)$ or $(w,r,u)$ with $r\in\bigcup_c W_c$} \Comment{both directions; \texttt{semantic\_link} never}
    \State \textbf{if} $\tau(w)\le t_q$ \textbf{then} add $(u,r,w)$ to $N$ \Comment{P1: predicate on the stored stamp}
  \EndFor
  \State for each category $c$: rank the lineages (or CRs) $w$ of $N$ with $r\in W_c$ and $w\models\Phi_c$ by $p_c\cdot\mathrm{sim}(w,q)$ and keep the top $b_c$
  \State $F_k\gets$ the union of the kept lineages (a lineage kept under two categories counts once, so $|F_k|\le B$)
  \State for each kept lineage: append to $\Pi$ its segments inside $\mathcal{G}_{\le t_q}$, one per Release, each listing its versions with $\tau\le t_q$ in $\tau$ order and joined to the lineage by the \texttt{evolves\_to} edge into its opening version; attach to every version the CRs that created it (\texttt{affects}) and to the lineage the traversed edge $(u,r,w)$
\EndFor
\State \Return the chains of $\Pi$, one per kept lineage, ordered by the $\tau$ of their earliest version, every version tagged with its version key and its role for $q$ \Comment{P2}
\end{algorithmic}
\end{algorithm}

\para{Chains and the P2 check.}
A chain emitted by line~12 is, by construction, a connected subgraph of $\mathcal{G}_{\le t_q}$: consecutive versions of a segment are joined by a \texttt{supersedes} edge, each later segment is joined by the \texttt{evolves\_to} edge into its opening version, each creating CR is attached to the version it creates by its \texttt{affects} edge, and the traversed cross-lineage edge joins the lineage to the lineage it was reached from. Sorting is applied only to the versions \emph{within} a segment, where $\tau$ order coincides with \texttt{supersedes} order (previous paragraph), and to the order in which whole chains are listed; it never places two versions next to each other across a missing edge. The mechanical P2 check run on every emitted chain verifies that every consecutive pair within a segment is a \texttt{supersedes} edge of $\mathcal{G}_{\le t_q}$, that every segment join is an \texttt{evolves\_to} edge of $\mathcal{G}_{\le t_q}$, that every attached CR has an \texttt{affects} edge to its version, that $\tau$ is non-decreasing along every segment, and that every node and edge has $\tau\le t_q$. Over all chains emitted for the 1,240 KGQA test questions and the 600 revision-boundary queries of \S\ref{s:results} no violation occurred. The motivating lineage of Fig.~\ref{fig:motivation} is reached from the seed CR by one \texttt{affects} hop and left by one \texttt{references} hop to \S6.7.2, so the chain uses two of the three cross-lineage hops; its segments are the Rel-8 segment opened by CR~0339, the Rel-9 to Rel-15 segments of TS~33.401, each joined by the \texttt{evolves\_to} edge into its opening copy, and the TS~33.501 segment joined by the cross-specification edge.

\para{Cost bound.}
Let $d_{\max}$ be the maximum number of cross-lineage edges (\texttt{affects} or \texttt{references}, either direction) incident to any one lineage; in \peg $d_{\max}=1{,}140$, the 99th percentile is 58 and the mean degree over all edge types is 5.93. Every layer holds at most $B$ lineages (line~9: the per-category shares sum to exactly $B$ and duplicates are merged), expanding a lineage scans at most $d_{\max}$ cross-lineage edges, and there are $K$ layers, so at most $B\,K\,d_{\max}=171{,}000$ edges are scanned and as many constant-time predicate tests are performed. Expanding a lineage's own versions is bounded separately by the lineage size (15.9 versions on average over the full graph, p90 36, maximum 118; a snapshot contains at most that many) and does not count as a hop. The measured number of cross-lineage edges scanned per query is 1,840 on average (p99 6,130), because few expanded lineages approach $d_{\max}$; the budgeted frontier, not the graph size, governs cost, which is why traversal time grows sub-linearly with the number of nodes (Table~\ref{tab:scale}). The bound is a budget guarantee, not a tight estimate; the measured scans and latencies are the operative numbers. $\mathrm{sim}(w,q)$ is the cosine similarity between the bge-large-en-v1.5 embedding of the candidate's text (CR: title and \emph{Reason}; lineage: clause heading and the first 64 tokens of its newest version in the snapshot) and the query embedding.

\begin{table}[!htb]
\centering
\caption{Category-conditioned expansion in full: edge types followed and entity-type filter per category. \texttt{semantic\_link} is never traversed; the entity filter keeps, at each hop, the lineages anchored to at least one entity of the listed types (CR nodes are always kept).}
\label{tab:routing}
\resizebox{\textwidth}{!}{\small\input{tables/supp_routing}}
\end{table}

\para{Connectivity reranking.}
For the top-100 fused candidates,
\[
\mathrm{Conn}_{\mathrm{PEG}}(d,q)=|\mathcal{E}_q|^{-1}\sum_{e_q\in\mathcal{E}_q}\max_{e_d\in\mathcal{E}_d}\exp(-\beta\,d_{\mathcal{G}}(e_q,e_d)),
\]
where $\mathcal{E}_q$ are the entity nodes linked to the query by the NER model and alias table and $\mathcal{E}_d$ those linked to the candidate. $d_{\mathcal{G}}$ is the unweighted shortest-path length in the \emph{undirected projection} of $\mathcal{G}_{\le t_q}$ in which every lineage is contracted to one node: direction is ignored, so an entity reaches, through its \texttt{semantic\_link} anchors, the CRs and lineages behind it, and a whole lineage counts as one step, consistent with the traversal. It is computed by a bidirectional breadth-first search truncated at $2K=6$ steps; unreachable pairs contribute $0$. Queries without entities receive $\mathrm{Conn}=0$ and are ranked by the hybrid score alone (3.1\% of test queries); candidates without entity links, which include the CRs of specifications outside the graph's clause coverage (\S\ref{s:peg}), likewise receive $\mathrm{Conn}=0$; ties are broken by the hybrid score. The reranking adds at most $100\cdot|\mathcal{E}_q|$ truncated searches per query and accounts for 0.07\,s of the 0.31\,s retrieval component.

\para{Storage and index semantics.}
\peg is stored in a property-graph database (Neo4j~5, community edition) with $\tau$ as a property of every node and relationship and $\iota$ as two date properties of clause-version nodes; range indices on $\tau$ and on the version key support the snapshot predicate, which is evaluated inside the traversal query rather than by materialising a per-query subgraph. Chains are serialised with $t^{\mathrm{close}}$ masked whenever it exceeds $t_q$. Text retrieval is designed so that the candidate set equals that of an index frozen at $t_q$: BM25 uses 18 Lucene shards partitioned by approval year; a query at $t_q$ searches the shards of earlier years in full and the shard of year$(t_q)$ under a doc-values range filter $\tau\le t_q$ applied \emph{before} scoring (a filtered query, not a post-filter on a truncated list), so no document dated after $t_q$ can occupy a candidate slot; the only approximation is that the IDF statistics of the current-year shard include the remainder of that year. Dense retrieval uses an exact inner-product search (a flat FAISS index without trained quantisation) over bge-large-en-v1.5 embeddings with an identifier selector $\tau\le t_q$ applied during the scan, which is exactly equivalent to searching an index that contains only the documents dated up to $t_q$. The fused list is cut to 100 candidates after fusion. To bound the IDF approximation we rebuilt exact indices at 12 monthly cut points of the test period and re-ran every retriever of Table~\ref{tab:retrfull}: NDCG@10 differs by at most 0.002 for every retriever and no ordering changes. SPARQL is used only offline, to compile the 5G-KGQA gold answers against an RDF export of the graph queried with Apache Jena; the online system never executes SPARQL. All dates are UTC calendar dates; a query whose $t_q$ falls on an approval date includes that date.

\begin{table}[!htb]
\centering
\caption{Traversal time per audit as the graph grows ($K=3$, $B=50$, one H200). The three smaller graphs are \peg restricted to the CRs and versions of Rel-8--10, Rel-8--12 and Rel-8--15; the last column is the full graph.}
\label{tab:scale}
{\small\input{tables/supp_scale}}
\end{table}

\section{Corpus Construction}
\label{s:corpus}

\para{Harvesting and parsing.}
We harvested every CR with status \emph{approved} from the 3GPP CR database (\url{https://www.3gpp.org/ftp/Information/Databases/}) for TSG meetings between 2008-01 and 2025-06-30: 212,406 files, between 6.1K and 18.4K per calendar year (mean 12.1K), touching 1,250 specifications. CRs are \texttt{.doc} or \texttt{.docx} files whose first pages carry the standard CR cover table. Legacy \texttt{.doc} covers use inconsistent table markup and embedded form fields, so we render the first two pages of every file to images and read the cover fields (\emph{Title}, \emph{Spec}, \emph{CR number}, \emph{Release}, \emph{old/new version}, \emph{Reason for change}, \emph{Summary of change}, \emph{Consequences if not approved}, \emph{Clauses affected}) with Qwen3-VL-30B; on a 1,000-CR expert-checked sample the field accuracy is 98.3\% (lowest: \emph{Clauses affected}, 96.1\%). The redlined body is read from the tracked-change markup of the \texttt{.docx} (\texttt{w:ins}/\texttt{w:del}); \texttt{.doc} files are converted with LibreOffice first and the converted markup is used when present (91\% of \texttt{.doc} files), otherwise the rendered pages are read by the same VL model. Publication dates of specification versions, used for valid time, are taken from the version histories on the 3GPP server.

\para{Deduplication and failures.}
A CR is submitted as a tdoc and may be revised several times before approval; the database lists the revisions. The 3,120 \emph{duplicates} are earlier revisions of a tdoc whose later revision is also listed as approved; we keep the final revision only. The 10,733 \emph{parse failures} (5.1\%) are files whose cover could not be parsed to the mandatory fields (corrupted archives, scanned covers, missing CR number); they are excluded from the corpus and from all counts below, and listed with their failure class in the released manifest.

\para{Security relevance and per-year counts.}
A CR is security-relevant if its \emph{Reason}/\emph{Consequences} fields invoke the 412-term TS~33-series vocabulary (compiled from the definitions and abbreviations clauses of TS~33.102, 33.401 and 33.501 and the TS~33-series titles, then pruned by two experts) and the rule/LLM adjudication of \S3.2 of the main text confirms it: 31,864 candidates, 26,932 confirmed. Over the 17.5 calendar years of the harvest the confirmed CRs average 1.5K per year (26,932 / 17.5), ranging from 0.9K (2008) to 2.9K (2019); the full harvest averages 12.1K per year (212,406 / 17.5, range 6.1K--18.4K). Filter recall is estimated from 800 discarded CRs labelled by two experts (6 security-relevant): $26{,}932/(26{,}932+166{,}689\cdot 6/800)=0.956$ with a Wilson interval of $[0.91,0.98]$ on the 6/800 rate; it is a small-sample estimate and is reported as such. As an independent coverage check, 1,181 (93.0\%) of the 1,270 CRs that CREEK's classifier labelled security-relevant at confidence above 0.99 are among the 26,932; CREEK reports a precision of 91.6\% for that set, so roughly one CR in twelve of it is, by CREEK's own estimate, not security-relevant, and the overlap a perfect filter could reach is bounded accordingly. The two figures therefore measure different things and are not in tension; the recall estimate rests on the expert-labelled discards.

\para{Mirror CRs and the split rule.}
3GPP Category-A CRs replicate a correction made in one Release into other Releases. 27.1\% of the security CRs belong to such mirror families (2.3 members on average). Each mirror is a distinct approved CR with its own tdoc, Release and version key and is therefore a distinct \peg node and a distinct retrievable document; mirrors are grouped by a family identifier (same specification, clause set and \emph{Summary of change} text). Splits are chronological by approval date with one stated exception that reconciles the two rules: a family whose members straddle a split boundary is assigned entirely to the split of its \emph{latest} member. 412 families (1,038 CRs) straddle the 2021-06-30 or 2022-06-30 boundary; their 626 earlier members are thereby withheld from the earlier split, i.e., they are never used to train any component and never serve as query anchors, so every training CR is approved on or before 2021-06-30, no training CR has a mirror in validation or test, and every test query keeps $t_q\ge$ 2022-07-01. The split counts of the main text (26,431 / 2,118 / 5,828) are the counts after this rule. Generated queries are assigned to the split of their latest gold CR; a gold set is built within one split, so no gold set mixes families across splits. Retrieval gold sets count a family at most once, and the pooled judgments treat mirrors of a judged CR as sharing its grade.

\para{Classification corpus.}
To the 26,932 security CRs we add 7,445 hard negatives (the 4,932 candidates rejected by adjudication and 2,513 near-misses whose vocabulary score fell just below the candidate threshold), giving 34,377 CRs split as above. Category counts: Protocol Logic \& State Machine 6,214; Authentication 5,470; Key Management 5,113; Encryption \& Integrity 4,884; Privacy 4,206; Service Availability (DoS) 4,102; Attack Surface \& Threats 3,908; Mobility Security 3,322; 68.2\% of CRs carry one label, 25.4\% two and 6.4\% three.

\para{Lexicon rule and adjudication.}
Each category lexicon $L_c$ lists 38--71 terms (412 in total). Three experts assigned each term a specificity weight in $\{1,2,3\}$ independently (ordinal Krippendorff's $\alpha=0.79$); the released weight is the majority vote. The rule score is $S_{\text{rule}}(x,c)=\min\{1,\tau_c^{-1}\sum_{w\in L_c}\omega_w\mathbf{1}[w\in\mathcal{R}(x)\cup\mathcal{Q}(x)]\}$ over the \emph{Reason} and \emph{Consequences} fields, where $\tau_c$ is the summed weight of the three most specific terms of $L_c$, hence $\tau_c\le 9$ and a weight-3 plus a weight-2 match always reaches $S_{\text{rule}}\ge 5/9>0.5$. The ambiguous band $0.3<S_{\text{rule}}<0.5$ (21.4\% of CR--category decisions) is adjudicated by Qwen3-8B with the category description, the CR's \emph{Reason} and \emph{Consequences} fields and three labelled examples per category (prompt in \S\ref{s:prompts}); experts did not adjudicate individual CRs but audited the stratified 1,500-CR sample reported in the main text.

\section{Protocol Evolutionary Graph: Construction, Coverage and Audit}
\label{s:peg}

\para{Node rules and coverage.}
A \textsf{ClauseVersion} node is created for every (specification, clause, Release, version) that at least one parsed CR lists in \emph{Clauses affected}, and for the copy of every such clause that is created when a later Release of the specification is opened (the opening copy carries the unchanged clause text and the opening version's approval date); clauses that no CR has ever touched have no node. Clause-version nodes are built for the 112 specifications that carry at least one security-relevant CR, which is where every security question of the benchmarks and of CREEK's sets lives; all 26,932 security CRs belong to these specifications. Of the 198,553 CR nodes, 101,204 belong to the 112 specifications and carry \texttt{affects} edges (2.0 per CR on average: 1.6 for the clauses listed on the cover, deterministically, plus the inferred ones), while the 97,349 CRs of the other 1,138 specifications are CR nodes and retrievable documents, so that text retrieval runs over the whole corpus, but carry no edges, are never traversed, and receive $\mathrm{Conn}=0$ in the reranking. This yields 121,277 clause-version nodes organised in 7,650 lineages, one per (specification, clause) with at least one version, with 24,300 Release segments between them: 15.9 versions and 3.2 segments per lineage on average, p90 36 versions, maximum 118; these are full-graph statistics, not snapshot fragments. Entity nodes (30,800) are canonicalised NER mentions (six types: key, identifier, procedure, message, algorithm, network function); specification (112) and Release (17) nodes are metadata. The recall gold standards of \S3.3 of the main text cover ten specifications that together hold 48\% of the clause-version nodes.

\para{Edge semantics and cardinality.}
\texttt{affects}: CR $\rightarrow$ the clause version it creates (one per listed clause; 162K deterministic, 44K inferred from clause identifiers in the body). \texttt{supersedes}: clause version $\rightarrow$ the version of the same clause and Release that it replaces (new to old; at most one per version; absent for the 24,300 versions that open a Release segment, i.e., the first version of each (specification, clause, Release)). \texttt{evolves\_to}: clause version $\rightarrow$ the earliest version of the same clause in the next Release approved on or after it (old to new; at most one same-specification edge per version). It is absent for 21,500 versions for one of three reasons: the lineage ends in that Release (the specification or clause is not continued), the next Release has not yet been opened at the corpus cut-off, or a late correction to the older Release postdates every version of the next Release and no mirror CR carries it forward, so that no later version of the next Release exists to point to; in the last case the correction is reachable through the earlier versions of its segment and through the \texttt{affects} edge of its CR. To these come 9K cross-specification edges for the LTE$\rightarrow$5G successor clauses, described below. \texttt{references}: clause version $\rightarrow$ clause version for explicit cross-references in the clause text (549K). \texttt{semantic\_link}: CR or clause version $\rightarrow$ entity for NER anchors, and entity $\rightarrow$ entity for co-reference judged by the LLM (79K). The counts reconcile: $121.3\text{K}-24.3\text{K}=97\text{K}$ \texttt{supersedes} edges and $121.3\text{K}-21.5\text{K}+9\text{K}\approx109\text{K}$ \texttt{evolves\_to} edges; the 7,650 lineages times 3.2 segments give the 24.3K segment-opening versions. Edge $\tau$ equals the later $\tau$ of its endpoints.

\para{Cross-specification succession edges.}
3GPP publishes no table that maps LTE clauses to their 5G successors; the 9K LTE$\rightarrow$5G \texttt{evolves\_to} edges are therefore inferred and are counted among the LLM-inferred edges throughout. Candidates come from two sources: the 5G clause's explicit cross-reference to its LTE counterpart in its own text (for example, TS~33.501 clauses that state that a procedure ``follows TS~33.401, clause~x'' or that a definition ``is as in TS~33.401''), and title and text similarity between an LTE clause and a 5G clause of the successor specification (BGE-M3 cosine $\ge0.82$, successor pairs such as TS~33.401$\to$33.501, 24.301$\to$24.501, 36.331$\to$38.331). Each candidate is judged by the succession prompt of \S\ref{s:prompts} with both texts and their metadata; the edge runs from the LTE version current at the creation date of the 5G clause to the 5G clause's first version. The edge manifest of the artifact lists, for every cross-specification edge, its candidate source (cross-reference or similarity) and the judge's one-sentence justification, so each can be checked against the specifications. Sixty such edges were audited by the three experts, 57 of them correct (precision 0.950); they are the cross-specification part of the 100 LLM-inferred \texttt{evolves\_to} edges of Table~\ref{tab:audit}, whose 40 within-specification edges score 35/40. The edge of the motivating case, TS~33.401 \S6.5$\to$TS~33.501 \S6.4.5, comes from the cross-reference source: the 5G clause defines the NAS COUNT handling of a partial native 5G security context in the same terms and refers to the LTE clause for the corresponding EPS context.

\para{Construction.}
(1)~The symbolic skeleton is extracted from $T_{\text{struct}}$ and the version histories: Release and version keys with their publication dates, listed \texttt{affects}, same-specification successions from version metadata, and \texttt{references} from clause-number patterns in the clause text. (2)~The NER model is RoBERTa-base trained on 3,000 expert spans (two annotators, span-level $\kappa=0.84$), expanded by two rounds of self-training on pre-2021-07 CRs in which Qwen3-32B accepted or rejected proposed spans (11,400 and 24,600 spans after rounds one and two); F1 is 0.912 on 500 held-out expert spans. (3)~Candidate implicit edges are proposed by title/text similarity (renumbered or cross-specification successions; BGE-M3 cosine $\ge 0.82$), by explicit cross-references, by clause identifiers in the CR body (\texttt{affects}) and by entity-embedding similarity (\texttt{semantic\_link}); Qwen3-32B judges each candidate with the two texts and the metadata (prompts in \S\ref{s:prompts}); 97.2\% of listed clause identifiers resolve directly, the rest through time-stamped aliases that record clause renumbering.

\para{Transaction time and public availability.}
$\tau$ is the TSG approval date (the date of the plenary that approved the CR pack, as recorded in the specification's change history). CR documents are public on the 3GPP file server from the working-group meeting at which they were first tabled, weeks to months before TSG approval, so a snapshot at $t$ may \emph{omit} a document that was already public but never contains one that was not. The conservative direction is the safe one for leakage control; it slightly disadvantages all systems equally.

\begin{figure}[!htb]
\centering
\includegraphics[width=\textwidth]{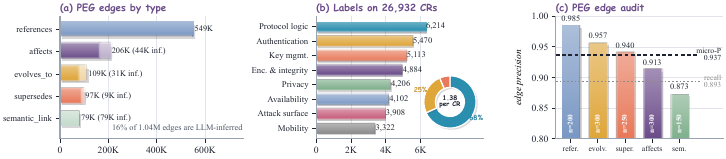}
\caption{Dataset overview. \textbf{(a)} \peg edges by type (deterministic vs.\ LLM-inferred). \textbf{(b)} Category labels on the 26,932 security CRs and label cardinality. \textbf{(c)} Edge-audit precision by type with the micro-precision and the recall of the gold reconstruction.}
\label{fig:data}
\end{figure}

\para{Audit by provenance.}
Table~\ref{tab:audit} gives the 1,200-edge audit broken down by edge type and provenance; the per-type precisions and the micro-average 0.937 of the main text are recovered exactly, and re-weighting by the full-graph edge counts gives 0.955. Fig.~\ref{fig:data} summarises the edge counts by type and provenance, the label distribution and the per-type audit precision. Of 200 random three-hop chains, 174 (0.87) are fully correct.

\begin{table}[!htb]
\centering
\caption{Edge audit (three experts, majority vote, Fleiss' $\kappa=0.81$) by edge type and provenance.}
\label{tab:audit}
\small\input{tables/supp_edge_audit}
\end{table}

\para{Recall gold standard.}
Two experts reconstructed, from the specification documents and change histories alone, the \texttt{supersedes}/\texttt{evolves\_to}/\texttt{affects} subgraph among the security-relevant clauses (clauses touched by at least one security CR) of five specifications across Rel-8--17; the scope excludes \texttt{references} and \texttt{semantic\_link}, which is why 2,145 gold edges suffice for five large specifications. The five were chosen before construction as the most CR-dense security-relevant specifications across layers (TS~24.301, 33.401, 33.501, 23.502, 38.331); \peg recovers 1,915 (0.893). Because CR density may favour recall, the same protocol was applied afterwards to five specifications drawn uniformly at random from those with at least 50 security CRs (TS~33.220, 33.210, 24.501, 36.413, 29.274): 1,148 of 1,318 gold edges (0.871). The ten specifications hold 48\% of all clause-version nodes; recall outside them is not measured and the audit precision, which is stratified over the whole graph, is the graph-wide quality figure.

\section{Benchmarks, Annotation and Judging Protocols}
\label{s:bench}

\para{Splits and query metadata.}
5G-Retrieval 1,610 / 321 / 719, 5G-KGQA 2,750 / 430 / 1,240 and 5G-SecQA 2,340 / 431 / 1,129 queries (train / validation / test, by the approval date of the latest gold CR at the corpus boundaries, with the family rule of \S\ref{s:corpus}); every test query has $t_q\ge$ 2022-07-01. Expert-written slices: 300, 150 and 100 items. Every query carries $t_q$ and $R_q$ as metadata. For generated queries $t_q$ is the approval date of the latest gold CR plus a uniform 30--180 day offset and $R_q$ is that CR's Release (5G-KGQA: the Release of the sampled template); for expert-written queries the expert states both when writing the query; for the CREEK cases they are the anchor date and the Release of the case CR. All systems receive the header line ``Audit date: $\langle t_q\rangle$. Target Release: $\langle R_q\rangle$.'' prepended to the question, so that no system lacks the temporal information the task needs. The training splits are used as follows: 5G-Retrieval's 1,610 queries fine-tune monoT5-3B, bge-reranker-v2 and TempRetriever in domain and, together with 20K weakly supervised title-to-cited-CR pairs, the dense retriever; 5G-KGQA's 2,750 questions train CronKGQA and TempoQR and tune the text-to-SPARQL few-shot pool; 5G-SecQA's 2,340 questions tune the $(\alpha,\beta,K)$ grid and every baseline's exposed hyper-parameters on validation. No generator is fine-tuned on any benchmark.

\para{5G-Retrieval.}
Gold sets of 1--5 CRs are sampled as connected sets in \peg (shared clause or an evolution edge); GPT-5 writes a query from the CR covers under concept-coverage control (each gold set must cover at least two of the security concepts named in its CRs, and no concept may appear in more than 3\% of queries), then a second GPT-5 pass rejects queries answerable from a single CR unless the type is \emph{factual}. Relevance grades: 2 for a primary gold CR, 1 for a supporting CR (linked by a shared clause or an evolution edge to a primary CR), 0 otherwise; NDCG@10 uses gain $2^g-1$ and the ideal ranking of the graded set. \emph{Judgment-independent pooled relevance}: the pool of a query is the union of the top-20 results of every retriever in Table~\ref{tab:retrfull} except \sys's own runs (so that pooling cannot favour our system), 38.2 unique CRs per query on average; two telecom-security engineers independent of the authors graded every pool document on the same three-level scale from the CR covers and the query ($\kappa=0.81$), without access to \peg edges or gold labels; the pools cover 94.3\% of \sys's top-10 and unjudged documents count as non-relevant, a bias against \sys. The margin over monoT5 is +2.7/+2.8/+2.7 points at pool depths 10/20/30; at depth 20 it passes both tests of the main text (95\% CI $[+1.2,+4.4]$, query-level $p=0.002$, seed-level $p=0.003$) and is therefore marked.

\para{The expert-written retrieval slice.}
The queries themselves remain generated from \peg-connected gold sets, which is why the paper additionally reports 300 expert-written queries whose query and gold inherit nothing from the graph. Each expert wrote a query from the CR database and the specifications, stating $t_q$ and $R_q$. Relevance was then graded on the same three-level scale over a pool formed by the union of the top-20 results of \emph{every} retriever of Table~\ref{tab:retrfull}, \sys's runs included, shuffled and shown without system identity (31 unique CRs per query on average), so that no retriever's results are unjudged; the authoring expert graded every query, and 100 of the 300 queries were graded independently by a second expert as well ($\kappa=0.79$ on the three-level grades). The slice has 110 comparative, 80 analytical, 70 temporal and 40 factual queries; \sys's margin over bge-reranker-v2 (the strongest baseline on this slice) is +1.9, +2.4, +6.1 and $-$1.5 NDCG@10 points by type and +2.6 overall (95\% CI $[+0.5,+4.7]$, paired bootstrap $p=0.02$, seed-level $p=0.01$); it passes the 0.05 level on both tests but not the 0.01 query-level threshold that the main text's $\dagger$ requires, so the main text reports it unmarked.

\para{5G-KGQA and gold verification.}
Subgraphs of 2--4 cross-lineage edges whose lineage segments span 1--11 Releases are sampled from \peg under four templates (temporal, relation, multi-hop, entity-linking), compiled to SPARQL over the RDF export and executed to obtain identifier-set answers; GPT-5 then paraphrases the template question into natural language, so the answering model (Qwen3-8B) is not involved in writing. EM is set equality after lower-casing, whitespace normalisation and alias resolution (clause renumbering, tdoc/CR-number equivalence). Because the gold inherits \peg errors, \emph{all} 1,240 test questions were re-verified by two annotators against the specification documents and change histories without \peg: 1,194 (96.3\%) agree with the PEG-derived gold and the 46 disagreements were corrected to the specification (disagreements between the two annotators were adjudicated by discussion). Every EM value in the main text and in this supplement is computed against the verified gold. Table~\ref{tab:kgqagold} reports, for the systems that use \peg, EM on the uncorrected PEG-derived gold next to the reported value: \sys loses 1.9 points and the strongest baseline 0.4, so the correction removes 1.5 points of the margin; the training and validation gold remain PEG-derived, which the main text states. The 150 expert-written questions follow the same answer format; their cross-lineage depth (1--5, mean 2.3) was set by the experts, not by our generator.

\begin{table}[!htb]
\centering
\caption{Effect of the gold verification on 5G-KGQA EM (test split, 1,240 questions).}
\label{tab:kgqagold}
\resizebox{0.9\textwidth}{!}{\small\input{tables/supp_kgqa_gold}}
\end{table}

\para{5G-SecQA and the counterfactual references.}
GPT-5 generates a question and a reference answer of at most 80 words from the CR text and the cited clauses (never from \peg chains) under three templates: \emph{factual} (what the CR changes), \emph{causal} (why, from \emph{Reason for change}) and \emph{counterfactual} (what would follow for the affected procedure had the CR not been approved). The \emph{Consequences if not approved} field is part of the indexed cover and is therefore available to every system; it is typically one line (``replay of the first NAS message is possible''). The counterfactual reference restates that consequence in terms of the affected procedure and its clause text (which message, which counter or state, which party is affected), so a report cannot obtain full credit by copying the cover field, and the low absolute F1 of all systems on this type (\sys 0.108; 0.161 with gold evidence) measures the difficulty of that restatement. A \emph{gold reasoning chain} is the expert-verified \peg chain from the CR to the clauses that realise its consequence (the clause it changes, the procedure clause that uses it and the message or state the consequence concerns); given that chain, F1 reaches 0.247, which is why the main text locates the bottleneck in reasoning over the chain rather than in retrieving it. Two senior engineers reviewed every question and answer double-blind (87 discarded, 214 corrected; $\kappa=0.78$).

\para{Metrics.}
CiteP: the fraction of cited (claim, evidence) pairs in which the evidence entails the claim. CiteR: the fraction of gold evidence items cited by at least one entailed claim. UCR: the fraction of atomic claims entailed neither by any item of the system's input evidence nor by any gold evidence item; UCR-G counts only gold evidence and is the variant comparable across systems whose input evidence differs in size. All claim-level metrics are computed on the delivered report with one gold set and scorer for every row. Entailment is DeBERTa-v3-large NLI with Qwen3-32B adjudication of pairs whose NLI confidence lies in $[0.35,0.65]$ (0.91 F1 against 300 expert-labelled claims; adjudication prompt in \S\ref{s:prompts}). Judge: DeepSeek-V3.1 scores correctness, groundedness and usefulness (1--5) from the question, the reference answer and the report; the Judge column of the main table is the mean over all 1,129 questions; the DeepSeek-V3.1 backbone is judged by GPT-5, which also wrote the questions, so its Judge values are listed for completeness and enter no comparison. Expert: three telecom-security experts score the same three dimensions for the 100 expert-written questions with systems hidden and order randomised (Krippendorff's $\alpha=0.74$); the expert score is the mean of the three dimensions and exists for the 18 non-ablation systems of the primary backbone (5,400 ratings, 1,800 reports), which is why it is reported in its own table in \S5.5 of the main text rather than as a column of the main table. Because the expert study has one seed, its differences are tested by the paired bootstrap over the 100 questions only.

\para{Blinding of the expert study.}
Before rating, every report of every system was re-rendered into one fixed template: numbered atomic claims (the output of the same splitter used for scoring), bracketed citations normalised to the form [tdoc] or [TS~xx.xxx \S c, v], and no system-specific headers, chain markup, role tags or Diagnosis block. Experts therefore saw the same surface form for every system; what differs between systems is which claims are made and which identifiers are cited, which is what the study is meant to measure. The expert-study table of the main text lists, for the 18 systems, citation precision on the same 100 questions, the judge mean over the same 100 questions and the expert mean (the judge mean and the citation precision over all 1,129 questions are columns of the main table; the ordering of the systems is the same under both CiteP values); the judge-minus-expert offset falls with citation precision (Spearman $\rho=-0.98$ over the 18 systems; it is not strictly monotone). Two readings are compatible with this pattern, a judge that discounts densely cited text or experts who credit it; the main text reports both columns, the ordering, which is identical under both, and the experts' correctness dimension alone, which does not ask about citations and ranks the systems identically ($\rho=0.97$ to the full score) with a lead of +0.41 for \sys over the long-context baseline.

\para{Expert-written slices.}
Three experts (two 3GPP SA3/CT1 delegates with more than eight years of standardisation experience and one academic security researcher), independent of the authors and paid at professional consulting rates, wrote the queries and questions from the CR database and the specifications; they were given the benchmark formats and the category definitions but neither \peg, the generation prompts nor any generated example.

\para{Classifier and the input parity of its baselines.}
RoBERTa-base; input = cover fields followed by the redlined body, tokenised into 512-token chunks with stride 128 (71\% of CRs fit in one chunk, 96\% in three); chunk-level token states are max-pooled before label-query attention; eight learnable label queries initialised from the encoded category descriptions; gate MLP on the pooled \texttt{[CLS]} state; losses: gate BCE on all 34,377 CRs, per-label BCE on gated CRs, multi-label supervised contrastive loss on the concatenated label vectors projected to 128 dimensions, where two gated CRs are positives when their label sets intersect, weighted by Jaccard overlap, temperature 0.07; AdamW, learning rate $2\times10^{-5}$, 5 epochs, batch 32, label threshold 0.5 tuned on validation; 5 seeds. Every encoder baseline (BERT-base, RoBERTa-base, SecureBERT, DeBERTa-v3-base, HGCLR) receives the identical chunked input with the same max-pooling and the same per-label BCE head, so the gap between plain RoBERTa-base (Macro-F1 0.718) and our module (0.849) is attributable to the gate, the label-query attention and the contrastive loss; the three ablations of the main text remove these one at a time ($-$0.020, $-$0.032, $-$0.063).

\para{Annotation workload and compensation.}
Table~\ref{tab:workload} lists every human activity of the project, including the gold-standard reconstructions, NER annotation, parsing and entailment audits, splitter audit, the expert-written slices with their relevance grading and double grading, the RFC probe and the revision-boundary pairs, with the number of judged items, annotators and hours. Annotators were compensated at local professional rates for domain experts; the total of 2,064 expert-hours was funded by the authors' institution. Guidelines (relevance grades, category definitions with positive and negative examples, strict-correctness rubric, judging dimensions) are released with the artifact.

\begin{table}[!htb]
\centering
\caption{Complete human annotation workload.}
\label{tab:workload}
\resizebox{\textwidth}{!}{\footnotesize\input{tables/supp_workload}}
\end{table}

\section{Baseline Configurations, Temporal Footing, Training Data and Deployment}
\label{s:baselines}

\para{Hardware, engines and budgets.}
Training used two NVIDIA H200 GPUs (141~GB each); serving used one unless stated. Table~\ref{tab:deploy} lists weights precision, inference engine, GPU count and token budgets for every generator. Every +PEG system receives its evidence under a 4K-token budget; audit reports are decoded with at most 512 new tokens (310 on average), the claim splitter with at most 128 (48 on average), KGQA answers with at most 64; the long-context baseline reads up to 32K input tokens. For 5G-KGQA every end-to-end system answers with a short-answer instruction (an identifier set) over the same evidence it would receive for an audit; the citation instruction and the cite-or-drop check act on 5G-SecQA reports only, so the two citation ablations of the main table leave EM and EM-H unchanged by construction. Qwen3-235B-A22B is served in FP8 across two H200s with tensor parallelism; DeepSeek-V3.1 and GPT-5 are accessed through their provider APIs and their per-audit costs are the API prices, which is why DeepSeek-V3.1 appears as a backbone and a judge but not in the GPU-based serving-cost table.

\begin{table}[!htb]
\centering
\caption{Deployment of every model used for generation, judging or extraction.}
\label{tab:deploy}
{\footnotesize\input{tables/supp_deploy}}
\end{table}

\para{Temporal footing of every baseline.}
Every system, ours and the baselines', answers each test query against exactly the CRs dated on or before $t_q$. For \sys and the text retrievers this is the pre-scoring date filter of \S\ref{s:algo}. The graph-RAG baselines (GraphRAG, LightRAG, HippoRAG~2, T-GRAG) cannot be filtered after the fact because their community summaries and personalised-PageRank graphs are built from whole corpora; they were therefore indexed once on the CRs approved up to 2022-06-30 and then advanced \emph{per query date}: the 3,088 test queries of the three benchmarks are processed in order of $t_q$, and before the queries of each distinct $t_q$ are answered, each baseline's own incremental update path (GraphRAG's incremental indexing, LightRAG's incremental insertion, HippoRAG~2's incremental graph update, T-GRAG's time-stamped insertion) inserts the CRs approved after the previous query date and up to $t_q$. The 3,088 queries carry 1,131 distinct $t_q$ values, so each baseline is advanced in 1,131 incremental steps; the generated $t_q$ lie between 2022-07-31 and 2025-12-27 (the earliest and latest test-period approval dates plus the 30- and 180-day offsets, 1,246 calendar days) and the expert-written $t_q$ inside the test period, so the count is bounded by 1,246 and is the number of distinct dates in the released query manifest, not the number of queries. A document-level $\tau\le t_q$ filter on the retrieved units is kept as a safeguard. Under this protocol the gold CRs of every test query are present in every baseline's index at query time (coverage 100\%, checked per query), so no comparison in the main text mixes retrieval quality with index freshness; the only remaining difference between the footing of the baselines and ours is that their community summaries reflect the corpus up to $t_q$ exactly as ours does. The long-context baseline reads the same $\mathcal{G}_{\le t_q}$ neighbourhood as \sys. All systems receive the $t_q$/$R_q$ header line.

\para{Retrievers and in-domain fine-tuning.}
\sys's dense retriever is off-the-shelf bge-large-en-v1.5; BGE-M3 is used only to compute SemSim, keeping retriever and metric disjoint. monoT5-3B, bge-reranker-v2 and TempRetriever are fine-tuned on the 1,610 training queries of 5G-Retrieval (graded gold, 3 epochs, validation early stopping) and the main-paper numbers are these in-domain versions; their off-the-shelf NDCG@10 is 0.392, 0.387 and 0.336 (Table~\ref{tab:retrfull}). The dense baseline fine-tuned in-domain (``bge-large, in-domain FT'') uses the same 1,610 queries together with 20K weakly supervised query--CR pairs (query = CR title, positive = the CR it explicitly cites) approved by 2022-06-30; no \peg clusters are used. Dense retrieval encodes each CR as its cover fields followed by the redlined body truncated to 512 tokens (the bge-large limit). Table~\ref{tab:retrfull} also reports a passage-level variant (512-token chunks, max-pooled scores), which lifts off-the-shelf NDCG@10 from 0.183 to 0.247 and, with in-domain fine-tuning, to 0.319, still below BM25 (0.377); the main-paper conclusion about dense retrieval on identifier-heavy queries does not rest on truncation. The gap persists on the expert-written queries (dense 0.164 against BM25 0.371), so it is a property of identifier-heavy telecom queries rather than only of the shared vocabulary between generated queries and covers.

\para{Version-metadata control (+VM).}
Standard-RAG + VM + cite receives the same retrieved CRs as Standard-RAG + cite and, for each of their listed affected clauses, the version of that clause current at $t_q$ for $R_q$, resolved from the cover's version step and the specification's version history without any graph edge; the versions are appended to the evidence as ``current text of clause $c$: version $v$''. It therefore has everything that deterministic version metadata provides but no \texttt{supersedes}, \texttt{evolves\_to} or \texttt{references} structure. It selects the right version in most cases (EM 0.441 against 0.290 without metadata) but cannot relate clauses to one another, which is where the remaining gap to the matched +PEG baseline (0.517) and to \sys (0.752) arises.

\para{Telco-RAG.}
Telco-RAG is run with its released pipeline (specification-aware chunking, query augmentation with 3GPP vocabulary, and its router) over the same 198,553 CRs, with our citation instruction and cite-or-drop post-processing appended to its generation step and Qwen3-8B as the generator; its chunk size and top-$k$ are tuned on the validation split.

\para{Graph-RAG baselines.}
GraphRAG, LightRAG, HippoRAG~2 and T-GRAG are indexed on the same 198,553 CRs with Qwen3-32B as the extraction LLM. For each system, chunk size (512 or 1024), community level (1 or 2, GraphRAG), temporal decomposition depth (1 or 2, T-GRAG) and top-$k$ (10, 20 or 50) are tuned on the validation split. ``+PEG'' replaces each system's extracted relations by the \peg records of the same documents; for T-GRAG the \peg records, stamped with $\tau$, replace its time-stamped facts while its temporal question decomposition and time-scoped subgraph retrieval are kept unchanged, so the comparison isolates specification-revision semantics from general temporal-graph machinery.

\para{KGQA-only baselines and oracle linking.}
The text-to-SPARQL baseline (Qwen3-32B, few-shot with the KGQA templates) is evaluated on 5G-KGQA only. CronKGQA (TComplEx embeddings of \peg's versions, CRs and dates, trained on the 2,750 KGQA training questions) and TempoQR (its question-time and entity-time embeddings over the same graph) are trained on the same split and answer with identifier sets. All three are run twice: with their own entity linking (the alias table alone resolves 71\% of the entity mentions in the test questions) and with \emph{oracle linking}, in which the gold mention-to-node links of the verified gold are supplied, so that the remaining errors are errors of query construction or scoring (Table~\ref{tab:kgqaonly}). Oracle linking lifts text-to-SPARQL from 0.641 to 0.712 and the embedding models by about 0.1, which places part but not all of their gap to \sys (0.752) in entity linking; \sys itself uses the same NER model and alias table without oracle links.

\begin{table}[!htb]
\centering
\caption{KGQA-only baselines over \peg with their own entity linking and with oracle mention-to-node links (5G-KGQA EM; EM-H on the 150 expert-written questions).}
\label{tab:kgqaonly}
\resizebox{0.8\textwidth}{!}{\footnotesize\input{tables/kgqa_only}}
\end{table}

\para{Long-context baseline.}
Qwen3-235B-A22B (22B active parameters) receives the question and the same \peg neighbourhood that \sys traverses, serialised as text (CR covers and clause versions with their keys, in document order) within 32K tokens, with the citation instruction and the identical cite-or-drop post-processing; it therefore shares \sys's evidence and citation rule and differs in model size and in reading text rather than typed chains; it is a long-context control, not a graph-RAG baseline, and ``the strongest graph-RAG baseline with \peg records'' in the main text refers to T-GRAG + PEG + cite. The 8B control in Table~\ref{tab:extra} (Qwen3-8B with the same 32K-token input) separates model size from evidence presentation.

\para{The matched control, the evidence budget and the conditions of every control row.}
The row printed as ``Std-RAG+PEG+cite (matched)'' is a \emph{matched control}, not a Standard-RAG run: for each question it receives exactly the final reranked CRs and \peg chains that \sys receives, under the same 4K-token budget, instruction and post-processing, serialised as an unordered set of records. Every record carries the same six fields in every presentation condition of the attribution ladder: (subject, relation, object, version key, approval date, role for $q$), where the role is \cur or \hist exactly as \sys's chain markup states it. The 4,096-token budget is applied once, to the record set: \sys's chains are added lineage by lineage in traversal order (highest-scored first) until the budget is reached, which happens on 11.6\% of questions (38.4 records per question on average); the resulting record set is then re-serialised for each ladder condition, so the final model inputs of the ladder differ only in ordering and grouping, never in content. Conditions in which a baseline keeps its \emph{own} retrieval are different rows with different numbers: HippoRAG~2 + PEG + cite and T-GRAG + PEG + cite replace the baseline's extracted relations by the \peg records of the documents it retrieved, and Standard-RAG's own retrieval with \peg records is reported only in the chain-format control of \S\ref{s:results} (Table~\ref{tab:chainfmt}), since the main-table Standard-RAG + PEG + cite row is the matched control. Table~\ref{tab:conditions} states, for every control row of the paper, the retriever and evidence set, the organisation of the evidence, whether the Diagnosis block is present and the post-processing.

\begin{table}[!htb]
\centering
\caption{Conditions of every control row: retriever and evidence set, organisation of the evidence, Diagnosis input and post-processing. Rows that share an evidence set differ only in organisation.}
\label{tab:conditions}
{\footnotesize\input{tables/supp_conditions}}
\end{table}

\section{Additional Results}
\label{s:results}

\begin{table}[!htb]
\centering
\caption{Full retrieval table (all retrievers and metrics), the monoT5\,+\,PEG combination, the dense document-representation variants and the off-the-shelf rerankers.}
\label{tab:retrfull}
\setlength{\tabcolsep}{3.5pt}
{\footnotesize\input{tables/supp_retrieval_full}}
\end{table}

\begin{table}[!htb]
\centering
\caption{Every Qwen3-8B configuration with SemSim, EM and UCR, followed by the four rows omitted from the main table (LightRAG without \peg, TempRetriever\,+\,PEG, HippoRAG~2\,+\,PEG without the citation prompt, 8B long-context control).}
\label{tab:extra}
\resizebox{0.95\textwidth}{!}{\footnotesize\input{tables/supp_extra_rows}}
\end{table}

\begin{table}[!htb]
\centering
\caption{Full multi-label classification table, including the weaker encoders (all on the identical chunked input). CREEK-25: category agreement with two-expert labels on the 25 security CRs of CREEK's manual testing set (majority vote over seeds).}
\label{tab:clffull}
\resizebox{0.95\textwidth}{!}{\footnotesize\input{tables/supp_classification_full}}
\end{table}

\para{Classifier results in full.}
Table~\ref{tab:clffull} gives every classifier on the chronological test split. \sys's module reaches Macro-F1 0.849$\pm$0.004 against encoders on the identical chunked input (strongest HGCLR 0.789; plain RoBERTa-base 0.718), R-GCN over \peg (0.801) and LoRA-tuned Qwen3-8B (0.826); removing the contrastive loss costs $-$0.063, the category attention $-$0.032 and the gate $-$0.020. Re-scored on the expert-labelled, test-period 1,500-CR audit set, trained systems drop by 1.4--2.9 points with the order preserved and the margin over LoRA widens to +3.6; on CREEK's 25 manual-set security CRs the order holds again (21/25 category agreement against 19/25 for LoRA and 18/25 for R-GCN).

\para{Significance tests.}
Two tests are run for every \sys-vs-strongest-baseline comparison: a paired bootstrap over queries (10K resamples of the seed-averaged per-query scores, Holm--Bonferroni within a table) and a paired $t$-test over the five training seeds, for which Table~\ref{tab:seed} reports the mean and the standard deviation of the five per-seed differences together with $t$ and $p$ (df~4), so that every $p$-value can be recomputed from the table; \S5.1 of the main text states which of the six comparisons pass. The main paper marks a gain only when both tests pass. Systems without a trained component (LLM-only, Standard-RAG with and without the citation prompt or version metadata, Telco-RAG, the graph-RAG baselines with their own retrieval, and the KGQA-only text-to-SPARQL baseline) are single deterministic runs at temperature 0; their EM and EM-H values are therefore exact fractions $k/1{,}240$ and $k/150$, and the shares of the revision-boundary test are fractions $k/300$. Systems whose evidence depends on a trained component (\sys, its ablations, the matched control, the long-context reader that reads \sys's neighbourhood, TempRetriever with its fine-tuned retriever, CronKGQA and TempoQR) are means over five training seeds. The token-F1 gain over the long-context baseline (+0.007, SD of the differences 0.004) passes both ($p=0.017$) and is marked, but it is below two seed standard deviations and is therefore described as small; the SemSim gain (+0.002, SD 0.003) passes the query-level test but not the seed-level one ($p=0.21$) and is unmarked. Table~\ref{tab:human} gives the paired-bootstrap deltas on the 100 expert-written 5G-SecQA questions.

\begin{table}[!htb]
\centering
\caption{Seed-level paired $t$-tests ($n=5$ seeds) for the \sys-vs-strongest-baseline gains.}
\label{tab:seed}
\resizebox{0.8\textwidth}{!}{\footnotesize\input{tables/supp_seed_tests}}
\end{table}

\begin{table}[!htb]
\centering
\caption{Paired-bootstrap tests on the 100 expert-written 5G-SecQA questions: \sys minus the strongest baseline (Qwen3-235B long-context + cite) and \sys minus the strongest chain-format control (T-GRAG retrieval, typed chains); the latter has no expert rating.}
\label{tab:human}
\resizebox{0.95\textwidth}{!}{\footnotesize\input{tables/supp_human_tests}}
\end{table}

\para{Attribution ladder (evidence held fixed).}
Every ladder row receives exactly \sys's final, category-routed evidence, cut once to the 4K budget as described in \S\ref{s:baselines}, as records with the six fields, roles included. The first three rows differ only in the ordering and grouping of identical records: unordered (EM 0.517, UCR 0.118; the matched control), sorted by approval date (0.603, 0.104), grouped into typed chains per lineage in \sys's chain layout (0.734, 0.071). The last row adds the Diagnosis input (categories with confidences), which is additional information rather than a re-arrangement, and is \sys (0.752, 0.062). Arrangement therefore accounts for 0.217 of the 0.235 gap and the Diagnosis input for 0.018. The Diagnosis gain is concentrated on relation (+0.034) and entity-linking (+0.027) questions, with +0.014 on multi-hop and +0.001 on temporal questions; the mechanism is that the predicted categories select the alias table and relation vocabulary with which the generator names identifiers in its answer set (for example, a Key Management category makes the generator resolve ``the derived key'' to the K$_{\mathrm{NASenc}}$ identifier rather than to the procedure that derives it), which is why the gain is small and lands on the question types where naming is the main source of error. Routing is measured separately by the ``w/o category routing'' ablation, which runs the full pipeline on unrouted evidence (EM 0.749, UCR 0.081): routing changes which evidence is reached and mostly affects faithfulness, whereas Diagnosis changes how the retained evidence is named. On the 150 expert-written questions the four ladder rows reach EM-H 0.447, 0.540, 0.671 and 0.687. Three paraphrases of the KGQA answer instruction change EM by at most 0.004, and three paraphrases of the audit instruction change F1 by at most 0.003 and UCR by at most 0.004. Flattening \sys's own chains to an unordered record set without Diagnosis gives EM 0.519.

\para{Chain-format control (retriever held fixed).}
To separate the arrangement effect from the retrieval itself, each baseline's \emph{own} retrieved CRs were expanded to their \peg records in the snapshot and given to the same generator under the same budget, instruction and check, once as unordered records and once in \sys's chain layout, without routing and without Diagnosis (Table~\ref{tab:chainfmt}). For HippoRAG~2 and T-GRAG the unordered condition is their +PEG+cite row of the main table; for Standard-RAG it is a separate run on its own top-100 (EM 0.486, UCR 0.124), which lies between the version-metadata control (0.441) and the matched control (0.517) and must not be confused with the latter. The chain layout lifts the three retrievers to 0.581, 0.688 and 0.701 (UCR 0.083, 0.075, 0.072; EM-H 0.507, 0.593, 0.613 on the 150 expert-written questions), a gain of 0.089--0.095 EM on each, and the matched evidence in the same layout reaches 0.734, so the organisation is a reusable component that carries most of the improvement to any retriever. What the full system adds beyond the strongest of these controls, T-GRAG retrieval with typed chains, is measured on the same questions and seeds in Table~\ref{tab:chainfmttests}: +0.051 EM (95\% CI $[+0.031,+0.071]$, query-level $p=0.001$), +0.074 EM-H ($[+0.012,+0.136]$, $p=0.02$), $-0.010$ UCR ($[-0.018,-0.002]$, $p=0.02$) and +0.038 CiteP ($[+0.016,+0.060]$, $p=0.003$), with seed-level $p\le0.005$ for all four; the UCR margin thus clears the 0.05 level but not the 0.01 level that the main table's $\dagger$ requires, which is why the main text describes it as small. On the 100 expert-written 5G-SecQA questions the same control scores token-F1 0.262, CiteP 0.797, UCR 0.079 and judge 3.76; \sys's margins there are significant for the judge (+0.28, $p=0.004$) and for CiteP (+0.029, $p=0.03$) and not for UCR ($-0.008$, $p=0.21$) or F1 (+0.019, $p=0.12$; Table~\ref{tab:human}). The chain-format controls were run after the expert study, which covers the 18 systems of the main table, so no expert rating exists for them and the main text claims no expert-judged superiority over them. The claim the paper makes for the full system beyond the organisation is therefore a modest but stable margin in EM, EM-H, citation precision and judged quality, and a small one in unsupported-claim rate that is significant on the full test set but not on the 100-question slice; token-F1 does not separate the two on the slice.

\begin{table}[!htb]
\centering
\caption{Paired comparison of \sys with the strongest chain-format control (T-GRAG retrieval, typed chains) on the full 5G-KGQA and 5G-SecQA test sets: query-level bootstrap and seed-level $t$-test.}
\label{tab:chainfmttests}
\setlength{\tabcolsep}{8pt}
\resizebox{0.95\textwidth}{!}{\footnotesize\input{tables/supp_chainfmt_tests}}
\end{table}

\begin{table}[!htb]
\centering
\caption{Chain-format control on 5G-KGQA and 5G-SecQA: each baseline's own retrieved CRs as unordered \peg records and as typed chains in \sys's layout (same generator, budget, instruction and check; no routing, no Diagnosis), next to the matched control that holds \sys's final evidence in the same two layouts.}
\label{tab:chainfmt}
\resizebox{0.85\textwidth}{!}{\footnotesize\input{tables/supp_chainfmt}}
\end{table}

\para{The two citation ablations and the claim-deletion accounting.}
The main table contains two citation ablations with different scope. \emph{w/o cite-or-drop check} keeps the citation instruction and the drafts (temperature 0, identical inputs) and only disables the deletion step; because a deletion-only check can neither add citations nor change which gold items are cited by entailed claims, its CiteR equals the full system's (0.769) and its F1 is unchanged, while UCR rises from 0.062 to 0.084 and CiteP falls from 0.842 to 0.812, which is exactly the share of uncited and P3-violating claims the check removes. \emph{w/o citation enforcement} removes the instruction as well, so its audit drafts differ: CiteR falls to 0.655 because the generator cites less and UCR rises to 0.121. EM and EM-H are identical to the full system's in both rows by construction, since the 5G-KGQA answers are produced by the separate short-answer prompt of \S\ref{s:baselines}, which carries no citation instruction and is not post-processed; the two rows therefore say nothing about reasoning, and the graph's contribution to EM is measured by the edge ablations (0.721 without LLM-inferred edges, 0.684 without \texttt{supersedes}/\texttt{evolves\_to}). Table~\ref{tab:claims} gives, for every +cite system of the primary backbone, the claims per report in the draft and in the delivered report, the share deleted, and the UCR of the draft and of the delivered report. The +cite baselines delete 10.1--12.0\% of their claims and deliver 13.9--14.3 claims per report; \sys deletes 7.9\% and delivers 14.2, and its draft UCR of 0.084 is already below the delivered UCR of every baseline of the main table, so the faithfulness margin is not produced by deleting more. The only delivered UCRs below 0.084 belong to the chain-format controls (0.072--0.083, Table~\ref{tab:chainfmt}), which adopt \sys's organisation of the evidence.

\begin{table}[!htb]
\centering
\caption{Claims per report before and after cite-or-drop, share deleted, and UCR of the draft and of the delivered report (5G-SecQA, same drafts; the check only deletes).}
\label{tab:claims}
\resizebox{0.95\textwidth}{!}{\footnotesize\input{tables/supp_claims}}
\end{table}

\para{Long lineages and the depth sweep.}
Table~\ref{tab:longlineage} reports EM on the 287 test questions of 5G-KGQA (23.1\% of the 1,240) whose gold chain spans at least six Releases; \sys's advantage over the long-context reader, T-GRAG+PEG+cite and the matched +PEG baseline is at least as large as on the full set. Table~\ref{tab:ksweep} gives the depth sweep: because the traversal expands whole lineages at no hop cost, $K$ counts cross-lineage hops only, and the sampled KGQA questions have 2--4 such hops by construction; the sweep is therefore also reported on the 150 expert-written questions, whose cross-lineage depth (1--5, mean 2.3) the experts set, and saturates at $K=3$ there as well.

\begin{table}[!htb]
\centering
\caption{Long-lineage slice of 5G-KGQA: EM on the questions whose gold chain spans at least six Releases.}
\label{tab:longlineage}
\resizebox{0.7\textwidth}{!}{\footnotesize\input{tables/supp_long_lineage}}
\end{table}

\begin{table}[!htb]
\centering
\caption{Depth sweep over cross-lineage hops $K$.}
\label{tab:ksweep}
\resizebox{0.8\textwidth}{!}{\footnotesize\input{tables/supp_ksweep}}
\end{table}

\para{Revision-boundary test (P3 measured directly).}
We selected 300 clause revisions of the test period whose CR changes normative text (editorial CRs excluded) and asked, for each, the same question one day before and one day after the publication date of the new version, with $R_q$ the revised Release; the gold answer is the current version at each date, taken from the version histories and re-verified by an engineer (297 confirmed, 3 corrected). Table~\ref{tab:boundary} reports, per system, the fraction of pairs in which the version cited as current is correct on both sides, the fraction in which the answer and its citation change across the boundary as they must, the fraction of \cur claims in the \emph{draft} that violate P3, and the residual violation rate after cite-or-drop. The check is clause-bound: a \cur claim that names clauses must cite a current version of \emph{every} clause it names, so a claim about clause A that cites only a CR touching A and B is not rescued by B's version remaining current. The residual for \sys (1.2\% of \cur claims) decomposes into 0.9\% role-tag errors of the splitter (a \cur claim tagged \hist, which the check therefore does not test) and 0.3\% claims that refer to a clause implicitly, without a clause number, so that clause binding fails and the claim falls into the unnamed-clause branch; neither is a clause-binding error of the kind described above. All residuals are shares of \cur claims, not of violations: after the check 1.2--1.9\% of \cur claims still violate P3 across the four systems. Relative to each system's draft rate (6.7\%, 15.8\%, 21.4\% and 31.2\%), the check removes 82\% of \sys's violations and 91--94\% of the baselines' (last column of Table~\ref{tab:boundary}); the smaller share for \sys reflects a floor of about one percent of \cur claims, splitter role-tag errors and implicit clause references that the check cannot see, which weighs more when the draft rate is low. The systems therefore differ in their draft-level rates and in the boundary behaviour, which the check cannot repair because it deletes rather than rewrites, not in the residual floor. P2 was checked mechanically on every chain emitted for the 1,240 KGQA test questions and the 600 boundary queries: no inversion of $\tau$ within a segment, no missing edge and no edge-type mismatch occurred. The one-year forward shift of $t_q$ reported below is a stability check (gold evidence remains current for 97\% of queries there) and is not a substitute for this boundary test.

\begin{table}[!htb]
\centering
\caption{Revision-boundary test on 300 question pairs straddling an actual revision.}
\label{tab:boundary}
\resizebox{\textwidth}{!}{\footnotesize\input{tables/supp_boundary}}
\end{table}

\para{Query-date controls.}
Redrawing $t_q$ uniformly one to three years after the latest gold CR (instead of 30--180 days) and re-freezing the snapshot and index gives an NDCG@10 margin over monoT5 of +3.2 points (95\% CI $[+1.6,+4.8]$) and KGQA EM 0.748, against +3.4 and 0.752 with the original offsets. Shifting $t_q$ one year later changes NDCG@10 by at most 0.004 and EM by 0.006 (gold evidence remains in the snapshot and remains current for 97\% of queries; the others are answered with the then-current successor version, which the graded scorer credits as supporting). Shifting $t_q$ earlier than the latest gold CR removes that CR from the snapshot by construction and is therefore not a meaningful robustness test; under a one-year-earlier shift, no gold CR approved after the shifted $t_q$ is retrieved by any system, as required by P1.

\para{Pre-training familiarity on the main benchmarks.}
The test period (2022-07 to 2025-06) lies inside the pre-training windows of every generator used. Three probes bound what that familiarity contributes. On 500 test-period CRs drawn at random from the gold sets of the three benchmarks, prompting each generator with only the CR identifier reproduces the CR content in none of the 500 cases, and continuing the first 50 tokens of each CR yields ROUGE-L at most 0.23 against the true continuation. Closed-book answering (no evidence, the LLM-only row) reaches EM 0.085 on 5G-KGQA and 0.067 on the expert-written questions, which is the ceiling of what memorised content contributes to the answer sets; every retrieval-augmented system is scored under P1 on evidence dated on or before $t_q$. The probes bound reproduction, not familiarity with the concepts, which the main text lists among the limitations.

\para{RFC transfer probe.}
The corpus is every RFC in the RFC Editor index published on or before 2025-06-30: 9,609 documents (the highest number published by then is RFC~9806; the 189 numbers marked ``Not Issued'' are excluded, as is anything that is not a published RFC), of which 3,864 carry at least one \texttt{Obsoletes}/\texttt{Updates} relation; these headers provide \texttt{supersedes}/\texttt{affects} edges. 150 queries were written by two independent annotators from errata reports with the headers hidden; pooled judgments on the three-level scale ($\kappa=0.77$); the index is frozen at each query's RFC date; $\alpha$ and $\beta$ are transferred without tuning. Supersession-aware reranking lifts NDCG@10 from 0.512 (monoT5) to 0.541 ($p<0.05$, paired bootstrap). The probe tests reranking only; it does not test traversal or synthesis, and the main text draws no conclusion beyond it.

\para{Error taxonomy.}
Of \sys's unsupported claims on 5G-SecQA, 41\% trace to a wrong hop in the traversed chain, 35\% to over-generalisation of a correctly cited clause and 24\% to citing a wrong clause version; 13\% of three-hop chains contain a wrong hop and 16\% of citations do not entail their claim. \sys leads in every security category; factual extraction is strongest (F1 0.512, UCR 0.031) and counterfactual inference weakest (F1 0.108).

\begin{figure}[!htb]
\centering
\includegraphics[width=\textwidth]{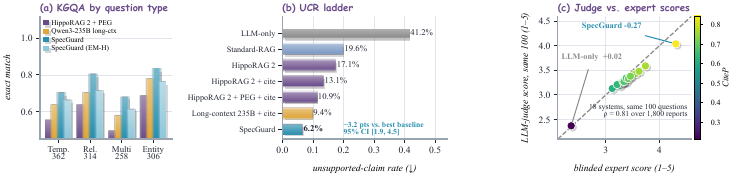}
\caption{\textbf{(a)} 5G-KGQA exact match by question type. \textbf{(b)} Unsupported-claim rate along a ladder of named systems (Qwen3-8B). \textbf{(c)} LLM-judge vs.\ expert scores on the same 100 expert-written questions, one point per system of the expert study (colour: CiteP).}
\label{fig:reasoning}
\end{figure}

\begin{figure}[!htb]
\centering
\includegraphics[width=\textwidth]{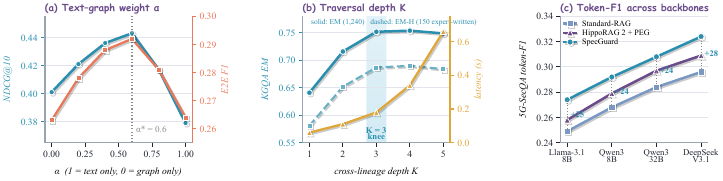}
\caption{Sensitivity. \textbf{(a)} Text--graph weight $\alpha$. \textbf{(b)} Cross-lineage depth $K$. \textbf{(c)} Token-F1 across backbones ($\Delta\times10^3$ over Standard-RAG).}
\label{fig:sens}
\end{figure}

\begin{figure}[!htb]
\centering
\includegraphics[width=\textwidth]{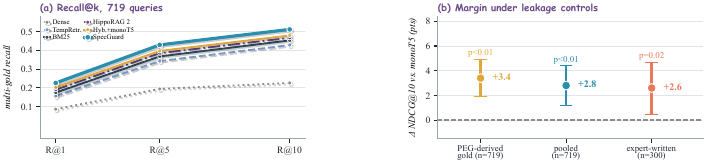}
\caption{Retrieval. \textbf{(a)} Recall@$k$. \textbf{(b)} NDCG@10 margin over the strongest learned reranker under PEG-derived gold, judgment-independent pooled relevance and the 300 expert-written queries (95\% CIs).}
\label{fig:retr}
\end{figure}

\begin{figure}[!htb]
\centering
\includegraphics[width=\textwidth]{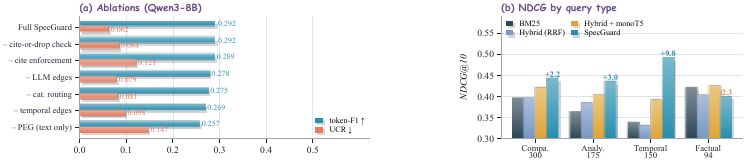}
\caption{\textbf{(a)} Ablations of \sys (token-F1 and UCR). \textbf{(b)} NDCG@10 by query type on the generated queries.}
\label{fig:abl}
\end{figure}

\section{Cost Accounting and Scaling}
\label{s:cost}
\begin{figure}[!htb]
\centering
\includegraphics[width=\textwidth]{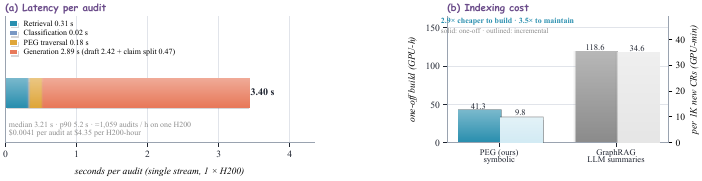}
\caption{Efficiency. \textbf{(a)} Latency per audit and its components. \textbf{(b)} Indexing cost against GraphRAG.}
\label{fig:eff}
\end{figure}

\begin{figure}[!htb]
\centering
\includegraphics[width=\textwidth]{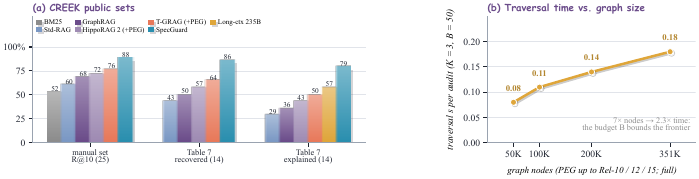}
\caption{\textbf{(a)} CREEK public sets: recovery on the 25 manual-set CRs, case-CR recovery and strict explanation on the 14 Table-7 cases, for \sys and the strongest baselines. \textbf{(b)} Traversal time against graph size.}
\label{fig:creek}
\end{figure}

Table~\ref{tab:cost} charges every offline component and states the operating conditions of every per-audit cost. The graph-specific part of \peg construction (NER inference, edge and link judging, graph load) is 41.3 GPU-h; parsing the 212K$\times$2 cover pages with the VL model is 38.0 GPU-h and category annotation 6.5 GPU-h, for a total of 85.8 GPU-h. GraphRAG's indexing of the same parsed text is 118.6 GPU-h, and with the shared 38.0 GPU-h of parsing its total is 156.6 GPU-h. Maintenance: 9.8 GPU-min per 1,000 new CRs (\peg) against 34.6 (GraphRAG). Online: \sys's 3.40\,s per audit (median 3.21, p90 5.2) decompose into retrieval 0.31, classification 0.02, traversal 0.18 and generation 2.89\,s. Generation is measured end to end at a single-stream decoding rate of 139 tokens/s on one H200: the draft costs 0.19\,s of prefill for the 4K-token input plus 2.23\,s for 310 output tokens (2.42\,s); the splitter costs 0.125\,s of prefill (report plus instruction) plus 0.35\,s for its 48 output tokens (0.47\,s). The splitter's output is small by design: for each of the 15.4 claims it emits only a sentence range and a role letter (``3--4 C''), about three tokens, and the bracketed identifiers and clause numbers of those sentences are copied programmatically from the draft, so the identifiers never pass through the decoder. Per-audit serving costs are single-stream latency times the number of H200s times \$4.35 per H200-hour; the long-context baseline runs on two H200s. Expert costs are the 2,064 hours of Table~\ref{tab:workload}; they are one-off resource-construction costs and do not recur per audit. Fig.~\ref{fig:eff} shows the latency components and the indexing cost; Fig.~\ref{fig:creek}b and Table~\ref{tab:scale} the scaling measurement on \peg restricted to Rel-8--10, Rel-8--12 and Rel-8--15 and on the full graph.

\begin{table}[!htb]
\centering
\caption{Offline GPU cost and per-audit serving cost with its operating conditions.}
\label{tab:cost}
\resizebox{0.95\textwidth}{!}{\footnotesize\input{tables/supp_cost}}
\end{table}

\section{External Validation on CREEK's Public Sets}
\label{s:creek}

\para{Sets and the semantics of Time2.}
CREEK~\cite{chen2022seeing} describes two sets that we use in full. (1)~The \emph{manual testing set} (CREEK, \S3.4): 25 CRs manually labelled security-relevant and 25 labelled non-security-relevant, listed on the CREEK project website; we use the 25 security CRs for recovery, gate recall and category agreement, and the 25 non-security CRs as gate negatives (all 25 are rejected by our gate). (2)~\emph{Table~7}: the 14 weaknesses reported by security CRs that were later confirmed in commercial devices or carrier networks. For each, CREEK gives Time1, the specification update date, and Time2, the earliest date at which the weakness could be documented in the field, which is one of three kinds: the release date of a device later confirmed vulnerable (ids 6, 7, 8, 10, 12, 13), the year of the reporting paper (ids 1--5, 9, 11) or CREEK's own device and carrier tests on 2021-10-01 (id~14). Time2 is therefore an \emph{anchor} rather than a confirmation date, and CREEK's windows are minimum estimates of how long a weakness persisted after the specification update; the main text uses this wording throughout. CREEK's 1,270 CRs labelled security-relevant by its classifier at confidence above 0.99 are used only for the coverage check of \S\ref{s:corpus}, not as a test set, since they are classifier outputs rather than manual labels. Table~\ref{tab:t7} reproduces Table~7 with the kind and the source of every anchor~\cite{kim2019touching,hussain2018lteinspector,chen2021bookworm,yu2019effects,shaik2016practical}; our two columns record, under the anchor-consistent protocol, whether the CR appears in \sys's top-10 for the case query and whether the weakness is explained strictly correctly.

\begin{table}[!htb]
\centering
\caption{CREEK Table~7 with our outcomes. Id, CR, Time1 (specification update), Time2, the minimum window and the 2021-test column (\ding{51} with the years since the update if CREEK's 2021-10-01 tests still found the weakness) are transcribed from CREEK; the working group is read from the tdoc prefix (C1 = CT1, S3 = SA3, R2 = RAN2), and the anchor kind and source annotate the origin of each Time2 from CREEK's text. Top-10 and Expl.\ are ours under the anchor-consistent protocol.}
\label{tab:t7}
\setlength{\tabcolsep}{3pt}
{\footnotesize\input{tables/cases_full}}
\end{table}

\begin{table}[!htb]
\centering
\caption{Recovery, gate recall and strict explanation on CREEK's public sets for \sys and the strongest baselines, under the protocol of the main text (manual set: snapshot at CREEK's publication date; Table~7: anchor-consistent).}
\label{tab:creekres}
\resizebox{0.95\textwidth}{!}{\footnotesize\input{tables/supp_creek}}
\end{table}

\para{Query writing.}
Two telecom engineers independent of the authors wrote one audit query per case from CREEK's per-case description (and, for the manual set, from the CR cover) with identifiers masked and wording paraphrased, then reconciled wording differences; agreement on the adjudicated outcomes (recovered / explained) was $\kappa=0.86$. Since the descriptions name the weakness, the exercise is closer to known-item retrieval than to discovery, which the main text states.

\para{Protocol for the manual testing set.}
The 25 security and 25 non-security CRs are removed from every training and validation split (retraining shifts test Macro-F1 by less than 0.002); no temporal claim is attached to this set, so the index and the \peg snapshot are frozen at CREEK's publication date, 2022-08-10, which postdates every CR in the set, and the standard models trained up to 2021-06-30 are used. Recovery is measured for \sys and for the retrievers of T-GRAG and HippoRAG~2 (both with \peg), GraphRAG, Standard-RAG and BM25 (Table~\ref{tab:creekres}).

\para{Anchor-consistent protocol for Table~7.}
Every number reported for the 14 cases, for \sys and for every baseline, is produced as follows. Year-only anchors are resolved to 1 January of that year, the smallest snapshot consistent with the anchor; the index and the \peg snapshot are frozen at Time2. Under the resolved anchors Time2 precedes our training cut-off of 2021-06-30 for ids 1--13 (ids 4 and 5, anchored to 2021, resolve to 2021-01-01) and only id~14 (2021-10-01) is later. For each of the 13 earlier cases every trained component, our gate, classifier and NER model as well as the fine-tuned rerankers and dense retriever used by the baselines, is retrained on CRs approved before that case's Time2, the graph-RAG baselines are re-indexed on the same CRs, the snapshot and index are rebuilt at Time2 and the case is re-run; id~14 uses the standard models. The 39 CRs (25 + 14) and their evolution descendants are removed from every training and validation split throughout; they remain \peg nodes because the graph uses public metadata, not labels. The long-context baseline, which has no trained component, reads the \peg neighbourhood retrieved at Time2 by the retrained \sys retriever, so its explanation count is a generator-only comparison on the same evidence.

\para{Outcomes, and why retraining changes none of them.}
Under this protocol \sys recovers the case CR in its top-10 for 12 of 14 cases and explains 11 strictly correctly; the recovered-but-not-explained case is id~8 and the two unrecovered cases are ids~9 and~12. Table~\ref{tab:gate} lists, per case, the security-gate score $p_{\text{sec}}$ of the case CR and its rank in \sys's final list under the standard models and under the retrained models. The retrained scores are lower by 0.02--0.08 for the retrained recovered cases, as expected from smaller training sets, but every recovered CR stays between 0.69 and 0.92, well above the 0.5 threshold, and its rank shifts by at most one place; the two gated-out CRs score 0.38/0.41 (id~9) and 0.44/0.46 (id~12) under the two protocols. The outcomes of every case are therefore identical under the anchor-consistent and the standard protocol, and the paper reports one set of numbers. Id~11 (C1-172658) is TS~24.301 CR~2871 revision~2, a Rel-14 CR approved on 2017-06-16 that takes the specification from version 14.3.0 to 14.4.0; its cover states both the missing integrity protection of the Detach Request and the resulting risk of denial of service, so the case CR is recovered (rank 5 and 6) and the weakness is explained from the cover alone, under both protocols. Id~8 (C1-112809) is recovered under both protocols (rank 6 and 7), and its report cites the changed NAS procedure with a valid citation; the consequence CREEK reports, however, arises in a procedure specified in another specification that no CR of either specification links to the changed clause, so \peg contains no path from the case CR to the clause that realises the consequence and no citable statement of it is reachable at any depth. This is the lateral-reasoning limitation named in the main text, not an artefact of the anchor date: the same chain is missing in the full graph.

\begin{table}[!htb]
\centering
\caption{Per-case security-gate score and rank of the case CR in \sys's final list under the standard models (trained up to 2021-06-30) and under the anchor-consistent retrained models; id~14 uses the standard models in both columns.}
\label{tab:gate}
\resizebox{0.85\textwidth}{!}{\footnotesize\input{tables/supp_creek_gate}}
\end{table}

\para{Exact tests.}
With $n=25$ and $n=14$ the comparisons are paired by CR or case, and Table~\ref{tab:mcnemar} reports exact two-sided McNemar tests computed from the discordant pairs: in every comparison the baseline recovers or explains no case that \sys misses, so the test reduces to a sign test on the number of cases that only \sys gets. The differences to BM25, Standard-RAG and LLM-only are significant ($p\le0.016$), the difference to GraphRAG is significant on the explained outcome ($p=0.031$) but not on the manual set ($p=0.063$), and those to T-GRAG, HippoRAG~2 and the long-context reader are not ($p\ge0.063$), as the main text states.

\begin{table}[!htb]
\centering
\caption{Exact McNemar tests of \sys against each baseline on the paired outcomes (manual set: CR in top-10; Table~7: weakness explained strictly correctly).}
\label{tab:mcnemar}
\resizebox{0.85\textwidth}{!}{\footnotesize\input{tables/supp_creek_mcnemar}}
\end{table}

\para{Follow-up control.}
Responsible disclosure typically precedes the publication that anchors Time2, so a Time2 snapshot could already contain CRs that 3GPP produced in response to a disclosure. For every case we listed the CRs approved in the twelve months before Time2 that touch any clause of the case CR: three cases (ids 4, 5, 14) have such CRs. Removing them from the snapshot and the index and re-running the three cases changes no outcome (recovered and explained before and after).

\para{Memorization probes.}
Prompting each LLM with only a CR identifier reproduces the CR content in 0 of 39 cases; continuing the first 50 tokens of each CR yields ROUGE-L $\le 0.21$ against the true continuation (the same probes on 500 test-period CRs of the main benchmarks are in \S\ref{s:results}). These probes cannot rule out familiarity with the attack mechanisms from pre-training on the security literature (CREEK itself dates from 2022), so the result is retrospective recovery under an anchor-date cutoff, not prospective discovery.

\para{Strict-correctness rubric.}
A report explains a weakness strictly correctly when (a) it names the affected mechanism as described by CREEK (e.g., the message, counter or state involved), (b) it states the real-world consequence CREEK reports (e.g., denial of service, tracking, downgrade), and (c) each of (a) and (b) carries a citation whose cited CR or clause version entails the statement; (d) no claim contradicts the CR. Two engineers adjudicated blind to the system identity; disagreements were resolved by discussion. \sys satisfies the rubric on 11 of 14 cases; the misses are ids 8 (consequence not reachable, above), 9 (R2-103442) and 12 (C1-154699). In ids 9 and 12 the weakness hinges on an interaction between a procedure specified in a RAN2 or CT1 specification and the security-context handling specified in TS~33.401, and no CR of either specification references the other for this interaction, so \peg contains no path between them at any depth; raising $K$ to 4 or 5 does not change this, and both CRs are gated out under either protocol because their covers carry little security vocabulary.

\section{Prompts}
\label{s:prompts}
All prompts are given verbatim below; fields filled at run time are in angle brackets. Temperature 0 unless stated. Every system, baseline or ours, receives the header line ``Audit date: $\langle t_q\rangle$. Target Release: $\langle R_q\rangle$.'' before the question.

\para{Query generation (5G-Retrieval, GPT-5).}
``You are a telecom security engineer writing a search query for an audit tool. You are given the cover pages of $\langle n\rangle$ related Change Requests (titles, specifications, Releases, reasons, consequences, clauses affected). Write one natural question of type $\langle$type$\rangle$ (comparative, analytical, temporal or factual) that an operator's security auditor could ask and that these CRs jointly answer. Mention at least two of the following security concepts: $\langle$concepts$\rangle$. Do not quote titles verbatim and do not mention CR numbers. Output the question only.''

\para{KGQA paraphrase (GPT-5).}
``Rewrite the following template question in fluent natural language without changing its meaning or its answer type ($\langle$type$\rangle$). Keep every specification, clause and Release identifier exactly as written. Template: $\langle$template question$\rangle$.''

\para{5G-SecQA generation (GPT-5).}
``From the following Change Request text and the cited clauses only, write one $\langle$factual$|$causal$|$counterfactual$\rangle$ audit question and a reference answer of at most 80 words. A factual question asks what the CR changes; a causal question asks why, based on the Reason for change; a counterfactual question asks what would follow for the affected procedure if the CR had not been approved, and its answer must name the procedure, message, counter or state concerned rather than repeat the Consequences field. Cite the CR identifier and clause numbers you rely on in square brackets. Do not use any information that is not in the text.''

\para{Audit synthesis (\sys and all +cite systems).}
``You are auditing a 3GPP security specification as of $\langle t_q\rangle$ for Release $\langle R_q\rangle$. Inputs: EVIDENCE (retrieved Change Requests), DIAGNOSIS (security categories with confidences), LOGIC (typed revision chains with version keys, dates and roles). Write an audit report answering the question. Every sentence must cite, in square brackets, a CR identifier or a clause version that appears in the inputs. State the rule in force for Release $\langle R_q\rangle$ at $\langle t_q\rangle$ only from versions marked current; when you cite a version marked historical, say so explicitly (for example, `as amended by CR \ldots in version \ldots'). Do not introduce facts that are not in the inputs.'' (Baselines without LOGIC or DIAGNOSIS receive the same instruction with those blocks absent; the +VM baseline receives its resolved current versions inside EVIDENCE; the ``w/o citation enforcement'' ablation omits the two citation sentences.)

\para{Claim splitting and role tagging (Qwen3-8B).}
``The report below is numbered by sentence. Output one line per atomic claim in the form `$\langle$first sentence$\rangle$--$\langle$last sentence$\rangle$ C' or `\ldots\ H', where C means the claim states a rule as in force at the audit date and Release and H means it describes an earlier state or a change. Output nothing else.'' (The bracketed identifiers and clause numbers of the listed sentences are then attached to the claim by a regular expression over the draft; the clause-binding audit of 200 reports measured 0.97 agreement with experts on which clauses a claim names.)

\para{LLM judge (DeepSeek-V3.1).}
``Rate the report on correctness (does it answer the question correctly with respect to the reference answer?), groundedness (is every statement supported by the cited evidence?) and usefulness (would a security auditor act on it?), each from 1 to 5, and give a one-sentence justification for each. Question: $\langle q\rangle$. Reference: $\langle a\rangle$. Report: $\langle r\rangle$. Evidence: $\langle$input evidence$\rangle$.''

\para{Succession judging (Qwen3-32B).}
``Two clause texts from 3GPP specifications are given with their specification, clause, Release and version. Decide whether the second is the successor of the first (the same normative content continued in a later Release or in the successor specification), answering YES or NO with one sentence of justification.''

\para{Affects judging (Qwen3-32B).}
``A Change Request is given with its cover fields and its redlined body, together with one clause identifier found in the body that is not listed under `Clauses affected'. Decide whether the CR changes the normative content of that clause (YES) or merely mentions it (NO). Answer YES or NO with one sentence of justification.''

\para{Co-reference judging (Qwen3-32B).}
``Two entity mentions are given with their types and their sentences. Decide whether they denote the same entity (the same key, identifier, procedure, message, algorithm or network function), answering YES or NO with one sentence of justification.''

\para{Entailment adjudication (Qwen3-32B).}
``A claim from an audit report and the text of the Change Request or clause version it cites are given. Decide whether the cited text entails the claim (ENTAILED), contradicts it (CONTRADICTED) or neither (NEUTRAL). Answer with one of the three labels and one sentence of justification.''

\para{Category adjudication (Qwen3-8B).}
``Category: $\langle$name and definition$\rangle$. Examples: $\langle$three labelled CRs$\rangle$. Decide whether the following Change Request belongs to this category, using only its Reason for change and Consequences if not approved. Answer YES or NO.''

\section{Public Record of the Motivating Case}
\label{s:case}
The tdoc list of the SA3 ad hoc meeting of September 2009 (Sophia Antipolis) records S3-091802 as ``CR to TS 33401 Partial Security Context Lifetime (R8)'', source ZTE Corporation, a revision of S3-091684, agreed under agenda item~6.1; the tdoc list is published on the 3GPP file server under the SA3 ad hoc meeting directory for September 2009 (\url{https://www.3gpp.org/ftp/tsg_sa/WG3_Security/ADHOCs/TSGS3_ADHOC_Sep09_Sophia/}). The change history of TS~33.401 lists, under 2009-12, TSG~SA\#46, SP-090811, CR~339 ``Partial native EPS security context NAS COUNT value'', from version 8.5.0 to 8.6.0 (ETSI TS~133~401 V8.6.0, published 2010-01). The two titles belong to one document: a tdoc list records the title under which a contribution is tabled at the working group, whereas the change history records the title on the CR cover, and the SA3 ad hoc forwarded the agreed CR to TSG~SA\#46 in the CR pack SP-090811, where it was approved as CR~0339; \peg keys the CR node by tdoc, CR number, Release and version step, so both identifiers resolve to it. The rule appears in clause~6.5 (``Before the MME sends the first NAS Security Mode Command for a given partial native security context, the MME sets the NAS COUNTs for the security context to 0'') and in the definition of a partial native EPS security context in clause~3.1; TS~33.501 carries the same definition for the 5G context and the NAS COUNT handling in clause~6.4.5, on which the NAS security mode command procedure (clause~6.7.2) relies. In \peg the Rel-8 version created by CR~0339 and the 5G clause TS~33.501 \S6.4.5 belong to one lineage (the Rel-9 to Rel-15 segments of TS~33.401, each joined by the \texttt{evolves\_to} edge into its opening copy, and the cross-specification edge of \S\ref{s:peg}), which Algorithm~\ref{alg:traverse} expands at no hop cost, so the chain of Fig.~\ref{fig:motivation} uses one \texttt{affects} hop from the seed CR and one \texttt{references} hop to \S6.7.2. CREEK's Table~7 (id~2) dates the specification update to 2009-12-18 and anchors Time2 to 2018, the year of the report by Hussain et al.~\cite{hussain2018lteinspector} on four major U.S. operators, an 8.0-year minimum window. The 11.7 years quoted in the introduction and in Table~\ref{tab:t7} belong to CREEK's ids~6 and~7 (C1-094446, C1-095712), found still present in CREEK's 2021 device tests.

%% file: tables/supp_routing.tex
\begin{tabular}{@{}l l l@{}}
\toprule
\rowcolor{RowHead}\textbf{Category} & \textbf{Edge types followed} & \textbf{Entity-type filter} \\
\midrule
Authentication & \texttt{affects, supersedes, evolves\_to, references} & procedure, message, key \\
Key Management & \texttt{affects, supersedes, evolves\_to} & key, algorithm, procedure \\
Encryption \& Integrity & \texttt{affects, supersedes, evolves\_to, references} & algorithm, key, message \\
Privacy & \texttt{affects, evolves\_to, references} & identifier, message, procedure \\
Protocol Logic \& State Machine & \texttt{affects, supersedes, evolves\_to, references} & procedure, message \\
Service Availability (DoS) & \texttt{affects, references} & procedure, message, network function \\
Attack Surface \& Threats & \texttt{affects, evolves\_to, references} & network function, message, procedure \\
Mobility Security & \texttt{affects, supersedes, evolves\_to, references} & procedure, key, identifier \\
\bottomrule
\end{tabular}

%% file: tables/supp_scale.tex
\setlength{\tabcolsep}{7pt}\renewcommand{\arraystretch}{1.16}
\begin{tabular}{lcccc}
\toprule
\rowcolor{RowHead}\textbf{Graph size (K nodes)} & \textbf{50} & \textbf{100} & \textbf{200} & \textbf{351} \\
\midrule
Traversal time per audit (s), $K=3$, $B=50$ & 0.08 & 0.11 & 0.14 & 0.18 \\
\bottomrule
\end{tabular}

%% file: tables/supp_edge_audit.tex
\begin{tabular}{@{}l r r r@{}}
\toprule
\rowcolor{RowHead}\textbf{Edge type (provenance)} & \textbf{Audited} & \textbf{Correct} & \textbf{Precision} \\
\midrule
references (symbolic) & 200 & 197 & 0.985 \\
evolves\_to (symbolic) & 200 & 195 & 0.975 \\
evolves\_to (LLM-inferred) & 100 & 92 & 0.920 \\
supersedes (symbolic) & 200 & 191 & 0.955 \\
supersedes (LLM-inferred) & 50 & 44 & 0.880 \\
affects (symbolic) & 150 & 144 & 0.960 \\
affects (LLM-inferred) & 150 & 130 & 0.867 \\
semantic\_link (LLM-inferred) & 150 & 131 & 0.873 \\
\midrule
all symbolic edges & 750 & 727 & 0.969 \\
all LLM-inferred edges & 450 & 397 & 0.882 \\
all audited edges & 1200 & 1124 & 0.937 \\
\bottomrule
\end{tabular}

%% file: tables/supp_kgqa_gold.tex
\begin{tabular}{@{}l ccc@{}}
\toprule
\rowcolor{RowHead}\textbf{System} & \textbf{EM, PEG-derived gold} & \textbf{EM, verified gold (reported)} & \textbf{$\Delta$} \\
\midrule
\rowcolor{RowOurs}SpecGuard & 0.771 & 0.752 & -0.019 \\
Qwen3-235B long-context + cite & 0.676 & 0.672 & -0.004 \\
T-GRAG + PEG + cite & 0.619 & 0.612 & -0.007 \\
HippoRAG 2 + PEG + cite & 0.598 & 0.594 & -0.004 \\
\bottomrule
\end{tabular}

%% file: tables/supp_workload.tex
\begin{tabular}{@{}p{6.3cm} r r r p{3.1cm}@{}}
\toprule
\rowcolor{RowHead}\textbf{Activity} & \textbf{Items} & \textbf{Annot.} & \textbf{Hours} & \textbf{Agreement / note} \\
\midrule
Pooled relevance judgments, 5G-Retrieval (719 queries x 38.2 unique CRs x 2) & 54,920 & 2 & 565 & 3-level scale, kappa 0.81 \\
Expert scoring of audit reports (18 systems x 100 questions x 3) & 5,400 & 3 & 360 & 3 dimensions, alpha 0.74 \\
5G-SecQA question and answer review (3,900 x 2) & 7,800 & 2 & 195 & 87 discarded, 214 corrected \\
5G-KGQA test-gold verification against the specifications (1,240 x 2) & 2,480 & 2 & 83 & 46 corrected \\
Category re-annotation of 1,500 test-period CRs (x 3) & 4,500 & 3 & 150 & Fleiss kappa 0.83 \\
Edge audit, 1,200 edges (x 3) and 200 three-hop chains (x 3) & 4,200 & 3 & 75 & kappa 0.81 \\
Recall gold standard, 10 specifications (3,463 edges x 2) & 6,926 & 2 & 120 & from change histories \\
NER spans: 3,000 training and 500 held-out spans (x 2) & 7,000 & 2 & 40 & span kappa 0.84 \\
Cover-parsing audit (1,000 CRs x 2) & 2,000 & 2 & 25 & field accuracy 98.3\% \\
Filter-recall labelling of discarded CRs (800 x 2) & 1,600 & 2 & 20 & 6 security-relevant \\
Lexicon specificity weights (412 terms x 3) & 1,236 & 3 & 12 & ordinal alpha 0.79 \\
Entailment audit (300 claim-evidence pairs x 2) & 600 & 2 & 10 & NLI F1 0.91 \\
Claim-splitting, role and clause-binding audit (200 reports x 2) & 400 & 2 & 14 & 0.93 / 0.95 / 0.97 agreement \\
Expert-written queries and questions (300 + 150 + 100) & 550 & 3 & 140 & written without PEG access \\
Expert-slice relevance grading (300 queries x 31 unique CRs by the authoring expert; 100 queries double-graded) & 12,400 & 2 & 128 & kappa 0.79 on the double-graded queries \\
CREEK: query writing, category labels and adjudication (39 x 2 + 25 x 2 + 39 x 2) & 206 & 2 & 20 & kappa 0.86 \\
RFC probe: 150 queries (x 2) and pooled judgments (150 x 30 x 2) & 9,300 & 2 & 95 & kappa 0.77 \\
Revision-boundary pairs: gold verification (300) and report checks (100 x 2) & 500 & 2 & 12 & 3 gold corrected \\
\midrule
\textbf{Total} & & & \textbf{2,064} & \\
\bottomrule
\end{tabular}

%% file: tables/supp_deploy.tex
\setlength{\tabcolsep}{3pt}\renewcommand{\arraystretch}{1.15}
\begin{tabular}{@{}>{\raggedright\arraybackslash}p{0.26\linewidth}p{0.08\linewidth}p{0.17\linewidth}p{0.10\linewidth}p{0.14\linewidth}p{0.17\linewidth}@{}}
\toprule
\rowcolor{RowHead}\textbf{Model (role)} & \textbf{Weights} & \textbf{Engine} & \textbf{GPUs} & \textbf{Max input} & \textbf{Max new tokens} \\
\midrule
Qwen3-8B (answerer, splitter, adjudicator) & bf16 & vLLM 0.8 & 1 x H200 & 8K (4K evidence budget) & 512 report / 128 splitter / 64 KGQA \\
Qwen3-32B (extraction, edge judging, NLI adjudication) & bf16 & vLLM 0.8 & 1 x H200 & 8K & 256 \\
Llama-3.1-8B (backbone sweep) & bf16 & vLLM 0.8 & 1 x H200 & 8K & 512 \\
Qwen3-235B-A22B (long-context baseline) & FP8 & vLLM 0.8, tensor parallel 2 & 2 x H200 & 32K & 512 \\
DeepSeek-V3.1 (backbone sweep and judge) & provider weights & provider API & provider-hosted & 32K & 512 \\
GPT-5 (writer, secondary judge) & provider weights & provider API & provider-hosted & 32K & 512 \\
\bottomrule
\end{tabular}

%% file: tables/kgqa_only.tex
\begin{tabular}{@{}l|cc|cc@{}}
\toprule
\rowcolor{RowHead}\textbf{KGQA-only baseline} & \multicolumn{2}{c|}{\textbf{own linking}} & \multicolumn{2}{c}{\textbf{oracle linking}} \\
\rowcolor{RowHead} & EM & EM-H & EM & EM-H \\
\midrule
Text-to-SPARQL (Qwen3-32B) & 0.641 & 0.553 & 0.712 & 0.633 \\
CronKGQA (TComplEx over PEG) & 0.402 & 0.311 & 0.498 & 0.402 \\
TempoQR (over PEG) & 0.428 & 0.337 & 0.531 & 0.433 \\
\bottomrule
\end{tabular}

%% file: tables/supp_conditions.tex
\setlength{\tabcolsep}{3pt}\renewcommand{\arraystretch}{1.15}
\begin{tabular}{@{}>{\raggedright\arraybackslash}p{0.25\linewidth}p{0.30\linewidth}p{0.17\linewidth}>{\centering\arraybackslash}p{0.10\linewidth}p{0.12\linewidth}@{}}
\toprule
\rowcolor{RowHead}\textbf{Row} & \textbf{Retriever / evidence set} & \textbf{Organisation} & \textbf{Diagnosis} & \textbf{Post-processing} \\
\midrule
Standard-RAG + cite & Standard-RAG top-100 (own) & CR text & no & cite-or-drop \\
Standard-RAG + VM + cite & own + current version of each listed clause & CR text + versions & no & cite-or-drop \\
Standard-RAG + PEG + cite (matched) & SpecGuard's final CRs and chains & unordered records & no & cite-or-drop \\
Ladder: sorted by approval date & SpecGuard's final CRs and chains & records in date order & no & cite-or-drop \\
Ladder: typed chains & SpecGuard's final CRs and chains & typed chains per lineage & no & cite-or-drop \\
\rowcolor{RowOurs}SpecGuard & routed traversal (own) & typed chains per lineage & yes & cite-or-drop \\
Standard-RAG, own + PEG records & Standard-RAG top-100 (own), PEG records & unordered records & no & cite-or-drop \\
Standard-RAG, own + PEG chains & Standard-RAG top-100 (own), PEG records & typed chains per lineage & no & cite-or-drop \\
HippoRAG 2 + PEG + cite & HippoRAG 2 retrieval (own), PEG records & unordered records & no & cite-or-drop \\
HippoRAG 2, own + PEG chains & HippoRAG 2 retrieval (own), PEG records & typed chains per lineage & no & cite-or-drop \\
T-GRAG + PEG + cite & T-GRAG time-scoped retrieval (own), PEG records & unordered records & no & cite-or-drop \\
T-GRAG, own + PEG chains & T-GRAG time-scoped retrieval (own), PEG records & typed chains per lineage & no & cite-or-drop \\
Qwen3-235B long-context + cite & SpecGuard's PEG neighbourhood as text (32K) & document order & no & cite-or-drop \\
\bottomrule
\end{tabular}

%% file: tables/supp_retrieval_full.tex
\setlength{\tabcolsep}{2.75pt}\renewcommand{\arraystretch}{1.05}
\begin{tabular}{@{}l@{\hspace{4pt}}>{\raggedright\arraybackslash}p{124pt}|ccccc|cc|c@{}}
\toprule
\rowcolor{RowHead}\multicolumn{2}{l|}{\textbf{Retriever}} & \textbf{R@1} & \textbf{R@5} & \textbf{R@10} & \textbf{MRR@10} & \textbf{NDCG@10} & \shortstack{\textbf{NDCG}\\\textbf{(pooled)}} & \shortstack{\textbf{NDCG}\\\textbf{(expert)}} & \shortstack{\textbf{R@10}\\\textbf{(temp.)}} \\
\midrule
\tagc{TagBase}{SPARSE} & BM25 & 0.174 & 0.367 & 0.452 & 0.450 & 0.377 & 0.370 & 0.371 & 0.401 \\
\tagc{TagBase}{SPARSE} & SPLADE-v3 & 0.183 & 0.379 & 0.465 & 0.457 & 0.389 & 0.381 & 0.352 & 0.412 \\
\tagc{TagRAG}{DENSE} & Dense (bge-large, doc-level, off-the-shelf) & 0.087 & 0.194 & 0.226 & 0.203 & 0.183 & 0.179 & 0.164 & 0.226 \\
\tagc{TagRAG}{DENSE} & Dense (bge-large, passage-level, in-domain FT) & 0.141 & 0.310 & 0.392 & 0.368 & 0.319 & 0.312 & 0.296 & 0.339 \\
\tagc{TagRAG}{DENSE} & HyDE + Dense & 0.112 & 0.243 & 0.291 & 0.262 & 0.228 & 0.221 & 0.221 & 0.263 \\
\tagc{TagTemp}{TEMP} & TempRetriever (in-domain FT) & 0.156 & 0.342 & 0.428 & 0.409 & 0.351 & 0.343 & 0.348 & 0.442 \\
\tagc{TagTemp}{HYBRID} & Hybrid (RRF) & 0.172 & 0.372 & 0.455 & 0.441 & 0.379 & 0.372 & 0.342 & 0.393 \\
\tagc{TagGraph}{GRAPH} & LightRAG retriever (+PEG) & 0.176 & 0.373 & 0.457 & 0.446 & 0.382 & 0.375 & 0.369 & 0.421 \\
\tagc{TagGraph}{GRAPH} & HippoRAG 2 retriever (+PEG) & 0.190 & 0.384 & 0.466 & 0.470 & 0.397 & 0.388 & 0.361 & 0.438 \\
\tagc{TagGraph}{GRAPH} & T-GRAG retriever (+PEG) & 0.193 & 0.388 & 0.471 & 0.474 & 0.403 & 0.392 & 0.372 & 0.461 \\
\tagc{TagLC}{RERANK} & Hybrid + bge-reranker-v2 (in-domain FT) & 0.198 & 0.391 & 0.472 & 0.478 & 0.405 & 0.396 & 0.393 & 0.427 \\
\tagc{TagLC}{RERANK} & Hybrid + monoT5-3B (in-domain FT) & 0.202 & 0.396 & 0.478 & 0.483 & 0.409 & 0.401 & 0.388 & 0.436 \\
\rowcolor{RowOurs}\tagc{TagOurs}{OURS} & Hybrid + PEG rerank (SpecGuard) & 0.226 & 0.428 & 0.511 & 0.524 & 0.443 & 0.429 & 0.419 & 0.548 \\
\midrule
\tagc{TagLC}{RERANK} & Hybrid + monoT5-3B + PEG connectivity & 0.229 & 0.431 & 0.513 & 0.527 & 0.446 & 0.431 & 0.420 & 0.551 \\
\tagc{TagLC}{RERANK} & Hybrid + monoT5-3B (off-the-shelf) & 0.191 & 0.381 & 0.462 & 0.462 & 0.392 & 0.385 & 0.374 & 0.418 \\
\tagc{TagLC}{RERANK} & Hybrid + bge-reranker-v2 (off-the-shelf) & 0.186 & 0.376 & 0.458 & 0.455 & 0.387 & 0.380 & 0.372 & 0.411 \\
\tagc{TagTemp}{TEMP} & TempRetriever (off-the-shelf) & 0.149 & 0.328 & 0.412 & 0.391 & 0.336 & 0.329 & 0.333 & 0.421 \\
\tagc{TagRAG}{DENSE} & Dense (bge-large, passage-level, off-the-shelf) & 0.118 & 0.262 & 0.312 & 0.278 & 0.247 & 0.241 & 0.228 & 0.286 \\
\bottomrule
\end{tabular}

%% file: tables/supp_extra_rows.tex
\begin{tabular}{@{}l@{\hspace{3pt}}l|ccc@{}}
\toprule
\rowcolor{RowHead}\multicolumn{2}{l|}{\textbf{System (Qwen3-8B)}} & \textbf{SemSim} & \textbf{EM} & \textbf{UCR}$\downarrow$ \\
\midrule
\tagc{TagBase}{BASE} & LLM-only & 0.737 & 0.085 & 0.412 \\
\tagc{TagRAG}{RAG} & Standard-RAG & 0.818 & 0.288 & 0.196 \\
\tagc{TagRAG}{RAG} & Standard-RAG + cite & 0.819 & 0.290 & 0.142 \\
\tagc{TagRAG}{RAG} & Standard-RAG + VM + cite & 0.821 & 0.441 & 0.131 \\
\tagc{TagRAG}{RAG} & Standard-RAG + PEG + cite & 0.822 & 0.517 & 0.118 \\
\tagc{TagRAG}{RAG} & Telco-RAG + cite & 0.820 & 0.305 & 0.138 \\
\tagc{TagTemp}{TEMP} & TempRetriever-RAG + cite & 0.816 & 0.302 & 0.149 \\
\tagc{TagGraph}{GRAPH} & GraphRAG & 0.813 & 0.485 & 0.183 \\
\tagc{TagGraph}{GRAPH} & GraphRAG + PEG & 0.817 & 0.556 & 0.158 \\
\tagc{TagGraph}{GRAPH} & LightRAG + PEG & 0.819 & 0.573 & 0.153 \\
\tagc{TagGraph}{GRAPH} & LightRAG + PEG + cite & 0.823 & 0.575 & 0.113 \\
\tagc{TagGraph}{GRAPH} & HippoRAG 2 & 0.821 & 0.512 & 0.171 \\
\tagc{TagGraph}{GRAPH} & HippoRAG 2 + cite & 0.822 & 0.514 & 0.131 \\
\tagc{TagGraph}{GRAPH} & HippoRAG 2 + PEG + cite & 0.825 & 0.594 & 0.109 \\
\tagc{TagGraph}{GRAPH} & T-GRAG + cite & 0.822 & 0.548 & 0.127 \\
\tagc{TagGraph}{GRAPH} & T-GRAG + PEG + cite & 0.826 & 0.612 & 0.104 \\
\tagc{TagLC}{LONG-CTX} & Qwen3-235B long-context + cite & 0.829 & 0.672 & 0.094 \\
\rowcolor{RowOurs}\tagc{TagOurs}{OURS} & SpecGuard & 0.831 & 0.752 & 0.062 \\
\rowcolor{RowVar}\tagc{TagVar}{ABL} & w/o cite-or-drop check & 0.831 & 0.752 & 0.084 \\
\rowcolor{RowVar}\tagc{TagVar}{ABL} & w/o citation enforcement & 0.830 & 0.752 & 0.121 \\
\rowcolor{RowVar}\tagc{TagVar}{ABL} & w/o LLM-inferred edges & 0.826 & 0.721 & 0.079 \\
\rowcolor{RowVar}\tagc{TagVar}{ABL} & w/o category routing & 0.823 & 0.749 & 0.081 \\
\rowcolor{RowVar}\tagc{TagVar}{ABL} & w/o temporal edges & 0.822 & 0.684 & 0.098 \\
\rowcolor{RowVar}\tagc{TagVar}{ABL} & w/o PEG (text only) & 0.813 & 0.301 & 0.147 \\
\midrule
\tagc{TagGraph}{GRAPH} & LightRAG & 0.815 & 0.498 & 0.179 \\
\tagc{TagTemp}{TEMP} & TempRetriever-RAG + PEG + cite & 0.820 & 0.520 & 0.116 \\
\tagc{TagGraph}{GRAPH} & HippoRAG 2 + PEG & 0.822 & 0.592 & 0.147 \\
\tagc{TagLC}{LONG-CTX} & Qwen3-8B long-context + cite (32K tokens, same neighbourhood) & 0.824 & 0.603 & 0.121 \\
\bottomrule
\end{tabular}

%% file: tables/supp_classification_full.tex
\begin{tabular}{@{}l@{\hspace{3pt}}l|cccc|c|c@{}}
\toprule
\rowcolor{RowHead}\multicolumn{2}{l|}{\textbf{Classifier}} & \textbf{Macro-F1} & \textbf{Micro-F1} & \textbf{P} & \textbf{R} & \textbf{Expert-F1} & \textbf{CREEK-25} \\
\midrule
\tagc{TagBase}{ENCODER} & BERT-base & 0.734 & 0.781 & 0.802 & 0.761 & 0.705 & 16/25 \\
\tagc{TagBase}{ENCODER} & RoBERTa-base & 0.718 & 0.769 & 0.791 & 0.748 & 0.701 & 16/25 \\
\tagc{TagBase}{ENCODER} & SecureBERT & 0.741 & 0.788 & 0.809 & 0.768 & 0.716 & 17/25 \\
\tagc{TagBase}{ENCODER} & DeBERTa-v3-base & 0.752 & 0.796 & 0.815 & 0.778 & 0.738 & 17/25 \\
\tagc{TagBase}{ENCODER} & HGCLR (hierarchical) & 0.789 & 0.822 & 0.838 & 0.807 & 0.761 & 18/25 \\
\tagc{TagGraph}{GNN} & R-GCN over PEG & 0.801 & 0.834 & 0.851 & 0.818 & 0.781 & 18/25 \\
\tagc{TagRAG}{LLM} & Qwen3-8B zero-shot & 0.298 & 0.403 & 0.289 & 0.663 & 0.307 & 11/25 \\
\tagc{TagRAG}{LLM} & Qwen3-8B few-shot CoT & 0.412 & 0.502 & 0.401 & 0.671 & 0.416 & 13/25 \\
\tagc{TagRAG}{LLM} & Qwen3-8B + LoRA & 0.826 & 0.851 & 0.868 & 0.835 & 0.799 & 19/25 \\
\rowcolor{RowOurs}\tagc{TagOurs}{OURS} & SpecGuard classifier & 0.849 & 0.876 & 0.894 & 0.858 & 0.835 & 21/25 \\
\rowcolor{RowVar}\tagc{TagVar}{ABL} & w/o contrastive loss & 0.786 & 0.821 & 0.842 & 0.801 & 0.764 & 18/25 \\
\rowcolor{RowVar}\tagc{TagVar}{ABL} & w/o category attention & 0.817 & 0.848 & 0.866 & 0.831 & 0.793 & 19/25 \\
\rowcolor{RowVar}\tagc{TagVar}{ABL} & w/o security gate & 0.829 & 0.856 & 0.871 & 0.842 & 0.812 & 20/25 \\
\bottomrule
\end{tabular}

%% file: tables/supp_seed_tests.tex
\begin{tabular}{@{}l r r r l@{}}
\toprule
\rowcolor{RowHead}\textbf{Metric} & \textbf{Mean $\Delta$ (5 seeds)} & \textbf{SD of $\Delta$} & \textbf{$t$ (df 4)} & \textbf{$p$} \\
\midrule
EM & +0.080 & 0.009 & 19.9 & $<$0.001 \\
UCR & -0.032 & 0.005 & 14.3 & $<$0.001 \\
CiteP & +0.081 & 0.010 & 18.1 & $<$0.001 \\
NDCG@10 & +0.034 & 0.006 & 12.7 & $<$0.001 \\
F1 & +0.007 & 0.004 & 3.9 & 0.017 \\
SemSim & +0.002 & 0.003 & 1.5 & 0.210 \\
\bottomrule
\end{tabular}

%% file: tables/supp_human_tests.tex
\begin{tabular}{@{}l l r l r@{}}
\toprule
\rowcolor{RowHead}\textbf{Comparison (100 human-written 5G-SecQA questions)} & \textbf{Metric} & \textbf{$\Delta$} & \textbf{95\% CI} & \textbf{$p$} \\
\midrule
SpecGuard $-$ Qwen3-235B long-context + cite & F1 & +0.023 & [-0.004, +0.051] & 0.090 \\
SpecGuard $-$ Qwen3-235B long-context + cite & CiteP & +0.082 & [+0.031, +0.133] & 0.004 \\
SpecGuard $-$ Qwen3-235B long-context + cite & UCR & -0.031 & [-0.058, -0.004] & 0.030 \\
SpecGuard $-$ Qwen3-235B long-context + cite & Expert & +0.57 & [+0.38, +0.76] & 0.001 \\
\midrule
SpecGuard $-$ T-GRAG, own + PEG chains & F1 & +0.019 & [-0.008, +0.046] & 0.120 \\
SpecGuard $-$ T-GRAG, own + PEG chains & CiteP & +0.029 & [+0.003, +0.055] & 0.030 \\
SpecGuard $-$ T-GRAG, own + PEG chains & UCR & -0.008 & [-0.021, +0.005] & 0.210 \\
SpecGuard $-$ T-GRAG, own + PEG chains & Judge (100) & +0.28 & [+0.10, +0.46] & 0.004 \\
\bottomrule
\end{tabular}

%% file: tables/supp_chainfmt_tests.tex
\begin{tabular}{@{}l r l r r l@{}}
\toprule
\rowcolor{RowHead}\textbf{Metric} & \textbf{$\Delta$ SpecGuard $-$ control} & \textbf{95\% CI (queries)} & \textbf{$p$ (queries)} & \textbf{SD of $\Delta$ (5 seeds)} & \textbf{$p$ (seeds)} \\
\midrule
EM & +0.051 & [+0.031, +0.071] & 0.001 & 0.007 & $<$0.001 \\
EM-H & +0.074 & [+0.012, +0.136] & 0.020 & 0.011 & $<$0.001 \\
UCR & -0.010 & [-0.018, -0.002] & 0.020 & 0.004 & 0.005 \\
CiteP & +0.038 & [+0.016, +0.060] & 0.003 & 0.008 & $<$0.001 \\
\bottomrule
\end{tabular}

%% file: tables/supp_chainfmt.tex
\begin{tabular}{@{}l l ccc ccc@{}}
\toprule
\rowcolor{RowHead}\textbf{Retriever} & \textbf{Evidence set} & \multicolumn{3}{c}{\textbf{unordered records}} & \multicolumn{3}{c}{\textbf{typed chains}} \\
\rowcolor{RowHead} & & EM & EM-H & UCR & EM & EM-H & UCR \\
\midrule
Standard-RAG & own top-$k$, PEG records & 0.486 & 0.433 & 0.124 & 0.581 & 0.507 & 0.083 \\
HippoRAG 2 & own top-$k$, PEG records & 0.594 & 0.500 & 0.109 & 0.688 & 0.593 & 0.075 \\
T-GRAG & own top-$k$, PEG records & 0.612 & 0.520 & 0.104 & 0.701 & 0.613 & 0.072 \\
\midrule
Matched control (Std-RAG+PEG+cite) & SpecGuard's final CRs and chains & 0.517 & 0.447 & 0.118 & 0.734 & 0.671 & 0.071 \\
\rowcolor{RowOurs}SpecGuard (+ Diagnosis) & routed traversal & \multicolumn{3}{c}{} & 0.752 & 0.687 & 0.062 \\
\bottomrule
\end{tabular}

%% file: tables/supp_claims.tex
\begin{tabular}{@{}l ccc cc@{}}
\toprule
\rowcolor{RowHead}\textbf{System (Qwen3-8B unless stated)} & \textbf{Claims, draft} & \textbf{Claims, delivered} & \textbf{Deleted} & \textbf{UCR, draft} & \textbf{UCR, delivered} \\
\midrule
Standard-RAG + cite & 15.9 & 14.0 & 12.0\% & 0.171 & 0.142 \\
Standard-RAG + VM + cite & 15.8 & 14.1 & 10.8\% & 0.158 & 0.131 \\
Standard-RAG + PEG + cite & 15.8 & 14.0 & 11.4\% & 0.146 & 0.118 \\
Telco-RAG + cite & 15.7 & 13.9 & 11.5\% & 0.166 & 0.138 \\
TempRetriever-RAG + cite & 15.6 & 13.9 & 10.9\% & 0.176 & 0.149 \\
LightRAG + PEG + cite & 15.9 & 14.2 & 10.7\% & 0.139 & 0.113 \\
HippoRAG 2 + cite & 15.7 & 14.0 & 10.8\% & 0.158 & 0.131 \\
HippoRAG 2 + PEG + cite & 15.8 & 14.1 & 10.8\% & 0.134 & 0.109 \\
T-GRAG + cite & 15.9 & 14.3 & 10.1\% & 0.152 & 0.127 \\
T-GRAG + PEG + cite & 15.6 & 14.0 & 10.3\% & 0.128 & 0.104 \\
Qwen3-235B long-context + cite & 16.1 & 14.3 & 11.2\% & 0.117 & 0.094 \\
\rowcolor{RowOurs}SpecGuard & 15.4 & 14.2 & 7.9\% & 0.084 & 0.062 \\
\bottomrule
\end{tabular}

%% file: tables/supp_long_lineage.tex
\begin{tabular}{@{}l c@{}}
\toprule
\rowcolor{RowHead}\textbf{System} & \textbf{EM, gold chain $\ge$ 6 Releases ($n=287$)} \\
\midrule
Standard-RAG + PEG + cite & 0.470 \\
T-GRAG + PEG + cite & 0.557 \\
Qwen3-235B long-context + cite & 0.613 \\
\rowcolor{RowOurs}SpecGuard & 0.731 \\
\bottomrule
\end{tabular}

%% file: tables/supp_ksweep.tex
\begin{tabular}{@{}lccccc@{}}
\toprule
\rowcolor{RowHead}\textbf{Cross-lineage depth $K$} & 1 & 2 & 3 & 4 & 5 \\
\midrule
EM, 5G-KGQA (1,240) & 0.641 & 0.716 & 0.752 & 0.754 & 0.749 \\
EM-H, 150 expert-written questions & 0.581 & 0.652 & 0.687 & 0.690 & 0.684 \\
Traversal time (s) & 0.06 & 0.11 & 0.18 & 0.34 & 0.66 \\
\bottomrule
\end{tabular}

%% file: tables/supp_boundary.tex
\begin{tabular}{@{}l ccccc@{}}
\toprule
\rowcolor{RowHead}\textbf{System (300 boundary pairs)} & \textbf{Current version cited} & \textbf{Answer changes at boundary} & \textbf{P3 violations, draft (share of \cur claims)} & \textbf{Residual after check (share of \cur claims)} & \textbf{Violations removed} \\
\midrule
\rowcolor{RowOurs}SpecGuard & 0.963 & 0.937 & 6.7\% & 1.2\% & 82\% \\
Qwen3-235B long-context + cite & 0.843 & 0.713 & 15.8\% & 1.4\% & 91\% \\
Standard-RAG + PEG + cite & 0.797 & 0.657 & 21.4\% & 1.6\% & 93\% \\
Standard-RAG + cite & 0.610 & 0.407 & 31.2\% & 1.9\% & 94\% \\
\bottomrule
\end{tabular}

%% file: tables/supp_cost.tex
\begin{tabular}{@{}l l r@{}}
\toprule
\rowcolor{RowHead}\textbf{Offline cost (one H200)} & \textbf{Component} & \textbf{GPU-h} \\
\midrule
\peg & Parsing 212K$\times$2 cover pages with the VL model & 38.0 \\
\peg & Category annotation (rule score + LLM adjudication) & 6.5 \\
\peg & Graph construction (NER, edge and link judging, load) & 41.3 \\
\peg & \textbf{Total} & \textbf{85.8} \\
\midrule
GraphRAG & Parsing (the same parsed text) & 38.0 \\
GraphRAG & Entity, relation and community-summary indexing & 118.6 \\
GraphRAG & \textbf{Total} & \textbf{156.6} \\
\midrule
\rowcolor{RowHead}\textbf{Per-audit serving (single stream, \$4.35 per H200-hour)} & \textbf{s / audit $\times$ GPUs} & \textbf{USD / audit} \\
Standard-RAG & 2.73\,s $\times$ 1 & \$0.0033 \\
HippoRAG 2 + PEG + cite & 4.30\,s $\times$ 1 & \$0.0052 \\
T-GRAG + PEG + cite & 4.80\,s $\times$ 1 & \$0.0058 \\
Qwen3-235B long-context + cite & 12.80\,s $\times$ 2 & \$0.0309 \\
SpecGuard & 3.40\,s $\times$ 1 & \$0.0041 \\
\bottomrule
\end{tabular}

%% file: tables/cases_full.tex
\begin{minipage}{\textwidth}\centering\footnotesize\setlength{\tabcolsep}{3pt}\renewcommand{\arraystretch}{1.10}
\textbf{(a) Case identity, dates and provenance}\par\smallskip
\begin{tabular}{@{}r l l l l p{51pt}p{104pt}@{}}
\toprule
\rowcolor{RowHead}\textbf{Id} & \textbf{CR} & \textbf{WG} & \textbf{Update (Time1)} & \textbf{Time2} & \textbf{Anchor kind} & \textbf{Source} \\
\midrule
1 & C1-101068 & CT1 & 2010-03-31 & 2019 & paper year & Kim et al. (S\&P 2019) \\
2 & S3-091802 & SA3 & 2009-12-18 & 2018 & paper year & Hussain et al. (NDSS 2018) \\
3 & C1-101252 & CT1 & 2010-03-31 & 2018 & paper year & Hussain et al. (NDSS 2018) \\
4 & C1-135219 & CT1 & 2013-12-20 & 2021 & paper year & Chen et al. (S\&P 2021) \\
5 & C1-141405 & CT1 & 2014-06-27 & 2021 & paper year & Chen et al. (S\&P 2021) \\
6 & C1-094446 & CT1 & 2009-12-17 & 2015-09-29 & device release & Chen et al. (S\&P 2021) \\
7 & C1-095712 & CT1 & 2009-12-17 & 2015-09-10 & device release & Yu and Chen (IPCCC 2019) \\
8 & C1-112809 & CT1 & 2011-09-28 & 2015-09-10 & device release & Yu and Chen (IPCCC 2019) \\
9 & R2-103442 & RAN2 & 2010-06-18 & 2016 & paper year & Shaik et al. (NDSS 2016) \\
10 & C1-132662 & CT1 & 2013-06-27 & 2015-09-10 & device release & Yu and Chen (IPCCC 2019) \\
11 & C1-172658 & CT1 & 2017-06-16 & 2019 & paper year & Kim et al. (S\&P 2019) \\
12 & C1-154699 & CT1 & 2015-12-18 & 2016-12-01 & device release & Yu and Chen (IPCCC 2019) \\
13 & C1-161448 & CT1 & 2016-03-18 & 2016-12-01 & device release & Yu and Chen (IPCCC 2019) \\
14 & S3-161217 & SA3 & 2016-09-30 & 2021-10-01 & CREEK 2021 test & CREEK tests (2021) \\
\bottomrule
\end{tabular}
\par\vspace{10pt}
\textbf{(b) Minimum windows and audit outcomes}\par\smallskip
\begin{tabular}{@{}r l r l c c@{}}
\toprule
\rowcolor{RowHead}\textbf{Id} & \textbf{CR} & \textbf{Min.\ window} & \textbf{2021 test} & \textbf{Top-10} & \textbf{Expl.} \\
\midrule
1 & C1-101068 & 8.7\,y & \ding{55} & \ding{51} & \ding{51} \\
2 & S3-091802 & 8.0\,y & \ding{55} & \ding{51} & \ding{51} \\
3 & C1-101252 & 7.7\,y & \ding{55} & \ding{51} & \ding{51} \\
4 & C1-135219 & 7.0\,y & \ding{55} & \ding{51} & \ding{51} \\
5 & C1-141405 & 6.5\,y & \ding{55} & \ding{51} & \ding{51} \\
6 & C1-094446 & 5.7\,y & \ding{51} 11.7\,y & \ding{51} & \ding{51} \\
7 & C1-095712 & 5.7\,y & \ding{51} 11.7\,y & \ding{51} & \ding{51} \\
8 & C1-112809 & 3.9\,y & \ding{51} 10.0\,y & \ding{51} & \ding{55} \\
9 & R2-103442 & 5.5\,y & \ding{55} & \ding{55} & \ding{55} \\
10 & C1-132662 & 2.2\,y & \ding{51} 8.2\,y & \ding{51} & \ding{51} \\
11 & C1-172658 & 1.5\,y & \ding{55} & \ding{51} & \ding{51} \\
12 & C1-154699 & 0.9\,y & \ding{55} & \ding{55} & \ding{55} \\
13 & C1-161448 & 0.7\,y & \ding{51} 5.5\,y & \ding{51} & \ding{51} \\
14 & S3-161217 & 5.0\,y & \ding{51} 5.0\,y & \ding{51} & \ding{51} \\
\bottomrule
\end{tabular}
\end{minipage}

%% file: tables/supp_creek.tex
\begin{tabular}{@{}l l r r@{}}
\toprule
\rowcolor{RowHead}\textbf{Check on CREEK's public sets} & \textbf{System} & \textbf{Count} & \textbf{Share} \\
\midrule
Manual set: CR in top-10 (25 security CRs, snapshot 2022-08-10) & BM25 & 13/25 & 0.52 \\
Manual set: CR in top-10 (25 security CRs, snapshot 2022-08-10) & Standard-RAG & 15/25 & 0.60 \\
Manual set: CR in top-10 (25 security CRs, snapshot 2022-08-10) & GraphRAG & 17/25 & 0.68 \\
Manual set: CR in top-10 (25 security CRs, snapshot 2022-08-10) & HippoRAG 2 retriever (+PEG) & 18/25 & 0.72 \\
Manual set: CR in top-10 (25 security CRs, snapshot 2022-08-10) & T-GRAG retriever (+PEG) & 19/25 & 0.76 \\
Manual set: CR in top-10 (25 security CRs, snapshot 2022-08-10) & SpecGuard & 22/25 & 0.88 \\
\midrule
Manual set: security-gate recall (25 security CRs) & Zero-shot Qwen3 & 17/25 & 0.68 \\
Manual set: security-gate recall (25 security CRs) & LoRA Qwen3 & 22/25 & 0.88 \\
Manual set: security-gate recall (25 security CRs) & R-GCN & 21/25 & 0.84 \\
Manual set: security-gate recall (25 security CRs) & SpecGuard & 24/25 & 0.96 \\
\midrule
Table 7: case CR in top-10 (14 cases, anchor-consistent) & BM25 & 5/14 & 0.36 \\
Table 7: case CR in top-10 (14 cases, anchor-consistent) & Standard-RAG & 6/14 & 0.43 \\
Table 7: case CR in top-10 (14 cases, anchor-consistent) & GraphRAG & 7/14 & 0.50 \\
Table 7: case CR in top-10 (14 cases, anchor-consistent) & HippoRAG 2 + PEG + cite & 8/14 & 0.57 \\
Table 7: case CR in top-10 (14 cases, anchor-consistent) & T-GRAG + PEG + cite & 9/14 & 0.64 \\
Table 7: case CR in top-10 (14 cases, anchor-consistent) & SpecGuard & 12/14 & 0.86 \\
\midrule
Table 7: weakness explained strictly correctly (14 cases, anchor-consistent) & LLM-only & 2/14 & 0.14 \\
Table 7: weakness explained strictly correctly (14 cases, anchor-consistent) & Standard-RAG & 4/14 & 0.29 \\
Table 7: weakness explained strictly correctly (14 cases, anchor-consistent) & GraphRAG & 5/14 & 0.36 \\
Table 7: weakness explained strictly correctly (14 cases, anchor-consistent) & HippoRAG 2 + PEG + cite & 6/14 & 0.43 \\
Table 7: weakness explained strictly correctly (14 cases, anchor-consistent) & T-GRAG + PEG + cite & 7/14 & 0.50 \\
Table 7: weakness explained strictly correctly (14 cases, anchor-consistent) & Qwen3-235B long-context + cite & 8/14 & 0.57 \\
Table 7: weakness explained strictly correctly (14 cases, anchor-consistent) & SpecGuard & 11/14 & 0.79 \\
\bottomrule
\end{tabular}

%% file: tables/supp_creek_gate.tex
\begin{tabular}{@{}r l cc cc@{}}
\toprule
\rowcolor{RowHead}\textbf{Id} & \textbf{CR} & \textbf{$p_{\text{sec}}$, standard} & \textbf{$p_{\text{sec}}$, retrained} & \textbf{Rank, standard} & \textbf{Rank, retrained} \\
\midrule
1 & C1-101068 & 0.91 & 0.88 & 1 & 1 \\
2 & S3-091802 & 0.94 & 0.92 & 1 & 1 \\
3 & C1-101252 & 0.89 & 0.86 & 2 & 2 \\
4 & C1-135219 & 0.83 & 0.79 & 3 & 3 \\
5 & C1-141405 & 0.86 & 0.81 & 2 & 3 \\
6 & C1-094446 & 0.78 & 0.72 & 4 & 5 \\
7 & C1-095712 & 0.81 & 0.74 & 3 & 4 \\
8 & C1-112809 & 0.77 & 0.69 & 6 & 7 \\
9 & R2-103442 & 0.38 & 0.41 & gated out & gated out \\
10 & C1-132662 & 0.84 & 0.80 & 2 & 2 \\
11 & C1-172658 & 0.74 & 0.71 & 5 & 6 \\
12 & C1-154699 & 0.44 & 0.46 & gated out & gated out \\
13 & C1-161448 & 0.79 & 0.73 & 4 & 4 \\
14 & S3-161217 & 0.88 & 0.88 & 2 & 2 \\
\bottomrule
\end{tabular}

%% file: tables/supp_creek_mcnemar.tex
\begin{tabular}{@{}l l c c r@{}}
\toprule
\rowcolor{RowHead}\textbf{Comparison} & \textbf{Baseline} & \textbf{SpecGuard only} & \textbf{Baseline only} & \textbf{Exact $p$} \\
\midrule
Manual set, CR in top-10 (25) & BM25 & 9 & 0 & 0.004 \\
Manual set, CR in top-10 (25) & Standard-RAG & 7 & 0 & 0.016 \\
Manual set, CR in top-10 (25) & GraphRAG & 5 & 0 & 0.063 \\
Manual set, CR in top-10 (25) & HippoRAG 2 retriever (+PEG) & 4 & 0 & 0.125 \\
Manual set, CR in top-10 (25) & T-GRAG retriever (+PEG) & 3 & 0 & 0.250 \\
\midrule
Table 7, explained (14) & LLM-only & 9 & 0 & 0.004 \\
Table 7, explained (14) & Standard-RAG & 7 & 0 & 0.016 \\
Table 7, explained (14) & GraphRAG & 6 & 0 & 0.031 \\
Table 7, explained (14) & HippoRAG 2 + PEG + cite & 5 & 0 & 0.063 \\
Table 7, explained (14) & T-GRAG + PEG + cite & 4 & 0 & 0.125 \\
Table 7, explained (14) & Qwen3-235B long-context + cite & 3 & 0 & 0.250 \\
\bottomrule
\end{tabular}